\documentclass[french, english,a4paper,11pt]{article}
\usepackage{jcappub} 

\usepackage[T1]{fontenc}
\usepackage[utf8]{inputenc}
\usepackage{geometry}
\usepackage{babel}
\makeatletter
\addto\extrasfrench{%
   \providecommand{\fg}{\ifdim\lastskip>\z@\unskip\fi~\frqq}%
}

\makeatother
\usepackage{mathrsfs}
\usepackage{amsmath}
\usepackage{amssymb}
\usepackage{esint}
\usepackage{color}
\usepackage{float} 
\usepackage{soul}

\makeatletter
\gdef\@fpheader{}
\makeatother

\title{Cosmic decoherence: primordial power spectra and non-Gaussianities}

\author[a,b]{Aoumeur Daddi Hammou,}
\author[b,c,d]{Nicola Bartolo}

\affiliation[a]{SUBATECH, IMT Atlantique, Nantes Universit\'e, IN2P3/CNRS, \\ Alfred Kastler 04, 44307 Nantes, France}
\affiliation[b]{Dipartimento di Fisica e Astronomia  ``G. Galilei'', Universit\`a degli Studi di Padova,\\Via Marzolo 8, I-35131 Padova, Italy}
\affiliation[c]{INFN, Sezione di Padova,\\Via Marzolo 8, I-35131 Padova, Italy}
\affiliation[d]{INAF - Osservatorio Astronomico di Padova, \\ Vicolo dell’Osservatorio 5, I-35122 Padova, Italy}

\emailAdd{daddiham@subatech.in2p3.fr}
\emailAdd{nicola.bartolo@pd.infn.it}

\vspace{2cm}
\abstract {We study the effect of quantum decoherence on the inflationary cosmological perturbations.  This process might imprint specific observational signatures revealing  the quantum nature of the inflationary mechanism being related to the longstanding issue of the quantum-to-classical  transition of inflationary fluctuations. 
 Several works have investigated the effect of quantum decoherence on the statistical properties of primordial fluctuations.  In particular, it has been shown that cosmic decoherence leads to corrections to the curvature power spectrum predicted by standard slow-roll inflation. 
Equally interesting,  a non zero curvature trispectrum has been shown to be purely induced by cosmic decoherence but,  surprisingly, decoherence seems not to generate any bispectrum. We further develop such an analysis by adopting a generalized form of the pointer observable, showing that decoherence does induce a non vanishing curvature bispectrum and providing a specific underlying concrete physical process. Present constraints on primordial bispectra allow to put an upper bound on the strength  of the environment-system interaction. In full generality, the decoherence-induced bispectrum can be scale dependent provided one imposes the corresponding correction to the power spectrum to be scale independent. Such scale dependence on the largest cosmological scales might represent a distinctive imprint of the quantum decoherence process taking place during inflation.
We also provide a criterion that allows to understand when cosmic decoherence induces scale independent corrections, independently of the type of environment considered. As a final result, we study the effect of cosmic decoherence on tensor perturbations and we derive the decoherence corrected tensor-to-scalar perturbation ratio. In specific cases,  decoherence induces a blue tilted correction to the standard tensor power spectrum.
 
}

\begin{document}

\maketitle

\section{ Introduction} 
{{Inflation theory was conceived to solve the shortcomings
of the standard Hot Big-Bang model, as the flatness and
horizon ``problems''~\cite{Guth:1980zm,Albrecht:1982wi,Linde:1983gd}. Further, it was realized that it provides  a
mechanism which explains the origin of all the anisotropies and inhomogeneities
we are observing nowadays, namely the Cosmic Microwave Background (CMB) anisotropies and the large-scale
structure of the Universe (LSS), tracing them back to tiny primordial quantum
fluctuations (of one or more scalar fields) during the early universe~\cite{Kirzhnits:1972ut, mukhanov1981quantum,Mukhanov:1982nu, Starobinsky:1982ee,Hawking:1982cz, Bardeen:1983qw, Guth:1982ec}.  Similarly, inflation predicts 
through the same mechanism a relic background of primordial
gravitational waves \cite{Starobinsky:1979ty,Grishchuk:1974ny, Rubakov:1982df,Abbott:1984fp,Fabbri:1983us,Allen:1987bk}. However, the adoption of the inflationary scenario of a quantum origin for the universe is still facing two puzzles that need to be cleaned away in order to have a self consistent  formalism for the study of primordial perturbations \cite{polarski1996semiclassicality,perez2006quantum,kiefer2022quantum}.

{It was shown in \cite{grishchuk1990squeezed} that the initial, Bunch-Davies, vacuum state of the primordial fluctuations evolve into a highly  squeezed quantum state. As consequence of the unitary dynamics that preserves the symmetries of initial state,  the squeezed quantum state still represents a homogeneous and isotropic universe \cite{,castagnino2017interpretations,martin2021collapse}. Therefore, we may ask how the inhomogeneous and anisotropic universe we observe nowadays emerged from an initial homogeneous and isotropic one?   

Second, we observe a classical universe rather than a quantum one, where we do not observe any of the distinctive properties of quantum mechanics, such as superposition of states implied by the highly quantum squeezed state into which the primordial fluctuations state evolve. Hence, how did the classical universe emerged? 

A quantitative answer to those questions, could shed light on  new aspects of the primordial fluctuations providing us with new probes of the primordial era, which consequently, would lead to a better understanding of it. For example, they could bring new specific observational signatures able to reveal some aspects of the quantum processes characterizing the emergence of cosmological perturbations from inflation.

In the quest for an answer to the raised questions above, it was shown that in the large squeezing limit,   the predictions   inferred   out of  quantum processes are indistinguishable from  those of  classical stochastic processes \cite{Guth:1985ya, albrecht1994inflation, polarski1996semiclassicality, kiefer2009cosmological}. However, such equivalence does only  justifies  the analysis of cosmological data by relying purely on classical techniques, without answering  the question of quantum to-classical transition in the early universe. 

In  the aim of studying  the classicalization of the primordial universe, and find  a signature of its quantum origin, many models have been suggested. Interestingly, cosmology offers a valubale arena to test and constrain
those models on completely different physical scales compared to
laboratories experiments \cite{Sakagami:1987mp,polarski1996semiclassicality,Lesgourgues:1996jc,Egusquiza:1997ez,kiefer1998quantum,kiefer1998emergence,perez2006quantum,Campo:2005sv,lombardo2005decoherence,bolis2016modifications,martineau2007decoherence,Burgess:2006jn,kiefer2009cosmological,Valentini:2008dq,Koksma:2010dt,sudarsky2011shortcomings,Bassi:2010ss,Pinto-Neto:2013npa,martin2012cosmological,Das:2013qwa,albrecht2014cosmological,maldacena2016model,Martin:2015qta,nelson2016quantum,martin2016bell,martin2016leggett,martin2017obstructions,martin2018observational,martin2018non,ye2018quantum,gong2019quantum,bolis2019non,Ashtekar:2020mdv,kiefer2022quantum, Gomez:2020xdb,Gomez:2020otn}.  Among the many phenomenological models suggested so far, quantum decoherence emerges as a leading one, since, it is a well-established and physically
tested concept \cite{Brune:1996zz,kwiat2000experimental,schlosshauer2006experimental, schlosshauer2007decoherence,takahashi2011decoherence, vepsalainen2020impact,kiefer2022quantum}.

Quantum decoherence lays on the fact that a quantum system is not a closed but, rather, an open system interacting with its environment \cite{Zurek:1981xq,Joos:1984uk,Schlosshauer:2003zy,breuer2002theory,joos2013decoherence,rivas2012open}.
 In cosmic decoherence, there are two main choices of the environment which have been investigated so far. The first is  to consider that super-horizon, or light,  modes of fluctuations to be decohered by sub-horizon, or heavy, modes  representing the environment \footnote{ Light modes represent our system, since they are the ones of observational interest for us. See however \cite{Burgess:2006jn, maldacena2016model}   for a critical discussions about some short-comings of this approach.}  \cite{lombardo2005decoherence,Burgess:2006jn,martineau2007decoherence,nelson2016quantum,gong2019quantum,burgess2022minimal}. The second choice is to consider an external field, such that  decoherence is caused by field-field interactions, rather than by self interactions of the same field as in first case (i.e. by a long-short splitting of the inflationary fluctuations).  Among the works which adopted such choice, we find \cite{boyanovsky2015effective,liu2016cosmic,martin2018observational,martin2018non} (see also \cite{Raveendran:2022dtb}), in addition to our current work. Such choice is motivated by the presence of more than one field in the early universe, at least, those  populating radiation era.
The system-environment interaction leads mainly to the suppression of off diagonal elements in the reduced density matrix  of the system. However,  decoherence   occurs over a finite time scale, therefore,
the environment does also affect the diagonal elements of the density
matrix. Consequently, it alters the probabilities associated to the
final outcomes of measurement~\cite{martineau2007decoherence,Burgess:2006jn, boyanovsky2015effective,martin2018observational}. 

Indeed,  several works  showed that cosmic decoherence does affect the statistical properties of the primordial perturbations. Such corrections are shown to be essential to account for the quantum to classical transition in the early universe, in addition, they carry observational signatures of a quantum origin for the primordial fluctuations \cite{lombardo2005decoherence,boyanovsky2015effective,nelson2016quantum,martin2018observational,martin2018non}. In particular, our work  builds up, extends, and to some extent completes, works done in \cite{boyanovsky2015effective, martin2018observational, martin2018non}.

In this work, we present a generalized model to deal with cosmic decoherence of both curvature and tensor perturbations. In particular, we study the  power spectra corrections in addition to non-Guassianities induced by decoherence. The results obtained for the different point correlation functions can be exploited to constrain the strength of the interaction between the system and the environment.\footnote{Some constraints were obtained, e.g., in \cite{martin2018observational, martin2018non} by using curvature power spectrum and tripsectrum.} Hence, this would allow to constrain the possible environments, and to reveal some properties of the inflation era.

The paper is organized as follows. In section
\ref{sec:The-Lindblad-equation} we present,  briefly, the Lindblad equation
which has been derived previously in several papers for cosmological applications. Then, we discuss
in section \ref{sec:J.Martin-et-al} the cosmic decoherence of curvature
fluctuations by providing a short summary of the approach followed by
J.Martin and V.Vennin, in \cite{martin2018observational, martin2018non}, and of their main results on both curvature
power spectrum and non-Gaussianities.  In section \ref{sec:Our-approach} we
 present our approach that generalizes the previous one, and present a concrete  physical process that  motivate it. In order to shed light on some of the many interesting results implied by our approach,  we  derive the bispectrum expression and we discuss its specific 
scale-dependence, emphasizing the fact that such bispectrum is induced purely by the decoherence phenomenon, rather than being merely  a  correction to the leading bispectrum derived within standard inflation theory.~\footnote{ In standard single field slow-roll inflation, where primordial fluctuations are considered as a closed system, curvature non-Gaussianities are a result of (gravitational) self-interaction of curvature fluctuations manifested through cubic, and higher order, terms in the system action \cite{bartolo2004non, Maldacena:2002vr}.  } In doing that, we also provide a general criterion to understand when one can achieve scale independence of the various decoherence induced corrections to the power spectrum, as well of the induced non-Gaussianities. We conclude this section by showing how our generalized approach can reproduce all the
results derived with previous approaches, while adding at the same time some corrections which were missed due to 
the restricted form of pointer observable previously adopted. Besides the analysis dealing with the curvature perturbations, in section
\ref{sec:Decoherence-of-tensor-1} we study the effect of cosmic decoherence
on tensor perturbations, and derive the induced correction to their
power spectrum and the corresponding modification to the tensor-to-scalar perturbation ratio $r$. 
Finally, we draw our conclusions in section~\ref{Conclusions}. Appendix \ref{sec:Appendix} contains various technical details of the bispectrum computation.


{\par}

\section{The Lindblad equation\label{sec:The-Lindblad-equation} }
{Since we are about to discuss inflationary fluctuations we need to
go beyond the homogeneous and isotropic Friedmann-Lemaitre-Robertson-Walker
(FLRW) metric. 
In what follows we will focus for the moment on scalar perturbations which, up to first order, decouple from the tensor ones. 
The perturbed FLRW metric is given by \cite{mukhanov1992theory} 

\begin{equation}
\mathrm{d}s^{2}=a^{2}\left(\eta\right)\left\{ -\left(1-2\phi\right)\mathrm{d}\eta^{2}+2\partial_{i}B\mathrm{d}x^{i}\mathrm{d}\eta+\left[\left(1-2\psi\right)\delta_{ij}+\partial_{ij}E\right]\mathrm{d}x^{i}\mathrm{d}x^{j}\right\}\, ,
\end{equation}
 where  $a$ is the FLRW scale factor, and  $\phi$, $B$, $\psi$, $E$ are functions of space and time.
Thanks to gauge freedom, the number of degrees of freedom can be
reduced, such that, e.g., the scalar gravitational perturbations can be encoded into two gauge-invariant Bardeen potentials~\cite{bardeen1980gauge},  
such as the one related to the spatial metric perturbation
\begin{equation}
\Psi=\psi -\frac{a'}{a} \left( B-E^{\prime}\right)\, . 
\end{equation}
where a prime denotes a derivative with respect to the conformal time $\eta$.

Besides the geometric part of perturbations, we turn now
to matter sector. The inflaton is decomposed into a classical background
$\varphi_{0}\left(\eta\right)$ plus a fluctuation $\delta\varphi\left(\eta,\boldsymbol{x}\right)$
to be quantized later. The latter, i.e $\delta\varphi\left(\eta,\boldsymbol{x}\right)$,
can be treated as a massless scalar field, which is an excellent approximation
when the inflaton field satisfies the slow-roll conditions \cite{kiefer2009cosmological}
\begin{equation}
\varphi\left(\eta,\boldsymbol{x}\right)=\varphi_{0}\left(\eta\right)+\delta\varphi\left(\eta,\boldsymbol{x}\right)\, .
\end{equation}
Similarly, the scalar field perturbations can be conveniently described by a gauge invariant variable, namely,
\begin{equation}
\delta\varphi^{gi}\left(\eta,\boldsymbol{x}\right)=\delta\varphi+\varphi_{0}^{\prime}\left(B-E^{\prime}\right)\, . 
\end{equation}
The Bardeen potential and the gauge invariant field can be combined into a single gauge-invariant scalar variable called Mukhanov-Sasaki (MS) variable  \cite{kodama1984cosmological,mukhanov1981quantum}
\begin{equation}
v\left(\eta,\boldsymbol{x}\right)=a\left[\delta\varphi^{gi}\left(\eta,\boldsymbol{x}\right)+\frac{\varphi_{0}^{\prime}}{\mathcal{H}}\Psi \right],
\end{equation}
where $\mathcal{H}=a^{\prime}/a$ represents the conformal Hubble parameter; the MS variable $v$ characterizes fully the scalar sector, being  related to the comoving curvature perturbation $\mathcal{\zeta}$ through \cite{mukhanov1992theory,Martin:1997zd,lyth1985large}
\textbf{
\begin{equation}
v\left(\eta,\boldsymbol{x}\right)\mathcal{=}- \frac{a\varphi_{0}^{\prime}}{\mathcal{H}}\mathcal{\mathcal{\zeta}}\, .\label{eq:MS variable curvature}
\end{equation}
}}

{In the simplest view of the standard inflation paradigm, the primordial quantum perturbations 
are considered as a closed system. The gravitational and matter
degrees of freedom are encoded in field operators on the Hilbert space
of physical states, and are governed by the standard quantum mechanics. }{\par}

{The curvature perturbations that describe jointly the inflaton
fluctuations in addition to the scalar metric perturbations are governed,
at the perturbative linear order, by the free Hamiltonian} \footnote{{ For the sake of studying the non-Gaussianities induced purely by cosmic decoherence, i.e by the interaction with an external
environment, we are considering single-field slow roll inflation, and neglecting self-interactions of the curvature perturbation field.}}{, written in Fourier space,
\begin{equation}
\hat{H}_{v}=\frac{1}{2}\int d^{3}\boldsymbol{k}\left[\hat{p}_{\boldsymbol{k}}\hat{p}_{\boldsymbol{k}}^{*}+\omega^{2}\left(\eta,k\right)\hat{v}_{\boldsymbol{k}}\hat{v}_{\boldsymbol{k}}^{*}\right]\,,\label{eq:1-1}
\end{equation}
where $\hat{v}$$_{\boldsymbol{k}}$ is the Fourier transform of the quantized Mukhanov-Sasaki variable~\eqref{eq:MS variable curvature} and $\hat{p}_{\boldsymbol{k}}$ is its conjugate momentum,
i.e $\hat{p}_{\boldsymbol{k}}=\hat{v}_{\boldsymbol{k}}^{\prime}$. {\par} $\hat{H}_{v}$ represents a collection of parametric oscillators (one
per each mode) with time dependent frequency given by 
\begin{equation}
\omega^{2}\left(\eta,k\right)=k^{2}-\frac{\left(\sqrt{\epsilon}a\right)^{\prime\prime}}{\sqrt{\epsilon}a}\label{eq:frequency}\, ,
\end{equation}
with $\epsilon$ being the first slow-roll parameter defined by $\epsilon=1-\mathcal{H}^{\prime}/\mathcal{H}^{2}$, which we will assume to be constant throughout this work. 
In what follows, it will prove to be convenient to decompose the MS variable into its real and imaginary part as $\hat{v}_{\boldsymbol{k}}=\left( \hat{v}_{\boldsymbol{k}}^{R}+\hat{v}_{\boldsymbol{k}}^{I}\right) /\sqrt{2}$. {\par}

{ Obviously the Hamiltonian (\ref{eq:1-1}) describes
a closed system, since it does not contain any interaction term. However,
it is unlikely in reality for the primordial perturbations to be non-interacting, 
at least gravitationally, with the other degrees of freedom
present in the early universe, possibly including the standard model fields,
needed to give rise to radiation era \cite{martin2018observational,boyanovsky2015effective,sudarsky2011shortcomings}.
Therefore, one should consider the inflationary perturbations rather as 
an open quantum system interacting with a given environment.  On the other hand, by 
doing so, we will face the problem of being unable to describe the exact properties of all the
possible degrees of freedom composing the environment. In addition,
we are interested, mainly, on the effect of the environment on the system
of interest, rather than the environment itself. It thus prove convenient to 
trace out the environmental degrees of freedom during the derivation
of the master equation for the system, such that the effect of the environment will be
encoded in a new term added to the Liouville equation, leading us
to the quantum master equation governing the evolution of the system, namely the Lindblad equation.}   {\par}


{The Lindblad equation is based on the Born and Markov approximations  \cite{Lindblad:1975ef,gorini1976completely,breuer2002theory,schlosshauer2007decoherence}
which could be summarized as follow \cite{sears1989neutron,Burgess:2006jn,martin2018observational}:}{\par}
\begin{enumerate}
\item {The environment evolves on a time scale that is much smaller
than that of the system.}{\par}
\item {The backreaction of the system on the environment is negligible.}{\par}
\item {The influence of the environment on the system, that is here
clearly crucial, can be treated perturbatively. }{\par}
\end{enumerate}

{Generally, within a quantum decoherence context, the total Hamiltonian
of the composite system, i.e system+environment, living in the Hilbert
space $\mathcal{E=E}_{sys}\otimes\mathcal{E}_{E}$ is given by
\begin{equation}
\hat{H}=\hat{H}_{0}+\hat{H}_{int}=\hat{H}_{sys}\otimes\hat{I}_{E}+\hat{I}_{sys}\otimes\hat{H}_{E}+g\hat{H}_{int}\,,\label{eq:3}
\end{equation}
where $\hat{H}_{sys}$ is the intrinsic Hamiltonian of the system
acting on the Hilbert space $\mathcal{E}_{sys}$, in our case given by 
$\hat{H}_{v}$ in (\ref{eq:1-1}), $\hat{H}_{E}$ is the free
evolution Hamiltonian of the environment that acts on $\mathcal{E}_{E}$
and it could be left unspecified. Finally, $\hat{H}_{int}$ is the
interaction Hamiltonian and $g$ is a dimensionless coupling constant.
{Following~ {\cite{Burgess:2006jn,martineau2007decoherence,Shandera:2017qkg} and, mainly, \cite{martin2018observational}} } 
we specialize to the case of interest for field theory, where interactions
are local. Suppose, then, that system and environment interact through
interactions of the local form such that $\hat{H}_{int}$ could be
written as 
\begin{equation}
\hat{H}_{int}\left(\eta\right)=\int d^{3}\boldsymbol{x}\hat{A}\left(\eta,\boldsymbol{x}\right)\otimes\hat{R}\left(\eta,\boldsymbol{x}\right)\,,\label{eq:135}
\end{equation}
where $\hat{A}$ denotes a local functional of the fields describing
the system sector, and similarly for $\hat{R}$ for the environment sector. According to the commutativity criterion, the
decoherence superselected pointer basis} (or preferred basis)}}{
is dictated by the system sector part of $\hat{H}_{int}\left(\eta\right)$,
where it states that the preferred basis is \footnote{This applies, at least, in the Quantum-Measurement limit of decoherence.}} the eigenstates of $\hat{A}\left(\eta,\boldsymbol{x}\right)$;
stated differently but equivalently, the pointer observables of the
system are those observables commuting with the interaction Hamiltonian~\cite{Zurek:1981xq,PhysRevD.26.1862,schlosshauer2007decoherence}.
However, in our case we will approach the problem in the reverse way, where
based on the fact that CMB map is localized in the field amplitude
of Mukhanov-Sasaki variable $\hat{v}$, we conclude that $\left|v_{\boldsymbol{k}}\right\rangle $
constitutes the pointer basis \cite{kiefer2007pointer}. Therefore,
using the commutativity criterion in the reverse way, we conclude
that $\hat{A}\left(\eta,\boldsymbol{x}\right)$ involves only $\hat{v}$
i.e 
\begin{equation}
\hat{A}\equiv\hat{A}\left[\hat{v}^{n}\right]\,,
\end{equation}
with $n$ an integer. However, the computations that will be made
all throughout this paper could also be applied for the case where
$\hat{A}$ is a function of field momentum $\hat{p}$. Actually, there
is an other reason behind neglecting a dependence of $\hat{A}$ on
the field momentum $\hat{p}$ and it consists of the fact that $\hat{p}$
is proportional to the decaying mode, therefore, any contribution
from $\hat{p}$ would be subdominant compared to that from $\hat{v}$
(see \cite{martin2018observational} for more details).{\par}

{The reduced density matrix of the system $\hat{\rho}_{sys}\left(t\right)$
is defined by tracing out the environmental degrees of freedom as }{\par}

{
\begin{equation}
\hat{\rho}_{sys}\left(t\right)={\rm Tr}_{E}\left[\hat{\rho}\left(t\right)\right]\,.
\end{equation}
 Exploiting the aforementioned assumptions, one can show that $\hat{\rho}_{sys}\left(t\right)$
evolves according to the Lindblad equation\footnote{Also called Gorini-Kossakowski-Sudarshan-Lindblad equation.} \cite{Lindblad:1975ef, Burgess:2006jn, pearle2012simple, brasil2013simple, Shandera:2017qkg, martin2018observational} }}{\par} 

{
\begin{equation}
\frac{d\hat{\rho}_{sys}}{d\eta}=i\left[\hat{\rho}_{sys},\hat{H}_{sys}\right]-\frac{\gamma}{2}\int d^{3}\boldsymbol{x}d^{3}\boldsymbol{y}C_{R}\left(\boldsymbol{x},\boldsymbol{y}\right)\left[\left[\hat{\rho}_{sys},\hat{A}\left(\boldsymbol{x}\right)\right],\hat{A}\left(\boldsymbol{y}\right)\right]\,,\label{eq:5}
\end{equation}
with\footnote{We refer to the  second term in the right hand side of (\ref{eq:5}) as Lindblad term. It is also called dissipator in some references.} $\gamma=2g^{2}t_{c}$. Here $t_{c}$ is the autocorrelation
time of the environment, which, in order to assure the Markovianity
of the process, should be small with respect to the time over which
the system of interest evolves. On the other hand, $C_{R}\left(\boldsymbol{x},\boldsymbol{y}\right)$
represents the equal-time environmental correlation function of $\hat{R}$, and is defined by 
\begin{equation}
C_{R}\left(\boldsymbol{x},\boldsymbol{y}\right)={\rm Tr}_{E}\left(\rho_{E}R\left(\boldsymbol{x}\right)R\left(\boldsymbol{y}\right)\right)\, , 
\end{equation}
{where $\rho_{E}$ is the density operator of the environment}.  The parameter $\gamma$ is generally time dependent so we adopt for
it a power law dependence on the scale factor \cite{martin2018observational}
\begin{equation}
\gamma=\gamma_{*}\left(\frac{a}{a_{*}}\right)^{p}\,,\label{eq:coupling}
\end{equation}
where $p$ represents a free parameter, and $*$ refers to a reference
time that is taken to be the time when the pivot scale $k_{*}=0.051{\rm Mpc}^{-1}$
crosses the Hubble radius, i.e $k_{*}=a_{*}H_*$.

We need also to adopt a convention for the correlator $C_{R}\left(\boldsymbol{x},\boldsymbol{y}\right)$,
so we assume the environment to be statistically homogeneous, i.e
$C_{R}\left(\boldsymbol{x},\boldsymbol{y}\right)\propto\boldsymbol{x}-\boldsymbol{y}$,
and be in an isotropic configuration, i.e $C_{R}\left(\boldsymbol{x},\boldsymbol{y}\right)\propto\left|\boldsymbol{x}-\boldsymbol{y}\right|$. In addition, if we assume that it is characterized by a physical correlation length scale $l_{E}$ then $C_{R}\left(\boldsymbol{x},\boldsymbol{y}\right)$
must be a function of }{$a\left|\boldsymbol{x}-\boldsymbol{y}\right|/l_{E}$.~\footnote{{Remember that $\left(\boldsymbol{x},\boldsymbol{y}\right)$
are comoving coordinates.}} For
convenience, we assume it to be a top hat function \cite{martin2018observational}
\begin{equation}
C_{R}\left(\boldsymbol{x},\boldsymbol{y}\right)=\bar{C}_{R}\Theta\left(\frac{a\left|\boldsymbol{x}-\boldsymbol{y}\right|}{l_{E}}\right)\,,\label{eq:190}
\end{equation}
with 
\begin{equation}
\Theta\left(x\right)=\begin{cases}
1 & {\rm if}\,x<1\\
\\
0 & \mathrm{otherwise}
\end{cases}\,,
\end{equation}
and $\bar{C}_{R}$ is a constant. Such generic form of the environmental correlation function serves to capture the most relevant properties of the environment necessary to derive the master equation, in addition to the assumption about the environment short-range correlations. However, these  requirements still allow some freedom on the choice of correlator generic form. For instance, a good alternative to the form suggested is 
\begin{equation}
C_{R}\left(\boldsymbol{x},\boldsymbol{y}\right)=\bar{C}_{R}\exp\left(-\frac{a\left|\boldsymbol{x-y}\right|}{l_{E}}\right),
\end{equation}
where we still have the property of deacying correlations within a characteristic length $l_E$. Needless to mention that such choice rends the computations more complicated, so we will stick to the top hat correlator form in our work. Another property that the correlator could reflect is whether the environmental correlations are local in time as in our case and in \cite{Burgess:2003zw,burgess2015eft, martin2018observational,martin2018non}, or non local in time as in \cite{boyanovsky2015effective}.~\footnote{In \cite{boyanovsky2015effective} it has been shown that non-local  time environmental correlation functions could cause a tiny decay in curvature power spectrum upon horizon crossing of perturbation modes.}

The precise form of the correlation function will be left unspecified in the following, since it depends on the type of field(s) constituting the environment,  some examples  can be found in~\cite{boyanovsky2015effective,martin2018observational,Burgess:2006jn}.

{Before proceeding with the application of the Lindblad equation
it is important to mention, as a general property of it, that it is
valid only at leading order in $\gamma$~\cite{schlosshauer2007decoherence}.
This remark will be crucial for the results obtained, both,  by us and  by the previous works in the literature, as in \cite{Burgess:2006jn,  Boyanovsky:2015xoa,martin2018observational, martin2018non} and the papers cited therein.

For  a ciritical analysis about the validity of Lindblad equation (\ref{eq:5}) in the cosmological context, in addition to its link to non-perturbative methods, we refer the reader to  \cite{colas2022benchmarking}.}{\par}

\section{Previous approaches\label{sec:J.Martin-et-al} {\bf  }}

{ In dealing with cosmic decoherence, many previous works, as \cite{Burgess:2006jn, martineau2007decoherence,Boyanovsky:2015xoa,martin2018observational,martin2018non},  assumed a monomial pointer observable of  the form $\hat{A}=\hat{v}^{n}$, with $n$ being a given integer. In particular, J.Martin et al.  considered  in \cite{martin2018observational, martin2018non} the cases of a linear and quadratic pointer observable, i.e $\hat{A}=\hat{v}$ or $\hat{A}=\hat{v}^{2}$, respectively; we will summarize their main results in the following section.\footnote{They also presented a diagrammatic approach to derive the cosmic decoherence induced correction to the curvature power spectrum for any order $n$ of $\hat{A}=\hat{v}^{n}$, see \cite{martin2018observational} for more details. }  However, as a key remark of our paper, we show that the  form of the  pointer observable is crucial in studying the cosmic decoherence effects. Indeed, our approach is based on adopting a pointer observable of the most general form $\hat{A}=\sum_{n}\alpha^{n-1}\hat{v}^{n}$, up to a certain order $n$ with $\alpha$ being an expansion parameter. Such generalized form of the pointer observable  will not only reproduce  the results obtained by ~\cite{martin2018observational,martin2018non}, but it leads also to  corrections to those results in addition to new interesting outcomes regarding primordial non-Gaussianities}. \footnote{{ Within our approach, we will stop at the
leading order that gives a non vanishing correction to the correlation
function considered, except for the case of the power spectrum for which we will
consider also the next to leading order to show that new contributions arise w.r.t. to the results obtained in the previous approach.}} 

{To compute the correlation functions within cosmic decoherence,  there are two methods:} 
\begin{enumerate}
\item {We can solve the evolution equation (\ref{eq:5}) and
obtain an explicit expression of the reduced system density matrix $\hat{\rho}_{sys}$
, so that we use it to compute the correlation function $O=f\left[\hat{v},\hat{p}\right]$
through its expectation value 
\begin{equation}
O\equiv\left\langle \hat{O}\right\rangle ={\rm Tr}_{sys}\left(\hat{O}\hat{\rho}_{sys}\right)\,.\label{eq:200}
\end{equation}
However, unfortunately, in most of the cases it is highly challenging
to solve (\ref{eq:5}) exactly, the only case doable with reasonable easiness being the linear case $\hat{A}\propto\hat{v}$.
In that case, the Lindblad term is quadratic in $v$ just as the free
Hamiltonian is, and we obtain a Gaussian density matrix with a width
controlled by the environment \cite{martin2018observational}. }{\par}
\item {Apart from the linear case it is better to pursue the
second method in computing the correlation functions, which consists
in solving directly the equation of motion governing $\left\langle \hat{O}\right\rangle $.
Therefore, using (\ref{eq:5}) and (\ref{eq:200}) we obtain 
\begin{equation}
\frac{d\left\langle \hat{O}\right\rangle }{d\eta}=\left\langle \frac{\partial\hat{O}}{\partial\eta}\right\rangle -i\left\langle \left[\hat{O},\hat{H_{v}}\right]\right\rangle -\frac{\gamma}{2}\int d^{3}\boldsymbol{x}d^{3}\boldsymbol{y}C_{R}\left(\boldsymbol{x},\boldsymbol{y}\right)\left\langle \left[\left[\hat{O},\hat{A}\left(\boldsymbol{x}\right)\right],\hat{A}\left(\boldsymbol{y}\right)\right]\right\rangle \,.\label{eq:201}
\end{equation}
}{\par}
\end{enumerate}

\subsection{Corrected curvature power spectrum}

{As we just mentioned, for the case of a linear interaction with environment, $\hat{A}\left(\boldsymbol{x}\right)=$$\hat{v}\left(\boldsymbol{x}\right)$,
there are two equivalent ways to determine the curvature power spectrum.
The first consists in solving equation (\ref{eq:5}) to get the explicit
expression of the reduced density matrix to be used in computing the curvature power spectrum 
\begin{equation}
P_{vv}\left(k\right)=\left\langle \left|\hat{v}_{\boldsymbol{k}}\right|^{2}\right\rangle =\left\langle \left(\hat{v}_{\boldsymbol{k}}^{s}\right)^{2}\right\rangle ={\rm Tr}_{sys}\left[\left(\hat{v}_{\boldsymbol{k}}^{s}\right)^{2}\hat{\rho}_{sys}\right]=\int\mathrm{d}\hat{v}_{\boldsymbol{k}}^{s}\left\langle v_{\boldsymbol{k}}^{s}\right|\hat{\rho}_{\boldsymbol{k}}^{s}\left|v_{\boldsymbol{k}}^{s}\right\rangle \left(\boldsymbol{\hat{v}}_{\boldsymbol{k}}^{\boldsymbol{s}}\right)^{2},
\end{equation}
with, $s=R,I$, standing for the real and imaginary part of $\hat{v}_{\boldsymbol{k}}$, and summation upon repeated indices is adopted. However this method is applicable only at linear order in ${\bf \hat{v}}$, where (\ref{eq:5})
could be solved exactly as shown in \cite{martin2018observational},
so for all higher order interactions we need the second method which
consists in solving directly (\ref{eq:201}) to get the correlator
expression, i.e. in this case $\left\langle \hat{O}\right\rangle =\left\langle \left|\hat{v}_{\boldsymbol{k}}\right|^{2}\right\rangle $. Hence, in what follows, we will restrict ourselves to the second method. The interested reader is referred to \cite{martin2018observational} to see the equivalence of the two methods at the linear order $\hat{A}\propto\hat{v}$. Using the second method yields \cite{martin2018observational} 
\begin{equation}
P_{vv}^{\prime\prime\prime}+4\omega^{2}P_{vv}^{\prime}+4\omega\omega^{\prime}P_{vv}=S_{1}\,,\label{eq:63-1}
\end{equation}
where $S_{1}$ is a source function  
\begin{equation}
S_{1}\left(k,\eta\right)=2\left(2\pi\right)^{3/2}\gamma\widetilde{C}_{R}\left(k\right)\,, 
\end{equation}
}{\par}
\noindent $\widetilde{C}_{R}\left(k\right)$ being the Fourier transform of the environmental correlation function and $\omega$ is given by Eq.~(\ref{eq:frequency})
{Needless to say that some approximations are needed to solve equation (\ref{eq:63-1}), among which we find the approximations
used previously to derive the Lindblad equation. In particular, we
solve (\ref{eq:201}) in the following two limits:}{\par}
\begin{itemize}
\item {The first limit uses the Markovian approximation that requires
the environment autocorrelation time $t_{c}$ to be very short compared
to the typical time scale over which the system evolves $\sim H^{-1}$.
Assuming the environment correlation time $t_{c}$ and length $l_{E}$
to be of the same order $t_{c}\sim l_{E}$ then 
\begin{equation}
Hl_{E}\ll1\,,
\end{equation}
}
\item {The second limit is to evaluate the corrections at the end
of inflation i.e $-k\eta\rightarrow0$, when the modes of observational
interest today are outside horizon  during inflation. }{\par}
\end{itemize}
{Considering now the case of a quadratic interaction $\hat{A}\left(\boldsymbol{x}\right)=\hat{v}^{2}\left(\boldsymbol{x}\right)$,
then the power spectrum is governed by}\footnote{{We will see later, in this paper, that our approach implies
the existence of an additional source function $S_{3}$ in this equation,
namely (\ref{eq:63-1-1}) will be given by $P_{vv}^{\prime\prime\prime}+4\omega^{2}P_{vv}^{\prime}+4\omega\omega^{\prime}P_{vv}=S_{2}+S_{3}$.}}{\large{} 
\begin{equation}
P_{vv}^{\prime\prime\prime}+4\omega^{2}P_{vv}^{\prime}+4\omega\omega^{\prime}P_{vv}=S_{2}\,,\label{eq:63-1-1}
\end{equation}
where the source function $S_{2}$ is now given by }\textbf{
\begin{equation}
S_{2}\left(k,\eta\right)=\alpha^{2}\frac{8\gamma}{\left(2\pi\right)^{3/2}}\int d^{3}\boldsymbol{k^{\prime}}\widetilde{C}_{R}\left(k^{\prime}\right)P_{vv}\left(\left|\boldsymbol{k^{\prime}}+\boldsymbol{k}\right|\right)\,.\label{eq:255}
\end{equation}
}{\par}

{In order to have a better comparison between the predictions
of the current model and those of standard inflation, i.e inflation
theory without cosmic decoherence, the curvature power spectrum is
written as 
\begin{equation}
\mathcal{P}_{\zeta}=\frac{k^{3}}{2\pi^{2}}\frac{1}{2a^{2}M_{\rm pl}\epsilon}P_{vv}\left(k\right)=\mathcal{P}_{\zeta}\left|_{\rm standard}\left(1+\bigtriangleup\mathcal{P}_{\boldsymbol{k}}\right)\right. \, .\label{eq:20-1}
\end{equation}
$\epsilon = - \dot{H}/H^{2}$ being the usual first slow-roll parameter, $H=\dot{a}/{a}$ the Hubble parameter and $M_{\rm pl}$ 
the Planck mass. }{\par} {$\mathcal{P}_{\zeta}\left|_{\rm standard}\right.$ in (\ref{eq:20-1})
is the standard curvature power spectrum obtained using standard inflation that
treats the primordial fluctuations as a closed system, while $\bigtriangleup\mathcal{P}_{\boldsymbol{k}}$
is the cosmic decoherence induced correction which will be of interest
for us. The explicit expressions of $\bigtriangleup\mathcal{P}_{\boldsymbol{k}}$,
as a function of the parameter $p$ characterizing the system-environment coupling as in Eq.~(\ref{eq:coupling}), were computed explicitly by J.Martin et al in \cite{martin2018observational}, so
 we limit ourselves to mention the main results of interest for the goals of 
this paper. In particular, we will write the expressions of $\bigtriangleup\mathcal{P}_{\boldsymbol{k}}$
for which there is a value of the free parameter $p$ that makes it
scale independent in accordance with the highly precise data collected
so far \cite{Aghanim:2018eyx}.

It could be shown that the inhomogenous part of the equations
governing the power spectrum, (\ref{eq:63-1}) and (\ref{eq:63-1-1}),
can be solved by the following ansatz\footnote{It is worth to recall that the functions $v_{\mathbf{k}}\left(\eta\right)$ to be used in evaluating (\ref{eq:19}) are the free modes obtained with the free Hamiltonian (\ref{eq:1-1}) } \cite{martin2018observational}
\begin{equation}
\mathcal{S}_{k}\left(\eta\right)=-\frac{2}{W^{2}}\intop_{\eta_{0}}^{\eta}d\eta^{\prime}S_{n}\left(k,\eta^{\prime}\right)\mathscr{\mathfrak{Im^{2}\left[\mathrm{v_{\mathbf{k}}\left(\eta^{\prime}\right)v_{\mathbf{k}}^{\ast}\left(\eta\right)}\right]}}\,,
\label{eq:19}
\end{equation}
 where, $n=1,2$, depending on whether we are considering a linear or
a quadratic interaction, and  $W$ is the Wronskian, $v_{\mathbf{k}}^{\prime}\left(\eta\right)v_{\mathbf{k}}^{\ast}\left(\eta\right)-v_{\mathbf{k}}\left(\eta\right)v_{\mathbf{k}}^{\ast\prime}\left(\eta\right)=i$
. The initial comoving time $\eta_{0}$ is chosen such that $\mathcal{S}_{\boldsymbol{k}}\left(\eta_{0}\right)=\mathcal{S}_{\boldsymbol{k}}^{\prime}\left(\eta_{0}\right)=\mathcal{S}_{\boldsymbol{k}}^{\prime\prime}\left(\eta_{0}\right)=0$.

The full solution is then obtained by adding to $\mathcal{S}_{k}\left(\eta\right)$,  the solution of the
homogeneous equation in~(\ref{eq:63-1}) and~(\ref{eq:63-1-1}), which gives the standard  power spectrum expression obtained in absence of quantum decoherence effects. Hence, the final expression of the power spectrum for scalar perturbations is given by

\begin{equation}
P_{vv}\left(k\right)=\left|\hat{v}_{k}\right|^{2}+\mathcal{S}_{k}.\label{equt:3.11}
\end{equation}

Adopting the Bunch Davies vacuum as initial state, then we set in \ref{eq:19} $\eta_{0}\rightarrow-\infty$ in order for $P_{vv}\left(k\right)$ to match  Bunch Davies result in the infinite past \cite{martin2018observational}. 

Implementing the full power spectrum expression~(\ref{equt:3.11}) into Eq.~(\ref{eq:20-1}) yields the following corrections}\footnote{These expressions are valid only for $3+\frac{1}{1+\epsilon_{*}}<p<3+\frac{5}{1+\epsilon_{*}}$
in the linear case, and for $2<p<6$ in the quadratic case. We refer the reader
to \cite{martin2018observational} for the power spectrum expressions
corresponding to other ranges of the free parameter $p$.}} $\bigtriangleup\mathcal{P}_{\boldsymbol{k}}${ 
\begin{itemize}
\item For the linear interaction
\begin{equation}
\bigtriangleup\mathcal{P}_{k}\left|\right.\simeq\mathcal{A}\left(k\right)\left[1+\mathcal{B}\epsilon^{*}+\mathcal{C}\epsilon_{2}^{*}+\left(\mathcal{D}\epsilon^{*}+\mathcal{E}\epsilon_{2}^{*}\right)\ln\left(\frac{k}{k_{*}}\right)\right],
\end{equation}
with 
\begin{equation}
\text{\ensuremath{\begin{array}{llc}
\mathcal{A}\left(k\right)=\left(\frac{k_{\gamma}}{k_{*}}\right)^{2}\left(\frac{k}{k_{*}}\right)^{p-5}\frac{\left(6-p\right)\pi}{2^{6-p}\left(p-2\right)\sin\left(\pi p/2\right)\Gamma\left(p-3\right)}\\
\\
\mathcal{B}=-2\frac{\left(p-1\right)\left(p-3\right)}{\left(p-4\right)\left(p-2\right)}-\frac{1}{2}\left(p-5\right)\psi\left(4-\frac{p}{2}\right)-\psi\left(-2+\frac{p}{2}\right)\\
\\
-\frac{1}{2}\left(p-3\right)\psi\left(-\frac{3}{2}+\frac{p}{2}\right)\\
\\
\mathcal{C}=\gamma_{E}+\ln\left(2\right)-2+\frac{6}{\left(p-2\right)\left(p-8\right)},\,\mathcal{D}=p-3,\,\mathcal{E}=0
\end{array}}},\label{equt:3.13}
\end{equation}
and 
\begin{equation}
k_{\gamma}=\sqrt{\frac{2}{\pi}\bar{C}_{R}\frac{\gamma_{*}l_{E}^{3}}{3a_{*}^{3}}}\,,\label{eq:3.14}
\end{equation}
where $k_{\gamma}$ is homogeneous to the dimension of wavenumber, and
 $\epsilon^{*}$ and $\epsilon_{2}^{*}$, are first and second
slow roll parameters, respectively, computed at Hubble exit time of
pivot scale $k^{*}$. The function $\psi(x)$ refers to digamma function
and $\gamma_{E}\simeq0.577$ is the Euler constant. 
\item For quadratic interaction 
\begin{equation}
\begin{array}{cll}
\Delta\mathcal{P}_{k} & \simeq & \frac{2^{p-1}\left(p-4\right)}{3\pi\varGamma\left(p-1\right)\text{sin}\left(\pi p/2\right)}\sigma_{\gamma}\left(\frac{k}{k_{*}}\right)^{p-3}\left[\ln\left(\frac{\eta_{IR}}{\eta_{*}}\right)+\frac{1}{p-4}-\frac{2\left(p-1\right)}{p\left(p-2\right)}\right.\\
\\
 &  & \left.-\frac{\pi}{2}\text{cot}\left(\frac{\pi p}{2}\right)+\text{ln}\left(2\right)-\psi\left(p-2\right)+\ln\left(\frac{k}{k_{*}}\right)\right],\label{equt:3.14}
\end{array}
\end{equation}
\end{itemize}

 with  the dimensionless coefficient $\sigma{}_{\gamma}$  being given by 
\begin{equation}
\sigma{}_{\gamma}=\bar{C}_{R}\frac{\gamma_{*}l_{E}^{3}}{a_{*}^{3}}\,.\label{eq:11-1}
\end{equation}
Notice that the difference of dimension between  the  scales $k_{\gamma}$ and $\sigma{}_{\gamma}$, in spite of the similar expressions,  is due to the fact that they correspond to two different types of interactions, namely, linear and quadratic.

The scales $k_{\gamma}$ and $\sigma{}_{\gamma}$  are very important to constrain the interaction
strength between the system and environment through the comparison
of decoherence induced corrections with observations. In the quadratic
case $\eta_{IR}$ refers to an IR cutoff in the integral (\ref{eq:19}),
with $\ln\left(\frac{\eta_{IR}}{\eta}\right)=N-N_{IR}$ giving
the number of e-folds elapsed since the beginning of inflation.}{\par}

{We notice that the values $p=3$ and $p=5$ are very peculiar,
since, as evident from (\ref{equt:3.13}) and (\ref{equt:3.14}) , they correspond to (almost) scale
independent corrections to the standard, well confirmed, quasi scale
independent power spectrum. An interesting example of the current
model was studied in detail in \cite{martin2018observational}, where
a massive scalar field was considered as an environment for
which the free parameter $p$ is given by 
\begin{equation}
p\simeq7-2n\,,\label{eq:23}
\end{equation}
$n$ being the order of the interaction, i.e the power of the Mukhanov-Sasaki operator $\hat{v}$ in the pointer observable $\hat{A}\left(\boldsymbol{x}\right)=\hat{v}^{n}\left(\boldsymbol{x}\right)$.
The interaction Hamiltonian considered in~\cite{martin2018observational} for such an environment is given by}\footnote{$\left\langle \cdots\right\rangle _{st}$ refers to the expectation
value with respect to the environmental stationary state, hence, the subscript "st". We remind
that we used the Markov approximation in the derivation of the Lindblad
equation.}{\large{} 
\begin{equation}
g\hat{H}_{int}=\lambda\mu^{4-n-m}a^{4}\int\mathrm{d^{3}}\boldsymbol{x}\hat{\varphi}^{n}\left(\hat{\psi}^{m}-\left\langle \hat{\psi}^{m}\right\rangle _{st}\right)\,,\label{eq:24-1}
\end{equation}
where} $\varphi$ {is a scalar field representing our system,
while $\psi$ is a heavy massive scalar field that represents the
environment. So in this example the action describing the total system is given by 
\begin{eqnarray}
S & = &-\int \mathrm{d^{4}}x\sqrt{-g}\left[ \frac{1}{2}g^{\mu\nu}\partial_{\mu}\varphi\partial_{\nu}\varphi+V\left(\varphi\right)+\lambda\mu^{4-n-m}\left\langle \psi^{m}\right\rangle _{st}\varphi^{n}\right.\\ \nonumber
 & &\left.+\frac{1}{2}g^{\mu\nu}\partial_{\mu}\psi\partial_{\nu}\psi+\frac{M^{2}}{2}\psi^{2}+\lambda\mu^{4-n-m}\varphi^{n}\left(\psi^{m}-\left\langle \psi^{m}\right\rangle _{st}\right)\right] \, . 
\label{eq:26-1}
\end{eqnarray}
}{\par}

{Indeed, this model is interesting because it exactly corresponds,
for a given interaction of order $n,$ to the values of $p$ which give
scale independent corrections to the power spectrum. On the other
hand, if we keep the environment unspecified and impose the corrections
to be scale independent, then we can obtain interesting constraints
on the interaction strength with the environment by constraining
$(k_{\gamma}/k_{*})$. In particular we obtain upper bounds on $(k_{\gamma}/k_{*})$, while a lower bound could be obtained
by imposing decoherence to take place before the end of inflation}\footnote{{Notice that it is not necessary for decoherence to take place
before end of inflation. However, it must take place before recombination
to give rise to the CMB classical fluctuations.}}{ as shown in \cite{martin2018observational}. }{\par}

{Another  interesting result found by J.Martin et
al. in~\cite{martin2018observational} was the computation of the decoherence corrected scalar spectral
index $n_{s}$ and the tensor-to-scalar perturbation ratio $r$, which represent two
important observables to remove the degeneracy among some inflation
models. Considering a massive scalar field as environment
and choosing $p=5$ (corresponding to a linear interaction) they showed that
the new scalar spectral index is given by 
\begin{equation}
n_{s}=n_{s}\left|_{\rm standard}\right.-\left(6m-2\right)\epsilon^{*}\,,\label{eq:27}
\end{equation}
where $\epsilon^{*}$ is the first slow roll parameter, and $m$
is the power of $\psi$ in (\ref{eq:26-1}). Similarly, they computed
the new tensor-to-scalar perturbation ratio
\begin{equation}
r=\frac{r\left|_{\rm standard}\right.}{1+\frac{\pi}{6}\frac{k_{\gamma}^{2}}{k_{*}^{2}}}\,,\label{eq:28}
\end{equation}
where by standard we refer as usual to the expressions obtained assuming
standard inflation (i.e when primordial perturbations are treated as a closed
system). }{\par}

{However it is important to mention that the new tensor-to-scalar perturbation ratio (\ref{eq:28}) was obtained assuming the tensor perturbations
to remain unaffected by the decoherence process. The latter assumption
will turn out to be inaccurate as we will show in the last section. }{\large\par}

{Such modified $n_{s}$ and $r$ could indeed improve the fit to data
for some inflationary models while worsening it for others \cite{martin2018observational}. }{\par}

\subsection{The cosmic decoherence induced non-Gaussianities}

{No wonder that even if we consider the system to be intrinsically
free, i.e with no self-interactions, then its interaction with the environment
will induce non-vanishing  higher-order correlation functions. Indeed, J.Martin
et al extended their approach to the study of primordial non-Gaussianities\footnote{It is important to notice that here we are considering the primordial
perturbations to be intrinsically (almost) Gaussian, i.e all higher-order correlation
functions are vanishing in the standard inflation scenario, and we will focus on the interactions with the environment as the main source
of primordial non-Gaussianity.}  following the
same line of thoughts applied previously to the power spectrum \cite{martin2018non}}. In the following we briefly summarize their main results.

{Considering the linear interaction, then, it is easy to see
that all the higher-order correlation functions vanish. Indeed, the
Lindblad term for a linear interaction is given by
\begin{equation}
-\frac{\gamma}{2}\int d^{3}\boldsymbol{k}\tilde{C}_{R}\left(k\right)\left\langle \left[\left[\hat{O},\hat{v}_{\boldsymbol{k}}\right],\hat{v}_{-\boldsymbol{k}}\right]\right\rangle \,,\label{eq:26}
\end{equation}
 so if we consider the three point correlator with less than two field momenta $\hat{p}_{\boldsymbol{k}}$, as $\left\langle \hat{v}_{\boldsymbol{\boldsymbol{k}}_{1}}\hat{v}_{\boldsymbol{\boldsymbol{k}}_{2}}\hat{v}_{\boldsymbol{\boldsymbol{k}_{3}}}\right\rangle $
or $\left\langle \hat{v}_{\boldsymbol{\boldsymbol{k}}_{1}}\hat{v}_{\boldsymbol{\boldsymbol{k}}_{2}}\hat{p}_{\boldsymbol{\boldsymbol{k}_{3}}}\right\rangle $,
then the term (\ref{eq:26}) vanishes because of the vanishing commutator
$\left[\hat{v}_{\boldsymbol{k}},\hat{v}_{\boldsymbol{k}^{\prime}}\right]=0$.
While if we consider correlators involving two  field momentum operators or three, i.e $\left\langle \hat{v}_{\boldsymbol{\boldsymbol{k}}_{1}}\hat{p}_{\boldsymbol{\boldsymbol{k}}_{2}}\hat{p}_{\boldsymbol{\boldsymbol{k}_{3}}}\right\rangle $
and $\left\langle \hat{p}_{\boldsymbol{\boldsymbol{k}}_{1}}\hat{p}_{\boldsymbol{\boldsymbol{k}}_{2}}\hat{p}_{\boldsymbol{\boldsymbol{k}_{3}}}\right\rangle$, then using the commutation relation }
\begin{equation}
\left[\hat{v}_{\boldsymbol{p}},\hat{p}_{\boldsymbol{k}}\right]=i\delta^{\left(3\right)}\left(\boldsymbol{p}+\boldsymbol{k}\right)\,,\label{eq:12-1}
\end{equation}
we will end up with either $\left\langle \hat{v}_{\boldsymbol{\boldsymbol{k}}}\right\rangle $
or $\left\langle \hat{p}_{\boldsymbol{\boldsymbol{k}}}\right\rangle $,
both of which gives zero due to $\left\langle \hat{a}_{\boldsymbol{k}}\pm\hat{a}_{-\boldsymbol{k}}^{\dagger}\right\rangle =0$, where  $\left(\hat{v},\hat{p}\right)$ are expressed as function of the
creation and annihilation operators $\left(\hat{a}_{\boldsymbol{k}},\hat{a}_{-\boldsymbol{k}}^{\dagger}\right)$
defined, as usual, by 
\begin{equation}
a_{\boldsymbol{k}}\left(\eta\right)=\frac{1}{\sqrt{2}}\left(\sqrt{k}\hat{v}_{\boldsymbol{k}}\left(\eta\right)+i\frac{1}{\sqrt{k}}\hat{p}_{\boldsymbol{k}}\left(\eta\right)\right).
\end{equation}

Regarding the trispectrum and when we consider solely the connected
terms, the various contributions cancel out and decoherence does not induce a non vanishing  trispectrum for a linear interaction \cite{martin2018non}. {\par}

{Therefore, one has to consider more complicated interactions, so
choosing the quadratic interaction $\hat{A}=\hat{v}^{2}$ and implementing it again in (\ref{eq:201}) leads to eight equations that represent
the various terms $\left\langle \hat{O}_{\boldsymbol{\boldsymbol{k}}_{1}}\hat{O}_{\boldsymbol{\boldsymbol{k}}_{2}}\hat{O}_{\boldsymbol{\boldsymbol{k}_{3}}}\right\rangle $
which could be built from $\hat{O}_{\boldsymbol{\boldsymbol{k}}}=\hat{v}_{\boldsymbol{\boldsymbol{k}}}$ or
$\hat{p}_{\boldsymbol{\boldsymbol{k}}}$. However now the correlators which
get modified by the Lindblad term, marked by$\underbrace{\left(\right)}$,
are proportional to the initial bispectrum coming from the system Hamiltonian, which is zero in our case since it corresponds to the free theory}\footnote{We remind the reader that we are  considering an intrinsic system Hamiltonian of quadratic form,  which does not contain cubic and higher-order self-interaction terms, see (\ref{eq:1-1}). \label{footnoe 18}}{\large{}, an example being}
\begin{equation}
\begin{array}{ccl}
\frac{d\left\langle \hat{v}_{\boldsymbol{\boldsymbol{k}}_{1}}\hat{p}_{\boldsymbol{\boldsymbol{k}}_{2}}\hat{p}_{\boldsymbol{\boldsymbol{k}_{3}}}\right\rangle }{d\eta} & = & \left\langle \hat{p}_{\boldsymbol{\boldsymbol{k}}_{1}}\hat{p}_{\boldsymbol{\boldsymbol{k}}_{2}}\hat{p}_{\boldsymbol{\boldsymbol{k}_{3}}}\right\rangle -\omega^{2}\left(k_{2}\right)\left\langle \hat{v}_{\boldsymbol{\boldsymbol{k}}_{1}}\hat{v}_{\boldsymbol{\boldsymbol{k}}_{2}}\hat{p}_{\boldsymbol{\boldsymbol{k}_{3}}}\right\rangle -\omega^{2}\left(k_{3}\right)\left\langle \hat{v}_{\boldsymbol{\boldsymbol{k}}_{1}}\hat{p}_{\boldsymbol{\boldsymbol{k}}_{2}}\hat{v}_{\boldsymbol{\boldsymbol{k}_{3}}}\right\rangle \\
\\
 & + & \underbrace{\frac{4\gamma}{\left(2\pi\right)^{3/2}}\int\mathrm{d}^{3}\boldsymbol{k}\widetilde{C}_{R}\left(\left|\boldsymbol{k}\right|\right)\left\langle \hat{v}_{\boldsymbol{\boldsymbol{k}}_{1}}\hat{v}_{\boldsymbol{\boldsymbol{k}}_{2}-\boldsymbol{k}}\hat{v}_{\boldsymbol{\boldsymbol{k}+\boldsymbol{k}_{3}}}\right\rangle }\, .
\end{array} \label{eq 3.25}
\end{equation}
{On the other hand, the Lindblad equation is valid only at
leading order in the coupling parameter $\gamma$, therefore, we cannot use~in the previous
equation the bispectrum generated by decoherence since the latter would arise 
at $\gamma^{2}$ order, where we can no more rely on the Lindblad equation.}{\par}

{Having found a vanishing bispectrum, the authors of~\cite{martin2018non} considered the trispectrum, showing that in this case a non-vanishing trispectrum is indeed generated by dechoerence effects. In particular,  they found a remarkable expression in the range $p<4$ }
\begin{equation}
g_{NL}\propto\frac{\sigma_{\gamma}}{\mathcal{P}_{\zeta}\left(k_{*}\right)}\left(\frac{k}{k_{*}}\right)^{p-3}\,,\label{eq:24}
\end{equation}
{where the non-linearity parameter $g_{NL}$ is defined as }
\begin{equation}
\left\langle \hat{\zeta}_{\boldsymbol{\boldsymbol{k}}_{1}}\hat{\zeta}_{\boldsymbol{\boldsymbol{k}}_{2}}\hat{\zeta}_{\boldsymbol{\boldsymbol{k}_{3}}}\hat{\zeta}_{\boldsymbol{\boldsymbol{k}_{4}}}\right\rangle _{c}=\frac{25}{54}g_{NL}\left[\mathcal{P}_{\zeta}\left(\boldsymbol{k}_{1}\right)\mathcal{P}_{\zeta}\left(\boldsymbol{k}_{2}\right)\mathcal{P}_{\zeta}\left(\boldsymbol{k}_{3}\right)+3\,\mathrm{permutations}\right]\delta^{(3)} \left(\boldsymbol{k}_{1}+\boldsymbol{k}_{2}+\boldsymbol{k}_{3}+\boldsymbol{k}_{4}\right)\,.
\end{equation}
{We notice through (\ref{eq:24}) that the trispectrum is
scale independent for $p=3$ which, again, corresponds to a massive
scalar field as an environment. Indeed, the fact that the bispectrum is vanishing lead the authors to state that decoherence is one of
the rare examples where the bispectrum is perturbatively suppressed compared to the trispectrum. 

{However, we will show that such a conclusion is not completely
accurate. Using   a generalized  
  form of the pointer observable $\hat{A},$ we will obtain
a non vanishing bispectrum which is dominant with respect to the trispectrum.  In addition, we will show that, at leading order in $\gamma$,  
the bispectrum turns out to be scale independent for $p=4$ (which again could corrrespond to a massive scalar field as environment, as will be explained in details below).} 
 Notice that, as an interesting consequence, one of our main results is that keeping the leading correction to the curvature power spectrum to be scale independent, in general does not guarantee the scale-independence of the bispectrum, rather it gives rise to a specific signature of the primordial bispectrum generated by decoherence effects, namely a linear-scale dependence of the primordial bispectrum.


Finally, we will give our own reasoning
on how to understand when does decoherence induce scale independent
corrections. We will show that scale independence is not related to a specific type of environment but, rather, a very general criterium holds 
that involves the power of $\alpha$ in the Lindblad equation.}{\par} 
\section{Our approach\label{sec:Our-approach}}

As mentioned, the results obtained previously in the literature were
based on restricted forms of the pointer observable $\hat{A}$,
namely monomial types, either  $\hat{v}$ or $\hat{v}^{2}$, or generally
$\hat{v}^{n}$.  Therefore, even if such a choice leads to interesting results we extend this approach to
get broader insights on the effect of cosmic decoherence on primordial perturbations. 
It is worth to mention that our generalization helps in shedding light
on some of the general conclusions in~\cite{boyanovsky2015effective,martin2018observational,martin2018non},
which can then be scrutinized further under the light of the results
obtained by our approach.

The approach we will adopt is to consider a pointer observable of
the from 
\begin{equation}
\hat{A}=\sum_{n}\alpha^{n-1}\hat{v}^{n}\,,\label{eq:352}
\end{equation}
up to a certain order $n$, with $\alpha$ being an expansion constant.

 One of the most interesting results we find is about the
scale-dependence of the cosmic decoherence induced bispectrum. In
general, guided by the observational constraints, and thus imposing
the leading corrections to the curvature power spectrum to be (almost)
scale independent does imply indeed a specific scale-dependence of
the bispectrum. The latter represents a specific prediction of our
scenario, starting from~(\ref{eq:352}). On the other hand, we will
show that there is a certain freedom in the free parameters in case
the underlying physical model is that of a massive scalar field as
environment, so that e.g. also a scale-independent bispectrum can
be obtained (similarly to what previously seen for the inflationary
power spectrum and trispectrum).

In order to discuss a concrete example, we provide some, possible,
physical processes that could yield the form (\ref{eq:352}).

The first example consists in adopting an interaction Hamiltonian
of the form 
\begin{equation}
g\hat{H}_{int}=\lambda\mu^{4-n-m}a^{4}\int\mathrm{d^{3}}\boldsymbol{x}\left(\hat{\varphi}-\sigma\hat{I}\right)^{n}\left(\hat\psi^{m}-\left\langle \hat\psi^{m}\right\rangle _{st}\right)\,,\label{eq:30}
\end{equation}
which leads us to a pointer observable of the form\footnote{It could be intuitively understood how (\ref{eq:30}) produces (\ref{eq:352})
if we expand it as follows 
\begin{equation}
g\hat{H}_{int}\propto\lambda\mu^{4-n-m}a^{4}\underset{\text{t}}{\sum}\int\mathrm{d^{3}}\boldsymbol{x}\varphi^{n-t}\sigma^{t}\left(\psi^{m}-\left\langle \psi^{m}\right\rangle _{st}\right)\,,
\end{equation}
and make the straightforward change of variables necessary to obtain
(\ref{eq:352}), starting by the change of variable $v\left(\eta,\boldsymbol{x}\right)\equiv a\left(\eta\right)\varphi\left(\eta,\boldsymbol{x}\right)$
which mimics the MS variable definition, and ending up by defining
$k=n-t$. Then, it easy to get the expansion constant that is to be
identified with $\alpha$ in (\ref{eq:352}).} (\ref{eq:352}), where       $\sigma$ is a constant of the same dimension of $\varphi$. The total action (\ref{eq:26-1}) now becomes 
\begin{equation}
\begin{array}{clc}
S & =- \int\mathrm{d^{4}}x\sqrt{-g}\left[\frac{1}{2}g^{\mu\nu}\partial_{\mu}\varphi\partial_{\nu}\varphi+V\left(\varphi\right)+\lambda\mu^{4-n-m}\left\langle \psi^{m}\right\rangle _{st}\varphi^{n}\right.\\
\\
 & \left.+\frac{1}{2}g^{\mu\nu}\partial_{\mu}\psi\partial_{\nu}\psi+\frac{M^{2}}{2}\psi^{2}+\lambda\mu^{4-n-m}\left(\varphi-\sigma\right)^{n}\left(\psi^{m}-\left\langle \psi^{m}\right\rangle _{st}\right)\right]\, . \\
\\
\end{array}\,
\end{equation}
Notice the similarity between (\ref{eq:30}) and the one considered
in~\cite{martin2018observational} given by (\ref{eq:24-1}). Indeed,
the physical process we suggest is a generalization of (\ref{eq:24-1}) where the environment is made of a massive scalar
field $\psi$, the only difference \footnote{Notice that with such choice of the $\hat{H}_{int}$, the parameter
$\gamma$ that characterizes the interaction strength with environment
in the Lindblad equation will be, obviously, modified with respect
to the interaction Hamiltonian given in (\ref{eq:24-1}), since it
depends on the model. In our case, it will be proportional, also,
to $\alpha$. However, for simplicity we will continue to use $\gamma$
instead of $\gamma^{\prime}$, for instance.} being the displacement of the field $\varphi$ by $\sigma$. 

Assuming small field values of the inflaton field, we can get more physical processes which could yield the
pointer observable form (\ref{eq:352}). As case in point, we can
have power law inflation potential \cite{lucchin1985power} as the system contribution part
to $\hat{H}_{int}$
\begin{equation}
g\hat{H}_{int}=\lambda\mu^{4-m}a^{4}\int\mathrm{d^{3}}\boldsymbol{x}\exp\left(-\frac{\alpha}{M_{pl}}\varphi\right)\left(\psi^{m}-\left\langle \psi^{m}\right\rangle _{st}\right)\,,
\end{equation}
and expanding the potential in the limit $\alpha\varphi\ll M_{pl}$,
we can easily recover the form (\ref{eq:352}). Similarly, for a Starobinsky-like \cite{starobinsky1980new} inflaton contribution in $\hat{H}_{int}$
\begin{equation}
g\hat{H}_{int}=\lambda\mu^{4-m}a^{4}\int\mathrm{d^{3}}\boldsymbol{x}\left[1-\exp\left(-\frac{\sqrt{2/3}}{M_{pl}}\varphi\right)\right]^{2}\left(\psi^{m}-\left\langle \psi^{m}\right\rangle _{st}\right)\,,
\end{equation}
 after double\footnote{By double, we mean to expand the exponential first, and then, expand
$\left[1-\sum_{k}\frac{\beta^{k}}{k!}x^{k}\right]^{2}$.} expanding the term $\left[1-\exp\left(-\frac{\sqrt{2/3}}{M_{pl}}\varphi\right)\right]^{2}$in
the small field inflation limit, we obtain the form (\ref{eq:352}).

For the sake of simplicity, we will pick up the physical process given
by (\ref{eq:30}) as reference in the following. However, as shown
above, any physical mechanism yielding the pointer observable form (\ref{eq:352})
would lead to the same general conclusions regarding the effect of
quantum decoherence on the inflationary observables. 

The free parameter $p$ expression within our generalized physical
process (\ref{eq:30}), instead of (\ref{eq:24-1}), could be found
following the computations done by J.Martin et al. in appendix B of
~\cite{martin2018observational}. Doing so, we get the relation
\begin{equation}
p=7-k-l\,,\label{32-1}
\end{equation}
that links the free parameter $p$ value to the parameters $\left(k,l\right)$
characterizing the physical process, where, $k$ and $l$ could be
understood from the following equation 
\begin{equation}
\frac{d\left\langle \hat{O}\right\rangle }{d\eta}=\left\langle \frac{\partial\hat{O}}{\partial\eta}\right\rangle -i\left\langle \left[\hat{O},\hat{H_{v}}\right]\right\rangle -\frac{1}{2}\underset{k,l}{\sum}\gamma\left(k,l\right)\alpha^{k+l-2}\int d^{3}\boldsymbol{x}d^{3}\boldsymbol{y}C_{R}\left(\boldsymbol{x},\boldsymbol{y}\right)\left\langle \left[\left[\hat{O},\hat{v}^{k}\right],\hat{v}^{l}\right]\right\rangle \,,\label{eq:4.8}
\end{equation}
where by writing $\gamma\equiv\gamma\left(k,l\right)$ and putting
it inside the summation, we are taking into account the dependence
of the parameter \footnote{We remind that the time dependence of $\gamma$ is given by \ref{eq:coupling}.} $p$ on $k$ and $l$ as given by (\ref{32-1}).

The relation (\ref{32-1}) generalizes the one obtained in~\cite{martin2018observational}
given by (\ref{eq:23}), where a monomial form for the pointer observable
was considered, $\hat{A}\left(\boldsymbol{x}\right)=\hat{v}^{n}\left(\boldsymbol{x}\right)$.
Indeed for $k=l\equiv n$ one recovers $p=7-2n.$

It is worth to mention that with our generalized form of the pointer
observable (\ref{eq:352}), given as an expansion in $\hat{v}$, the coupling
system-environment in the Lindblad equation, represented by  $\gamma$  might  
vary from one term to another in that expansion.   Such variation of $\gamma$ is captured by the fact
that the free parameter $p$ might not be the same, but it might depend on
which term of (\ref{eq:352}) is contributing to the non-unitary term of the Lindblad
equation. A case in point of such variation of $\gamma$  could be seen through the suggested physical
processes above, where $p$ is a function of $\left(k,l\right)$ rather
than being the same (i.e constant) for all the terms, see (\ref{32-1}) and (\ref{eq:4.8}).

We will soon see that the leading term giving a non zero bispectrum
is 
\begin{equation}
\frac{\gamma}{2}\alpha\int d^{3}\boldsymbol{x}d^{3}\boldsymbol{y}C_{R}\left(\boldsymbol{x},\boldsymbol{y}\right)\left\langle \left[\left[\hat{O},\hat{v}^{2}\right],\hat{v}\right]\right\rangle \,,
\end{equation}
and it will turn out to be scale independent for $p=4.$ Therefore,
from (\ref{32-1}) we see that this value of $p$ could correspond
to a massive scalar field as environment, since for $k=2 (1)$ and $l=1 (2)$
equation (\ref{32-1}) yields $p=4$.

However, it is very important to mention that the results derived
below, and those obtained in \cite{martin2018observational,martin2018non},
are independent from the type of environment adopted as long as it
satisfies the Born and Markov approximations.
{\par}

{Finally, we remark that it is possible to have a pointer observable
of the form (\ref{eq:352}) by considering a decoherence scenario
where short wavelengths of primordial perturbations spectrum decohere
the long wavelengths ones, which are those of observational interest. For
more details on that, see \cite{Burgess:2006jn,martineau2007decoherence,boyanovsky2015effective}.}

{For the bispectrum computation, we will stop at the order $n=2$ in (\ref{eq:352}), which represents the leading order for which decoherence does
induce a non-vanishing bispectrum.  After dealing with the computation of the bispectrum, we will go back to the power spectrum and show that, if one adopts as the pointer observable Eq.~(\ref{eq:352}), then an additional contribution to the  
Lindblad equation arises w.r.t to the results obtained in \cite{martin2018observational} which used a quadratic  interaction term. 
These new contributions are indeed of the same order in $\alpha$ as the contribution coming from a pure quadratic interaction. This last observation is also applicable to the trispectrum computations in \cite{martin2018non}, however we will restrict ourselves to the computation details of the power spectrum case.}{\par}

\subsection{Computation of the cosmic decoherence induced bispectrum \label{subsec:Computation-of-bispectrum}}

{First let us discuss briefly how our choice of $\hat{A}$ as in Eq.~(\ref{eq:352}) does 
reproduce and generalize the results obtained by J.Martin et al. in
\cite{martin2018observational,martin2018non}, and most importantly
the reason behind obtaining in this case a non vanishing bispectrum.}{\par}

{The starting point is the previously derived Lindblad equation
}
\begin{equation}
\frac{d\left\langle \hat{O}\right\rangle }{d\eta}=\left\langle \frac{\partial\hat{O}}{\partial\eta}\right\rangle -i\left\langle \left[\hat{O},\hat{H_{v}}\right]\right\rangle -\frac{\gamma}{2}\int d^{3}\boldsymbol{x}d^{3}\boldsymbol{y}C_{R}\left(\boldsymbol{x},\boldsymbol{y}\right)\left\langle \left[\left[\hat{O},\hat{A}\left(\boldsymbol{x}\right)\right],\hat{A}\left(\boldsymbol{y}\right)\right]\right\rangle \,,\label{eq:1}
\end{equation}
{where, now, we use the pointer observable $\hat{A}$ }
\begin{equation}
\hat{A}\left(\boldsymbol{x}\right)=\hat{v}+\alpha\hat{v}^{2}\label{eq:2}\, ,
\end{equation}
{where $\alpha$ is a coupling constant, with dimension $\left[ {\rm momentum}\right]^{-1}$
which is introduced for the sake of dimensions homogeneity. }{\par}
\noindent
{Substituting (\ref{eq:2}) in (\ref{eq:1}) gives}
\begin{equation}
\begin{array}{cllcc}
\frac{d\left\langle \hat{O}\right\rangle }{d\eta} & = & \left\langle \frac{\partial\hat{O}}{\partial\eta}\right\rangle -i\left\langle \left[\hat{O},\hat{H_{v}}\right]\right\rangle -\frac{\gamma}{2}\int d^{3}\boldsymbol{x}d^{3}\boldsymbol{y}C_{R}\left(\boldsymbol{x},\boldsymbol{y}\right)\left\langle \left[\left[\hat{O},\hat{v}\left(\boldsymbol{x}\right)+\alpha\hat{v}^{2}\left(\boldsymbol{x}\right)\right],\hat{v}\left(\boldsymbol{y}\right)+\alpha\hat{v}^{2}\left(\boldsymbol{y}\right)\right]\right\rangle \\
\\
 & = & \left\langle \frac{\partial\hat{O}}{\partial\eta}\right\rangle -i\left\langle \left[\hat{O},\hat{H_{v}}\right]\right\rangle -\frac{\gamma}{2}\int d^{3}\boldsymbol{x}d^{3}\boldsymbol{y}C_{R}\left(\boldsymbol{x},\boldsymbol{y}\right)\left\{ \left\langle \left[\left[\hat{O},\hat{v}\left(\boldsymbol{x}\right)\right],\hat{v}\left(\boldsymbol{y}\right)\right]\right\rangle \right.\\
\\
 & + & \left.\alpha^{2}\left\langle \left[\left[\hat{O},\hat{v}^{2}\left(\boldsymbol{x}\right)\right],\hat{v}^{2}\left(\boldsymbol{y}\right)\right]\right\rangle +\underbrace{\alpha\left\langle \left[\left[\hat{O},\hat{v}\left(\boldsymbol{x}\right)\right],\hat{v}^{2}\left(\boldsymbol{y}\right)\right]+\left[\left[\hat{O},\hat{v}^{2}\left(\boldsymbol{x}\right)\right],\hat{v}\left(\boldsymbol{y}\right)\right]\right\rangle }\right\} .\\
\\\label{356}
\end{array}
\end{equation}

{From this equation we can see how the interaction terms we are considering do contain
all the results derived in \cite{martin2018non,martin2018observational}.
Indeed, in order to obtain their results, it is sufficient for their linear
case to set $\alpha=0$, where for this type of interaction only power
spectrum receives correction while all the other correlation functions
remain untouched. Then, to recover the pure quadratic contribution to the 
power spectrum correction we can add a dimensionless constant in front
of $\text{\ensuremath{\hat{v}}}$ in $\hat{A}\left(\boldsymbol{x}\right)$
and set it to zero for that purpose, though, at least within a perturbative approach, 
it is more legitimate to consider the quadratic order along with the linear
one, since the latter usually gives the dominant contribution}.\footnote{{We will see in the next section that the pointer observable 
 $\hat{A}\left(\boldsymbol{x}\right)=\hat{v}+\alpha\hat{v}^{2}+\alpha^{2}\hat{v}^{3}$,
will induce a correction to both the power spectrum and the trispectrum,
of the same order in $\alpha$ as that of a pure quadratic contribution
$\alpha^{2}\left\langle \left[\left[\hat{O},\hat{v}^{2}\left(\boldsymbol{x}\right)\right],\hat{v}^{2}\left(\boldsymbol{y}\right)\right]\right\rangle $,
namely of order $\alpha^{2}$. Those additional corrections come from
$\alpha^{2}\left[\left[\hat{O},\hat{v}^{3}\left(\boldsymbol{x}\right)\right],\hat{v}\left(\boldsymbol{y}\right)\right]$
and $\alpha^{2}\left[\left[\hat{O},\hat{v}\left(\boldsymbol{x}\right)\right],\hat{v}^{3}\left(\boldsymbol{y}\right)\right]$
and were not considered in \cite{martin2018observational} because
they show up only if we consider the most general form of the pointer
observable mentioned above, thus we must consider also those extra
corrections to get a complete and accurate result. We will come back
to this point soon.}}{In case of the trispectrum we do not need this constant since
there is no correction to it from the linear order in $\hat{v}$. Similarly, there is no correction
from the last term, marked with $\underbrace{\cdots}$,  of the real and non unitary part of (\ref{356}), given by 
\begin{equation}
\propto\alpha\left\langle \left[\left[\hat{O},\hat{v}\left(\boldsymbol{x}\right)\right],\hat{v}^{2}\left(\boldsymbol{y}\right)\right]+\left[\left[\hat{O},\hat{v}^{2}\left(\boldsymbol{x}\right)\right],\hat{v}\left(\boldsymbol{y}\right)\right]\right\rangle \,,\label{285}
\end{equation}
since it gives for the power spectrum case, after contractions,  terms  proportional to $\left\langle v_{\boldsymbol{k}}\right\rangle 
$ or $\left\langle p_{\boldsymbol{k}}\right\rangle 
$
which are zero with respect to the Bunch-Davies vacuum, while, for the trispectrum case it gives terms proportional to the initial bispectrum coming from the system Hamiltonian, which is zero in our
case since it corresponds to the free theory, see footnote \ref{footnoe 18} and explanations below (\ref{eq 3.25}) }. { However, it is this last term, given by (\ref{285}), that  generates a purely quantum decoherence induced bispectrum, at leading order in $\gamma$,
and such a term is there thanks to the new form of the pointer observable $\hat{A}$ which enables cross terms of the type $\left[\left[\hat{O},\hat{v}^{k}\left(\boldsymbol{x}\right)\right],\hat{v}^{l}\left(\boldsymbol{y}\right)\right]$
to show up with $k\neq l$. We remind that in \cite{martin2018non}
it was found that the first term of the third line in (\ref{356}), which
refers to a pure quadratic interaction,  would generate at leading order in $\gamma$ a bispectrum, 
 but this would turn out to be proportional to the initial bispectrum which is zero.
On the other hand, our pointer observable choice leads to a bispectrum that is proportional
to the power spectra, thus representing a direct consequence of the decoherence effects.

Having said that and since we are focusing on the bispectrum,
in this section we will keep only the new term in the Lindblad equation, 
so that we start the Fourier transform operation from}
\begin{eqnarray}
\frac{d\left\langle \hat{O}\right\rangle }{d\eta}&=&\left\langle \frac{\partial\hat{O}}{\partial\eta}\right\rangle -i\left\langle \left[\hat{O},\hat{H_{v}}\right]\right\rangle -\frac{\gamma\alpha}{2}\int\mathrm{d}^{3}\boldsymbol{x}\mathrm{d}^{3}\boldsymbol{y}C_{R}\left(\boldsymbol{x},\boldsymbol{y}\right)\left\{ \left\langle \left[\left[\hat{O},\hat{v}\left(\boldsymbol{x}\right)\right],\hat{v}^{2}\left(\boldsymbol{y}\right)\right]+ \right. \right. \nonumber \\ 
&+&  \left. \left. \left[\left[\hat{O},\hat{v}^{2}\left(\boldsymbol{x}\right)\right],\hat{v}\left(\boldsymbol{y}\right)\right] \right\rangle \right\} \,.\label{equ:4.14}
\end{eqnarray}
Using

{
\begin{equation}
\hat{v}^{2}\left(\boldsymbol{x}\right)=\frac{1}{\left(2\pi\right)^{3}}\int\mathrm{d}^{3}\boldsymbol{k}^{\prime}\mathrm{d}^{3}\boldsymbol{p}\hat{v}_{\boldsymbol{k}^{\prime}}\hat{v}_{\boldsymbol{p}-\boldsymbol{k}^{\prime}}e^{i\boldsymbol{p}.\boldsymbol{x}}\,,
\end{equation}
 we get in Fourier space\footnote{To arrive to this form, we use  the Fourier transform $C_{R}\left(\boldsymbol{x},\boldsymbol{y}\right)=\frac{1}{\left(2\pi\right)^{3/2}}\int\mathrm{d}^{3}\boldsymbol{k}\widetilde{C}_{R}\left(\left|\boldsymbol{k}\right|\right)e^{i\boldsymbol{k}.\left(\boldsymbol{x}-\boldsymbol{y}\right)}$
in the first term $\propto\left[\left[\hat{O},\hat{v}\left(\boldsymbol{x}\right)\right],\hat{v}^{2}\left(\boldsymbol{y}\right)\right]$
of the non unitary part of (\ref{equ:4.14}). Then, we make a variable change
$\boldsymbol{k}\rightarrow\boldsymbol{-k}$ in the Fourier transform of
$C_{R}\left(\boldsymbol{x},\boldsymbol{y}\right)$ in the second term
$\propto\left[\left[\hat{O},\hat{v}^{2}\left(\boldsymbol{x}\right)\right],\hat{v}\left(\boldsymbol{y}\right)\right]$,
and we take advantage of the invariance of the environment correlation
function $\widetilde{C}_{R}\left(\left|\boldsymbol{k}\right|\right)$
under such transformation, because of its dependence on the modulus
$\left|\boldsymbol{k}\right|$. }
\begin{eqnarray}
\frac{d\left\langle \hat{O}\right\rangle }{d\eta} & = & \left\langle \frac{\partial\hat{O}}{\partial\eta}\right\rangle -i\left\langle \left[\hat{O},\hat{H_{v}}\right]\right\rangle -\frac{\gamma\alpha}{2}\int\mathrm{d}^{3}\boldsymbol{k}\mathrm{d}^{3}\boldsymbol{p}\widetilde{C}_{R}\left(\left|\boldsymbol{k}\right|\right)\left\{ \left\langle \left[\left[\hat{O},\hat{v}_{\boldsymbol{\boldsymbol{-k}}}\right],\hat{v}_{\boldsymbol{p}}\hat{v}_{\boldsymbol{\boldsymbol{k}-p}}\right]+\right.\right. \nonumber \\
 & + & \left.\left.\left[\left[\hat{O},\hat{v}_{\boldsymbol{p}}\hat{v}_{\boldsymbol{\boldsymbol{k}-p}}\right],\hat{v}_{\boldsymbol{\boldsymbol{-k}}}\right]\right\} \right\rangle \, . 
\label{eq:9}
\end{eqnarray}

{Now, using the commutators property for $A,\,B,\,C$ being three
operators}
\begin{equation}
\left[\left[A,B\right],C\right]=\left[\left[A,C\right],B\right]\,\,\,\,if\,\,\,\left[B,C\right]=0\,,\label{eq:9 real}
\end{equation}
{equation (\ref{eq:9}) becomes}
\begin{eqnarray}
\frac{d\left\langle \hat{O}\right\rangle }{d\eta}=\left\langle \frac{\partial\hat{O}}{\partial\eta}\right\rangle -i\left\langle \left[\hat{O},\hat{H_{v}}\right]\right\rangle -\gamma\alpha\int\mathrm{d}^{3}\boldsymbol{k}\mathrm{d}^{3}\boldsymbol{p}\widetilde{C}_{R}\left(\left|\boldsymbol{k}\right|\right)\left\langle \left[\left[\hat{O},\hat{v}_{\boldsymbol{p}}\hat{v}_{\boldsymbol{\boldsymbol{k}-p}}\right],\hat{v}_{\boldsymbol{\boldsymbol{-k}}}\right]\right\rangle . \label{eq:10}
\end{eqnarray}
{The latter is the main equation that will enable us to get
the eight differential equations satisfied by each bispectrum. Before
listing them, notice that a particular comoving scale appears in the
interaction term, indeed, in order for (\ref{356}) (or equivalently
(\ref{eq:10})) to have the correct dimension, $\gamma\widetilde{C}_{R}\left(\left|\boldsymbol{k}\right|\right)$
must be homogeneous to the square of a comoving wavenumber which we
define as~ in Eq.~(\ref{eq:3.14})~\cite{martin2018observational}.}
{Using the commutation relation (\ref{eq:12-1}) we get,
after some straightforward computations, the following set of equations
satisfied by the various bispectra }{\par}

\begin{equation}
\begin{array}{ccl}
\frac{d\left\langle \hat{v}_{\boldsymbol{\boldsymbol{k}}_{1}}\hat{v}_{\boldsymbol{\boldsymbol{k}}_{2}}\hat{v}_{\boldsymbol{\boldsymbol{k}_{3}}}\right\rangle }{d\eta} & = & \left\langle \hat{v}_{\boldsymbol{\boldsymbol{k}}_{1}}\hat{v}_{\boldsymbol{\boldsymbol{k}}_{2}}\hat{p}_{\boldsymbol{\boldsymbol{k}_{3}}}\right\rangle +\left\langle \hat{v}_{\boldsymbol{\boldsymbol{k}}_{1}}\hat{p}_{\boldsymbol{\boldsymbol{k}}_{2}}\hat{v}_{\boldsymbol{\boldsymbol{k}_{3}}}\right\rangle +\left\langle \hat{p}_{\boldsymbol{\boldsymbol{k}}_{1}}\hat{v}_{\boldsymbol{\boldsymbol{k}}_{2}}\hat{v}_{\boldsymbol{\boldsymbol{k}_{3}}}\right\rangle \\
\\
\frac{d\left\langle \hat{v}_{\boldsymbol{\boldsymbol{k}}_{1}}\hat{v}_{\boldsymbol{\boldsymbol{k}}_{2}}\hat{p}_{\boldsymbol{\boldsymbol{k}_{3}}}\right\rangle }{d\eta} & = & \left\langle \hat{v}_{\boldsymbol{\boldsymbol{k}}_{1}}\hat{p}_{\boldsymbol{\boldsymbol{k}}_{2}}\hat{p}_{\boldsymbol{\boldsymbol{k}_{3}}}\right\rangle +\left\langle \hat{p}_{\boldsymbol{\boldsymbol{k}}_{1}}\hat{p}_{\boldsymbol{\boldsymbol{k}}_{2}}\hat{v}_{\boldsymbol{\boldsymbol{k}_{3}}}\right\rangle -\omega^{2}\left(k_{3}\right)\left\langle \hat{v}_{\boldsymbol{\boldsymbol{k}}_{1}}\hat{v}_{\boldsymbol{\boldsymbol{k}}_{2}}\hat{v}_{\boldsymbol{\boldsymbol{k}_{3}}}\right\rangle \\
\\
\frac{d\left\langle \hat{v}_{\boldsymbol{\boldsymbol{k}}_{1}}\hat{p}_{\boldsymbol{\boldsymbol{k}}_{2}}\hat{v}_{\boldsymbol{\boldsymbol{k}_{3}}}\right\rangle }{d\eta} & = & \left\langle \hat{v}_{\boldsymbol{\boldsymbol{k}}_{1}}\hat{p}_{\boldsymbol{\boldsymbol{k}}_{2}}\hat{p}_{\boldsymbol{\boldsymbol{k}_{3}}}\right\rangle +\left\langle \hat{v}_{\boldsymbol{\boldsymbol{k}}_{1}}\hat{p}_{\boldsymbol{\boldsymbol{k}}_{2}}\hat{p}_{\boldsymbol{\boldsymbol{k}_{3}}}\right\rangle -\omega^{2}\left(k_{2}\right)\left\langle \hat{v}_{\boldsymbol{\boldsymbol{k}}_{1}}\hat{v}_{\boldsymbol{\boldsymbol{k}}_{2}}\hat{v}_{\boldsymbol{\boldsymbol{k}_{3}}}\right\rangle \\
\\
\frac{d\left\langle \hat{p}_{\boldsymbol{\boldsymbol{k}}_{1}}\hat{v}_{\boldsymbol{\boldsymbol{k}}_{2}}\hat{v}_{\boldsymbol{\boldsymbol{k}_{3}}}\right\rangle }{d\eta} & = & \left\langle \hat{p}_{\boldsymbol{\boldsymbol{k}}_{1}}\hat{p}_{\boldsymbol{\boldsymbol{k}}_{2}}\hat{v}_{\boldsymbol{\boldsymbol{k}_{3}}}\right\rangle +\left\langle \hat{p}_{\boldsymbol{\boldsymbol{k}}_{1}}\hat{p}_{\boldsymbol{\boldsymbol{k}}_{2}}\hat{v}_{\boldsymbol{\boldsymbol{k}_{3}}}\right\rangle -\omega^{2}\left(k_{1}\right)\left\langle \hat{v}_{\boldsymbol{\boldsymbol{k}}_{1}}\hat{v}_{\boldsymbol{\boldsymbol{k}}_{2}}\hat{v}_{\boldsymbol{\boldsymbol{k}_{3}}}\right\rangle \\
\\
\frac{d\left\langle \hat{v}_{\boldsymbol{\boldsymbol{k}}_{1}}\hat{p}_{\boldsymbol{\boldsymbol{k}}_{2}}\hat{p}_{\boldsymbol{\boldsymbol{k}_{3}}}\right\rangle }{d\eta} & = & \left\langle \hat{p}_{\boldsymbol{\boldsymbol{k}}_{1}}\hat{p}_{\boldsymbol{\boldsymbol{k}}_{2}}\hat{p}_{\boldsymbol{\boldsymbol{k}_{3}}}\right\rangle -\omega^{2}\left(k_{2}\right)\left\langle \hat{v}_{\boldsymbol{\boldsymbol{k}}_{1}}\hat{v}_{\boldsymbol{\boldsymbol{k}}_{2}}\hat{p}_{\boldsymbol{\boldsymbol{k}_{3}}}\right\rangle -\omega^{2}\left(k_{3}\right)\left\langle \hat{v}_{\boldsymbol{\boldsymbol{k}}_{1}}\hat{p}_{\boldsymbol{\boldsymbol{k}}_{2}}\hat{v}_{\boldsymbol{\boldsymbol{k}_{3}}}\right\rangle \\
\\
 & + & 2\gamma\alpha\left(\widetilde{C}_{R}\left(\left|\boldsymbol{k}_{2}\right|\right)+\widetilde{C}_{R}\left(\left|\boldsymbol{k}_{3}\right|\right)\right)\left\langle \hat{v}_{\boldsymbol{\boldsymbol{k}}_{1}}\hat{v}_{\boldsymbol{\boldsymbol{k}}_{2}+\boldsymbol{\boldsymbol{k}}_{3}}\right\rangle \\
\\
\frac{d\left\langle \hat{p}_{\boldsymbol{\boldsymbol{k}}_{1}}\hat{v}_{\boldsymbol{\boldsymbol{k}}_{2}}\hat{p}_{\boldsymbol{\boldsymbol{k}_{3}}}\right\rangle }{d\eta} & = & \left\langle \hat{p}_{\boldsymbol{\boldsymbol{k}}_{1}}\hat{p}_{\boldsymbol{\boldsymbol{k}}_{2}}\hat{p}_{\boldsymbol{\boldsymbol{k}_{3}}}\right\rangle -\omega^{2}\left(k_{1}\right)\left\langle \hat{v}_{\boldsymbol{\boldsymbol{k}}_{1}}\hat{v}_{\boldsymbol{\boldsymbol{k}}_{2}}\hat{p}_{\boldsymbol{\boldsymbol{k}_{3}}}\right\rangle -\omega^{2}\left(k_{3}\right)\left\langle \hat{p}_{\boldsymbol{\boldsymbol{k}}_{1}}\hat{p}_{\boldsymbol{\boldsymbol{k}}_{2}}\hat{v}_{\boldsymbol{\boldsymbol{k}_{3}}}\right\rangle \\
\\
 & + & 2\gamma\alpha\left(\widetilde{C}_{R}\left(\left|\boldsymbol{k}_{1}\right|\right)+\widetilde{C}_{R}\left(\left|\boldsymbol{k}_{3}\right|\right)\right)\left\langle \hat{v}_{\boldsymbol{\boldsymbol{k}}_{2}}\hat{v}_{\boldsymbol{\boldsymbol{k}}_{1}+\boldsymbol{\boldsymbol{k}}_{3}}\right\rangle \\
\\
\frac{d\left\langle \hat{p}_{\boldsymbol{\boldsymbol{k}}_{1}}\hat{p}_{\boldsymbol{\boldsymbol{k}}_{2}}\hat{v}_{\boldsymbol{\boldsymbol{k}_{3}}}\right\rangle }{d\eta} & = & \left\langle \hat{p}_{\boldsymbol{\boldsymbol{k}}_{1}}\hat{p}_{\boldsymbol{\boldsymbol{k}}_{2}}\hat{p}_{\boldsymbol{\boldsymbol{k}_{3}}}\right\rangle -\omega^{2}\left(k_{1}\right)\left\langle \hat{v}_{\boldsymbol{\boldsymbol{k}}_{1}}\hat{p}_{\boldsymbol{\boldsymbol{k}}_{2}}\hat{v}_{\boldsymbol{\boldsymbol{k}_{3}}}\right\rangle -\omega^{2}\left(k_{2}\right)\left\langle \hat{p}_{\boldsymbol{\boldsymbol{k}}_{1}}\hat{v}_{\boldsymbol{\boldsymbol{k}}_{2}}\hat{v}_{\boldsymbol{\boldsymbol{k}_{3}}}\right\rangle \\
\\
 & + & 2\gamma\alpha\left(\widetilde{C}_{R}\left(\left|\boldsymbol{k}_{1}\right|\right)+\widetilde{C}_{R}\left(\left|\boldsymbol{k}_{2}\right|\right)\right)\left\langle \hat{v}_{\boldsymbol{\boldsymbol{k}}_{3}}\hat{v}_{\boldsymbol{\boldsymbol{k}}_{1}+\boldsymbol{\boldsymbol{k}}_{2}}\right\rangle \\
\\
\frac{d\left\langle \hat{p}_{\boldsymbol{\boldsymbol{k}}_{1}}\hat{p}_{\boldsymbol{\boldsymbol{k}}_{2}}\hat{p}_{\boldsymbol{\boldsymbol{k}_{3}}}\right\rangle }{d\eta} & = & -\omega^{2}\left(k_{1}\right)\left\langle \hat{v}_{\boldsymbol{\boldsymbol{k}}_{1}}\hat{p}_{\boldsymbol{\boldsymbol{k}}_{2}}\hat{p}_{\boldsymbol{\boldsymbol{k}_{3}}}\right\rangle -\omega^{2}\left(k_{2}\right)\left\langle \hat{p}_{\boldsymbol{\boldsymbol{k}}_{1}}\hat{v}_{\boldsymbol{\boldsymbol{k}}_{2}}\hat{p}_{\boldsymbol{\boldsymbol{k}_{3}}}\right\rangle -\omega^{2}\left(k_{3}\right)\left\langle \hat{p}_{\boldsymbol{\boldsymbol{k}}_{1}}\hat{p}_{\boldsymbol{\boldsymbol{k}}_{2}}\hat{v}_{\boldsymbol{\boldsymbol{k}_{3}}}\right\rangle \\
\\
 & + & 2\gamma\alpha\widetilde{C}_{R}\left(\left|\boldsymbol{k}_{1}\right|\right)\left[\left\langle \hat{p}_{\boldsymbol{\boldsymbol{k}}_{2}}\hat{v}_{\boldsymbol{\boldsymbol{k}}_{1}+\boldsymbol{\boldsymbol{k}}_{3}}\right\rangle +\left\langle \hat{p}_{\boldsymbol{\boldsymbol{k}}_{3}}\hat{v}_{\boldsymbol{\boldsymbol{k}}_{1}+\boldsymbol{\boldsymbol{k}}_{2}}\right\rangle \right]+2\gamma\alpha\widetilde{C}_{R}\left(\left|\boldsymbol{k}_{2}\right|\right)\\
\\
 & \times & \left[\left\langle \hat{p}_{\boldsymbol{\boldsymbol{k}}_{1}}\hat{v}_{\boldsymbol{\boldsymbol{k}}_{2}+\boldsymbol{\boldsymbol{k}}_{3}}\right\rangle +\left\langle \hat{p}_{\boldsymbol{\boldsymbol{k}}_{3}}\hat{v}_{\boldsymbol{\boldsymbol{k}}_{1}+\boldsymbol{\boldsymbol{k}}_{2}}\right\rangle \right]+2\gamma\alpha\widetilde{C}_{R}\left(\left|\boldsymbol{k}_{3}\right|\right)\left[\left\langle \hat{p}_{\boldsymbol{\boldsymbol{k}}_{1}}\hat{v}_{\boldsymbol{\boldsymbol{k}}_{2}+\boldsymbol{\boldsymbol{k}}_{3}}\right\rangle +\left\langle \hat{p}_{\boldsymbol{\boldsymbol{k}}_{2}}\hat{v}_{\boldsymbol{\boldsymbol{k}}_{1}+\boldsymbol{\boldsymbol{k}}_{3}}\right\rangle \right]\, ,
\end{array}  \label{eq:12}
\end{equation}
{where $\omega^2$ is given by Eq.~(\ref{eq:frequency})}. 

Combining those eight equations into a single one for
$\left\langle \hat{v}_{\boldsymbol{\boldsymbol{k}}_{1}}\hat{v}_{\boldsymbol{\boldsymbol{k}}_{2}}\hat{v}_{\boldsymbol{\boldsymbol{k}_{3}}}\right\rangle $
yields an equation of order eight which is not too illuminating. So
to get a simpler equation of lower order we adopt the equilateral
configuration limit in which all the momenta $\boldsymbol{\boldsymbol{k}}_{1}$,
$\boldsymbol{\boldsymbol{k}}_{2}$, $\boldsymbol{\boldsymbol{k}}_{3}$
have the same modulus $k=\left|\boldsymbol{k}\right|$ so that $\omega^{2}\left(k_{i}\right)\equiv\omega^{2}$$\left(k\right)$
and $\widetilde{C}_{R}\left(\left|\boldsymbol{k}_{i}\right|\right)=\widetilde{C}_{R}\left(\left|\boldsymbol{k}\right|\right)=\widetilde{C}_{R}\left(k\right)$.
Doing so, the order of equation can be reduced into an equation
of order four as we will see now. Moreover in this way we will be able to compute the overall amplitude $f_{\rm NL}$ of the decoherence induced bispectrum. In the rest of this part, we will
adopt the notation $\left\langle \hat{v}_{\boldsymbol{\boldsymbol{k}}_{1}}\hat{v}_{\boldsymbol{\boldsymbol{k}}_{2}}\hat{v}_{\boldsymbol{\boldsymbol{k}_{3}}}\right\rangle \equiv B_{vvv}$,
$\left\langle \hat{v}_{\boldsymbol{\boldsymbol{k}}_{1}}\hat{v}_{\boldsymbol{\boldsymbol{k}}_{2}}\hat{p}_{\boldsymbol{\boldsymbol{k}_{3}}}\right\rangle =B_{vvp}$ and similar expressions.
Since we are interested in the bispectrum of curvature perturbations
$B_{vvv}$, then, the strategy now is to differentiate the first equation of  (\ref{eq:12}) and use the other equations in (\ref{eq:12}) in order to obtain a
closed differential equation in $B_{vvv}$. Let us show how it works}\footnote{In these equations primes denote derivatives w.r.t. conformal time in order to simplify the notations.} 
\begin{equation}
\begin{array}{lll}
\\
\frac{d^{2}B_{vvv}}{d\eta^{2}} & =2\left[B_{vpp}+B_{pvp}+B_{ppv}\right]-3\omega^{2}B_{vvv}\\
\\
\frac{d^{3}B_{vvv}}{d\eta^{3}} & =6B_{ppp}-4\omega^{2}\left[B_{vvp}+B_{vpv}+B_{pvv}\right]-3\omega^{2}\frac{dB_{vvv}}{d\eta}-6\omega\omega^{\prime}B_{vvv}\\
\\
 & +4\gamma\alpha\widetilde{C}_{R}\left(k\right)\left[\left\langle \hat{v}_{\boldsymbol{\boldsymbol{k}}_{1}}\hat{v}_{\boldsymbol{\boldsymbol{k}}_{2}+\boldsymbol{\boldsymbol{k}}_{3}}\right\rangle +\left\langle \hat{v}_{\boldsymbol{\boldsymbol{k}}_{2}}\hat{v}_{\boldsymbol{\boldsymbol{k}}_{1}+\boldsymbol{\boldsymbol{k}}_{3}}\right\rangle +\left\langle \hat{v}_{\boldsymbol{\boldsymbol{k}}_{3}}\hat{v}_{\boldsymbol{\boldsymbol{k}}_{1}+\boldsymbol{\boldsymbol{k}}_{2}}\right\rangle \right]\\
\\
\frac{d^{4}B_{vvv}}{d\eta^{4}} & =-14\omega^{2}\left[B_{vpp}+B_{pvp}+B_{ppv}\right]-8\omega\omega^{\prime}\left[B_{vvp}+B_{vpv}+B_{pvv}\right]\\
\\
 & +12\omega^{2}B_{vvv}-12\omega\omega^{\prime}\frac{dB_{vvv}}{d\eta}-3\omega^{2}\frac{d^{2}B_{vvv}}{d\eta^{2}}-6\omega^{\prime2}B_{vvv}-6\omega\omega^{\prime\prime}B_{vvv}\\
\\
 & +24\gamma\alpha\widetilde{C}_{R}\left(k\right)\left[\left\langle \hat{p}_{\boldsymbol{\boldsymbol{k}}_{1}}\hat{v}_{\boldsymbol{\boldsymbol{k}}_{2}+\boldsymbol{\boldsymbol{k}}_{3}}\right\rangle +\left\langle \hat{p}_{\boldsymbol{\boldsymbol{k}}_{2}}\hat{v}_{\boldsymbol{\boldsymbol{k}}_{1}+\boldsymbol{\boldsymbol{k}}_{3}}\right\rangle +\left\langle \hat{p}_{\boldsymbol{\boldsymbol{k}}_{3}}\hat{v}_{\boldsymbol{\boldsymbol{k}}_{1}+\boldsymbol{\boldsymbol{k}}_{2}}\right\rangle \right]\\
\\
 & +4\alpha\left(\gamma\widetilde{C}_{R}\left(k\right)\right)^{\prime}\left[\left\langle \hat{v}_{\boldsymbol{\boldsymbol{k}}_{1}}\hat{v}_{\boldsymbol{\boldsymbol{k}}_{2}+\boldsymbol{\boldsymbol{k}}_{3}}\right\rangle +\left\langle \hat{v}_{\boldsymbol{\boldsymbol{k}}_{2}}\hat{v}_{\boldsymbol{\boldsymbol{k}}_{1}+\boldsymbol{\boldsymbol{k}}_{3}}\right\rangle +\left\langle \hat{v}_{\boldsymbol{\boldsymbol{k}}_{3}}\hat{v}_{\boldsymbol{\boldsymbol{k}}_{1}+\boldsymbol{\boldsymbol{k}}_{2}}\right\rangle \right]\\
\\
 & +4\gamma\alpha\widetilde{C}_{R}\left(k\right)\frac{d}{d\eta}\left[\left\langle \hat{v}_{\boldsymbol{\boldsymbol{k}}_{1}}\hat{v}_{\boldsymbol{\boldsymbol{k}}_{2}+\boldsymbol{\boldsymbol{k}}_{3}}\right\rangle +\left\langle \hat{v}_{\boldsymbol{\boldsymbol{k}}_{2}}\hat{v}_{\boldsymbol{\boldsymbol{k}}_{1}+\boldsymbol{\boldsymbol{k}}_{3}}\right\rangle +\left\langle \hat{v}_{\boldsymbol{\boldsymbol{k}}_{3}}\hat{v}_{\boldsymbol{\boldsymbol{k}}_{1}+\boldsymbol{\boldsymbol{k}}_{2}}\right\rangle \right] \, ,
\end{array} \label{eq:13}
\end{equation}
{and notice that from the the first equation in (\ref{eq:13}) 
}
\begin{equation}
B_{vpp}+B_{pvp}+B_{ppv}=\frac{1}{2}\left(\frac{d^{2}B_{vvv}}{d\eta^{2}}+3\omega^{2}B_{vvv}\right)\,. 
\end{equation}
{Therefore, substituting it in the fourth derivative equation of $B_{vvv}$, in addition to using the first equation in (\ref{eq:12}),  we get our
final differential equation}
\begin{equation}
\frac{d^{4}B_{vvv}}{d\eta^{4}}+10\omega^{2}\frac{d^{2}B_{vvv}}{d\eta^{2}}+20\omega\omega^{\prime}\frac{dB_{vvv}}{d\eta}+\left[9\omega^{2}+6\omega^{\prime2}+6\omega\omega^{\prime\prime}\right]B_{vvv}=S\left(k,\eta\right)\,,\label{15}
\end{equation}
{where at leading order in $\gamma$, the source function
$S\left(k,\eta\right)$ is given by }
\begin{equation}
\begin{array}{rlc}
\\
S\left(k,\eta\right)= & 24\gamma\alpha\widetilde{C}_{R}\left(k\right)\left[\mathfrak{Re}\left\langle \hat{p}_{\boldsymbol{\boldsymbol{k}}_{1}}\hat{v}_{\boldsymbol{\boldsymbol{k}}_{2}+\boldsymbol{\boldsymbol{k}}_{3}}\right\rangle +\mathfrak{Re}\left\langle \hat{p}_{\boldsymbol{\boldsymbol{k}}_{2}}\hat{v}_{\boldsymbol{\boldsymbol{k}}_{1}+\boldsymbol{\boldsymbol{k}}_{3}}\right\rangle +\mathfrak{Re}\left\langle \hat{p}_{\boldsymbol{\boldsymbol{k}}_{3}}\hat{v}_{\boldsymbol{\boldsymbol{k}}_{1}+\boldsymbol{\boldsymbol{k}}_{2}}\right\rangle \right]\\
\\
 & +4\alpha\left(\gamma\widetilde{C}_{R}\left(k\right)\right)^{\prime}\left[\left\langle \hat{v}_{\boldsymbol{\boldsymbol{k}}_{1}}\hat{v}_{\boldsymbol{\boldsymbol{k}}_{2}+\boldsymbol{\boldsymbol{k}}_{3}}\right\rangle +\left\langle \hat{v}_{\boldsymbol{\boldsymbol{k}}_{2}}\hat{v}_{\boldsymbol{\boldsymbol{k}}_{1}+\boldsymbol{\boldsymbol{k}}_{3}}\right\rangle +\left\langle \hat{v}_{\boldsymbol{\boldsymbol{k}}_{3}}\hat{v}_{\boldsymbol{\boldsymbol{k}}_{1}+\boldsymbol{\boldsymbol{k}}_{2}}\right\rangle \right]\\
\\
 & +4\gamma\alpha\widetilde{C}_{R}\left(k\right)\frac{d}{d\eta}\left[\left\langle \hat{v}_{\boldsymbol{\boldsymbol{k}}_{1}}\hat{v}_{\boldsymbol{\boldsymbol{k}}_{2}+\boldsymbol{\boldsymbol{k}}_{3}}\right\rangle +\left\langle \hat{v}_{\boldsymbol{\boldsymbol{k}}_{2}}\hat{v}_{\boldsymbol{\boldsymbol{k}}_{1}+\boldsymbol{\boldsymbol{k}}_{3}}\right\rangle +\left\langle \hat{v}_{\boldsymbol{\boldsymbol{k}}_{3}}\hat{v}_{\boldsymbol{\boldsymbol{k}}_{1}+\boldsymbol{\boldsymbol{k}}_{2}}\right\rangle \right] \,. 
\end{array}\label{eq:15}
\end{equation}
{An important remark is that the terms involving $\gamma$
are not explicitly proportional to $\delta^{(3)}\left(\boldsymbol{\boldsymbol{k}}_{1}+\boldsymbol{\boldsymbol{k}}_{2}+\boldsymbol{\boldsymbol{k}}_{3}\right)$,
where the presence of the Dirac delta function ensures the three Fourier modes
of the bispectrum forming a triangle. However, remember that the system is
solved through a perturbative expansion in $\gamma$ such that, during
the first iteration, the Lindblad term containing the correlators
$\left\langle \hat{p}_{\boldsymbol{\boldsymbol{k}}_{1}}\hat{v}_{\boldsymbol{\boldsymbol{k}}_{2}+\boldsymbol{\boldsymbol{k}}_{3}}\right\rangle ...,$$\left\langle \hat{v}_{\boldsymbol{\boldsymbol{k}}_{1}}\hat{v}_{\boldsymbol{\boldsymbol{k}}_{2}+\boldsymbol{\boldsymbol{k}}_{3}}\right\rangle ...$ should be evaluated in the free theory. Therefore, we retrive the expectation that the correlators 
turn out to be proportional indeed to $\delta^{(3)}\left(\boldsymbol{\boldsymbol{k}}_{1}+\boldsymbol{\boldsymbol{k}}_{2}+\boldsymbol{\boldsymbol{k}}_{3}\right)$, implying $\boldsymbol{\boldsymbol{k}}_{2}+\boldsymbol{\boldsymbol{k}}_{3}=-\boldsymbol{k}_{1}\equiv-\boldsymbol{k}$
so that we can write $\left\langle \hat{p}_{\boldsymbol{\boldsymbol{k}}}\hat{v}_{-\boldsymbol{\boldsymbol{k}}}\right\rangle ,\left\langle \hat{v}_{\boldsymbol{\boldsymbol{k}}}\hat{v}_{-\boldsymbol{\boldsymbol{k}}}\right\rangle $.
Equation (\ref{eq:15}) becomes}
\begin{equation}
\begin{array}{rlc}
\\
S\left(k,\eta\right)= & 72\gamma\alpha\widetilde{C}_{R}\left(k\right)\mathfrak{Re}\left\langle \hat{p}_{\boldsymbol{\boldsymbol{k}}}\hat{v}_{\boldsymbol{-\boldsymbol{k}}}\right\rangle +12\alpha\left(\gamma\widetilde{C}_{R}\left(k\right)\right)^{\prime}\left\langle \hat{v}_{\boldsymbol{\boldsymbol{k}}}\hat{v}_{\boldsymbol{-\boldsymbol{k}}}\right\rangle +12\gamma\alpha\widetilde{C}_{R}\left(k\right)\frac{d}{d\eta}\left\langle \hat{v}_{\boldsymbol{\boldsymbol{k}}}\hat{v}_{\boldsymbol{-\boldsymbol{k}}}\right\rangle \\
\\
\end{array}\,.\label{eq:16}
\end{equation}
{Our result (\ref{15}) endorses the suggestion made in \cite{martin2018non}
according to which one can even conjecture that any correlator must obey
a linear differential equation with a source term arising from the
interaction with the environment.}{\par}

{The analogy between our equation (\ref{15}) and the one for the power spectrum and trispectrum derived in ~\cite{martin2018observational,martin2018non},
makes us suggest that an exact solution is given by}
\begin{equation}
B_{vvv}=\frac{4}{3W^{3}}\intop_{-\infty}^{\eta}d\eta^{\prime}S\left(k,\eta^{\prime}\right)\mathscr{\mathfrak{Im^{3}\left[\mathrm{v_{\mathbf{k}}\left(\eta^{\prime}\right)v_{\mathbf{k}}^{\ast}\left(\eta\right)}\right]}}\,,\label{eq:17}
\end{equation}
{$W$ being the Wronskian, as in Eq.~(\ref{eq:19}) .  Indeed, using the Mukhanov-Sasaki equation $v_{\mathbf{k}}^{\prime\prime}+\omega^{2}v_{\mathbf{k}}=0$
in addition to $v_{k}=\left(v_{k}^{R}+iv_{k}^{I}\right)/\sqrt{2}$,
we can check with a straightforward but lengthy computation that (\ref{eq:17})
provides an exact solution to (\ref{15}). It is worth to mention
that the bispectrum we obtained is purely caused by decoherence, because
we started with the quadratic Hamiltonian for the system which does not generate  
primordial non-Guassianities.}{\par}

{Computing the integral (\ref{eq:17}) will enable us to discuss
the non linearity-parameter $f_{NL}$ characterizing the amplitude
of the curvature bispectrum, which is defined as the ratio between the bispectrum
in the equilateral configuration to the square of power spectrum of
curvature perturbation \cite{bartolo2004non,chen2010primordial}}
\begin{equation}
f_{NL}=\frac{5}{18}\frac{B_{\zeta\zeta\zeta}\left(k,k,k\right)}{P_{\zeta}^{2}(k)}\,,\label{eq:19-1}
\end{equation}
where $P_{\zeta}(k)=H^{2}/(4\epsilon M_{pl}^{2}k^{3})=P_{vv}/(2\epsilon M_{pl}^{2}a^{2})$
is the dimensionless power spectrum of the curvature perturbation
related to Mukhanov-Sasaki variable by $\zeta=- v/(\sqrt{2\epsilon}M_{pl}a)$
. Therefore, our $f_{NL}$ could be expressed as~\footnote{Notice that we are neglecting corrections at second-order in the perturbations connecting the Mukhanov-Sasaki variable with $\zeta$, since these are of gravitational strength and hence slow-roll suppressed w.r.t. to the contribution we are computing here.}
\begin{equation}
f_{NL}=\frac{5}{18}\frac{\sqrt{2\epsilon}M_{pl}aB_{vvv}\left(k,k,k\right)}{P_{vv}^{2}(k)}=\frac{10}{27W^{3}}\frac{\sqrt{2\epsilon}M_{pl}a}{P_{vv}^{2}(k)}\intop_{-\infty}^{\eta}d\eta^{\prime}S\left(k,\eta^{\prime}\right)\mathscr{\mathfrak{Im^{3}\left[\mathrm{v_{\mathbf{k}}\left(\eta^{\prime}\right)v_{\mathbf{k}}^{\ast}\left(\eta\right)}\right]}}\,.\label{eq:Fnl}
\end{equation}

{ Needless to say that computing the above integral  is not
straightforward, and some approximations are needed to rend
it tractable. The details of the computation
are gathered in the appendix (see \ref{sec:Appendix}), and here we
will quote directly the final results we obtain for the $f_{NL}$ expressions as a function of the free
parameter $p$.\footnote{We are using in the following expressions of $f_{NL}$ the dimensionless curvature power
spectrum $\mathcal{P_{\zeta}}=H^{2}/\left(8\pi^{2}M_{\text{pl}}^{2}\epsilon\right)\simeq2.2\times10^{-9}$. \label{footnote 27}}}{\par}
\begin{itemize}
\item $p=1$, in this case the contribution coming from\footnote{ The expressions corresponding to $B_{vvv}^{\left(1\right)}$, $B_{vvv}^{\left(2\right)}$, and  $B_{vvv}^{\left(3\right)}$ are given in the Appendix \ref{sec:Appendix}. } $B_{vvv}^{\left(1\right)}$
is negligible compared to the other two
\begin{equation}
\begin{alignedat}{1}f_{NL}= & \frac{5}{9\pi\sqrt{\mathcal{P}_{\zeta}}}\frac{\alpha k_{\gamma}^{2}}{k_{*}}\left(\frac{k}{k_{*}}\right)^{-3}\left[\frac{5}{3}\left(Hl_{E}\right)^{-1}\sin\left(Hl_{E}\right)^{-1}\left(\frac{k}{k_{*}}\right)^{-3}e^{3\left(N-N_{*}\right)}\right.\\
\\
 & \left.-\frac{32}{9}\cos\left(Hl_{E}\right)^{-1}\left(\frac{k}{k_{*}}\right)^{-3}e^{\left(N-N_{*}\right)}+\left(Hl_{E}\right)^{-1}\cos^{3}\left(Hl_{E}\right)^{-1}\right]\\
\\
\end{alignedat}
\end{equation}
\item $p=2$, in this case the contribution coming from $B_{vvv}^{\left(1\right)}$
is negligible compared to the other two
\begin{equation}
f_{NL}=\frac{5}{9\pi\sqrt{\mathcal{P}_{\zeta}}}\frac{\alpha k_{\gamma}^{2}}{k_{*}}\left(\frac{k}{k_{*}}\right)^{-2}\left[\cos^{3}\left(Hl_{E}\right)^{-1}+\frac{3}{4}Si\left(-\left(Hl_{E}\right)^{-1}\right)-\frac{11}{16}Ci\left(\frac{k}{k_{*}}e^{-\left(N-N_{*}\right)}\right)\right]
\end{equation}
\item $3\leq p\leq5$, in this case the contribution coming from both $B_{vvv}^{\left(1\right)}$
and $B_{vvv}^{\left(3\right)}$ is negligible compared to the $B_{vvv}^{\left(2\right)}$,
thus we get
\begin{equation}
\label{4.30newtext}
f_{NL}=\frac{5}{9\pi\sqrt{\mathcal{P}_{\zeta}}}\frac{\alpha k_{\gamma}^{2}}{k_{*}}\left(\frac{k}{k_{*}}\right)^{p-4}\left(Hl_{E}\right)^{p-2}\cos^{3}\left(Hl_{E}\right)^{-1}
\end{equation}
\item $p=6$

\begin{equation}
\begin{alignedat}{1}f_{NL}= & \frac{5}{9\pi\sqrt{\mathcal{P}_{\zeta}}}\frac{\alpha k_{\gamma}^{2}}{k_{*}}\left[\left(\frac{k}{k_{*}}\right)^{2}\left(Hl_{E}\right)^{4}\cos^{3}\left(Hl_{E}\right)^{-1}+\frac{49}{30}\left(\frac{k}{k_{*}}\right)^{3}e^{-\left(N-N_{*}\right)}\right.\\
\\
 & \left.-\frac{17}{60}\frac{k}{k_{*}}e^{-\left(N-N_{*}\right)}-3\left(\frac{k}{k_{*}}\right)^{3}e^{-3\left(N-N_{*}\right)}\left(\text{ln}\left(\frac{k}{k_{*}}\right)+N_{*}-N\right)-\frac{3}{2}\left(\frac{k}{k_{*}}\right)^{3}e^{-3\left(N-N_{*}\right)}\right]\\
\\
\\
\end{alignedat}
\end{equation}

\item for $p>6$, taking into account all the terms, including the subleading,
we get 

\begin{equation}
\begin{alignedat}{1}f_{NL}= & \frac{5}{9\pi\sqrt{\mathcal{P}_{\zeta}}}\frac{\alpha k_{\gamma}^{2}}{k_{*}}\left(\frac{k}{k_{*}}\right)^{3}\left[\left(Hl_{E}\right)^{p-2}\left(\frac{k}{k_{*}}\right)^{p-7}\cos^{3}\left(Hl_{E}\right)^{-1}-7\frac{\left(3p-4\right)e^{-\left(7-p\right)\left(N-N_{*}\right)}}{\left(p-4\right)\left(p-3\right)\left(p-2\right)\left(p-1\right)}\right.\\
\\
 & \left.-6\left(\frac{k}{k_{*}}\right)^{2}\frac{e^{-\left(9-p\right)\left(N-N_{*}\right)}}{\left(p-6\right)\left(p-5\right)\left(p-4\right)}-6\frac{e^{-\left(7-p\right)\left(N-N_{*}\right)}}{\left(p-6\right)\left(p-5\right)\left(p-4\right)}\right]\\
\\
\\
\end{alignedat}
\end{equation}
\end{itemize}
 Some comments are in order here. First, at this level of the results, if we start from Eq.~(\ref{eq:1}) where in full generality we adopt the same power of $p$ as in (\ref{eq:coupling}) for the various contributions in the Lindblad term, then it is interesting to notice that, imposing (almost) scale-invariance of the power spectrum and negligible (scale-dependent) decoherence induced corrections to the power-spectrum on large (CMB) scales, guided by observational constraints, then the resulting bispectra would be strongly scale-dependent. This, e.g., could be the case of $p=5$ (see Eq.~(\ref{equt:3.13}) and Eq.~(\ref{4.30newtext}),  and the discussion there), leading to $f_{\rm NL} \propto k$.  Notice that this remains true even when accounting for the additional corrections to the power-spectrum discussed in details in Sec.~\ref{subsec:Additional-correction-to}.~\footnote{\label{footnote26}Indeed imposing (nearly) scale-invariance of the decoherence-induced corrections to the scalar power-spectrum might be a too restrictive condition, since the measured (almost) scale independent scalar power spectrum could be merely provided by the leading order, 
and if we increase the measurements sensitivity we might be able to detect a non-trivial weak scale-dependence which reflects the decoherence effect for $p\neq 5$. \label{f26}}

{On the other hand it is quite remarkable that we can obtain a scale invariant
bispectrum for $p=4$. The latter result, as already mentioned, might
correspond to a massive scalar field as environment as can be checked
using the relation (\ref{32-1}) and remembering that the bispectrum was
induced by the Lindblad term
\begin{equation}
\frac{\gamma}{2}\alpha\int d^{3}\boldsymbol{x}d^{3}\boldsymbol{y}C_{R}\left(\boldsymbol{x},\boldsymbol{y}\right)\left\langle \left[\left[\hat{O},\hat{v}^{2}\right],\hat{v}\right]\right\rangle \, ,
\end{equation}
thus corresponding to $k=2$ and $l=1$ in the massive scalar field scenario discussed previously.}{\par}

{The scale invariance obtained in \cite{martin2018observational,martin2018non}
for the power spectrum and trispectrum was also corresponding to a
massive scalar field as environment,  and indeed within this model, one has enough freedom that the various correlators induced by decoherence effects turn out to be indeed scale-invariant.{\par}

{In spite of these results, we think that the scale independence
of various correlation functions is related to a more general factor
that is independent of the type of environment considered. In the
case of a linear interaction, $\hat{A}\left(\boldsymbol{x}\right)=\hat{v}$,
for which the Lindblad term is given by}
\begin{equation}
\frac{\gamma}{2}\int d^{3}\boldsymbol{x}d^{3}\boldsymbol{y}C_{R}\left(\boldsymbol{x},\boldsymbol{y}\right)\left\langle \left[\left[\hat{O},\hat{v}\left(\boldsymbol{x}\right)\right],\hat{v}\left(\boldsymbol{y}\right)\right]\right\rangle \,,\label{eq:54}
\end{equation}
{the scale independence of the curvature power spectrum}\footnote{{We remind that for the linear interaction only power spectrum
receives corrections while all the other higher order correlation
functions remain unaffected by the cosmic decoherence.}}{ was obtained for a value of $p=5$. In terms of the expansion
parameter $\alpha$, it correspond to the zeroth order, i.e to $\alpha^{0}$.
Through our choice of the form of pointer observable (\ref{eq:352}),
we considered $\hat{A}\left(\boldsymbol{x}\right)=\hat{v}+\alpha\hat{v}^{2}$,
and we noticed that the leading contribution to the curvature bispectrum was coming
from a term in (\ref{eq:54}) that is $\left(\hat{A}\left(\boldsymbol{x}\right)\right)^{2}\propto\alpha\hat{v}^{3}$
as could be seen in (\ref{eq:10}), and scale independence was found
for $p=4$. While, for the case of a pure quadratic interaction where
the Lindblad term (\ref{eq:54}) is $\left(\hat{A}\left(\boldsymbol{x}\right)\right)^{2}\propto\alpha^{2}\hat{v}^{4}$,
the scale invariance (of the power spectrum and trispectrum) was obtained in \cite{martin2018observational,martin2018non}  for $p=3$. In terms
of the expansion parameter $\alpha$, it correspond to the second
order, i.e to $\alpha^{2}$. Therefore we summarize our reasoning as follows
\begin{equation}
\begin{array}{c}
\mathrm{Value\,of\,p\,corresponding}\\
\mathrm{\,to\,scale\, independent\,}\\
\mathrm{corrections}
\end{array}\begin{cases}
\left(\hat{A}\left(\boldsymbol{x}\right)\right)^{2}\propto\alpha^{0}\rightarrow p=5\\
\\
\left(\hat{A}\left(\boldsymbol{x}\right)\right)^{2}\propto\alpha\rightarrow p=4\\
\\
\left(\hat{A}\left(\boldsymbol{x}\right)\right)^{2}\propto\alpha^{2}\rightarrow p=3
\end{cases}\,.\label{eq:61-1}
\end{equation}
}{\large\par}

{Based on this analysis and reasoning we conjecture a relation
that enables us to understand when decoherence does induce scale
independent  (contributions to the) correlation functions. Such a relation is simply linked to the order
of $\alpha$ in the Lindblad term 
\begin{equation}
\frac{\gamma}{2}\int d^{3}\boldsymbol{x}d^{3}\boldsymbol{y}C_{R}\left(\boldsymbol{x},\boldsymbol{y}\right)\left\langle \left[\left[\hat{O},\hat{A}\left(\boldsymbol{x}\right)\right],\hat{A}\left(\boldsymbol{y}\right)\right]\right\rangle \,,
\end{equation}
from which the decoherence induced contributions are coming, independently from the physical
process and the type of environment considered. We can take further the analysis
summarized in (\ref{eq:61-1}) and claim that the scale invariance
of a contribution coming from the Lindblad term in $d\left\langle \hat{O}\right\rangle /d\eta$
of the order} $\alpha$$^{n}$, with $n=k+l-2$,
\begin{equation}
\frac{\gamma}{2}\alpha^{k+l-2}\int d^{3}\boldsymbol{x}d^{3}\boldsymbol{y}C_{R}\left(\boldsymbol{x},\boldsymbol{y}\right)\left\langle \left[\left[\hat{O},\hat{v}^{k}\right],\hat{v}^{l}\right]\right\rangle \,,\label{eq:63-2}
\end{equation}
{\large{}corresponds to the value of $p$ given by}
\begin{equation}
p=5-n\,.\label{eq:73}
\end{equation}

{The latter claim is valid, at least, in all the computations
made so far, namely power spectrum, bispectrum and trispectrum , as we explained in (\ref{eq:61-1}). Since the order of $\alpha$ in (\ref{eq:63-2}), i.e $n$,  is related to $k$ and $l$,  the relation (\ref{eq:73}) could be written, equivalently, as
\begin{equation}
p=7-k-l\ ,\label{eq:73-1}
\end{equation}
surprisingly,
this relation turns to be just identical to that of the physical process
we suggested of a massive scalar field as environment, see equation
(\ref{32-1}). However, any physical process, or type of environment,
which produces a pointer observable of the form (\ref{eq:352}), would
produce corrections to the power spectrum or induce non-Gaussianities which are scale independent for a value of $p$ that is dictated by (\ref{eq:73}), according to the order of $\alpha$ in (\ref{eq:63-2})
from which the corrections were derived. For example, imposing the bispectrum to be scale independent, regardless of the type of environment, then will allow to get upper 
bounds on the interaction strength with the environment as a function of $p.$ }{\par}

 Conversely, our general result (\ref{eq:73}) shows once more that in the case where it is possible to have a unique value of $p$ (see~(\ref{eq:coupling})) for the interaction strength $\gamma$ in the Lindblad term~(\ref{eq:201}) such 
that, e.g., the power spectrum is scale-invariant, then necessarily this would imply that all the other correlators will unavoidably feature a specific and predictable scale-dependence.

{Notice also that $f_{NL}$ is suppressed by the square root  \footnote{To see this, we remind  that we expressed $f_{NL}$ in terms of the dimensionless curvature power
spectrum $\mathcal{P_{\zeta}}$, see footnote \ref{footnote 27}. }
of the slow-roll parameter $\epsilon$, which is to be compared with
the standard bispectrum obtained in the standard single field slow-roll inflation that is slow-roll suppressed by the first and second slow roll parameters $\epsilon$ and  $\epsilon_{2}$, respectively. Therefore the bispectrum induced by cosmic decoherence is parametrically greater w.r.t to the standard one.~\footnote{ It might well be that however the two bispectra are characterized by different {\it shapes}. A full analysis of this aspect is beyond the scope of this paper, where we have focused on the possibility of generating a non-vanishing bispectrum from decoherence effects, on its overall amplitude and scale dependence.} 
Also, $M_{Pl}/H$ and specific scale dependencies contribute to increase the amplitude and to compensate the suppression by $Hl_{E}\ll1$, present for some of the bispectra found, and by the coupling $\alpha$ which is supposed to be small.}

We conclude this section by presenting some illustrating  plots for the overall non-Gaussianity amplitude $f_{NL}$. 
In Fig.~\ref{fig:1} we present $f_{NL}$ as a function
of $k/k_{*}$ for various values of the free parameter $p$.  As mentioned before, two cases are of particular interest: $p$=4 corresponds to a scale-invariant bispectrum, in agreement with our general criterium given in~(\ref{eq:73}); $p=5$ would correspond instead to a scale-invariant decoherence-induced contribution to the power spectrum (and negligible scale-dependent contributions on large CMB scales), resulting here in a non-linearity parameter scaling as $f_{\rm NL} \propto k$. 
In Fig.~\ref{fig:2} we present the plots of $f_{NL}$ for different values of $Hl_{E}$, i.e. different values of environment correlation length compared to Hubble radius. We can see that this factor plays a major role for $2 < p \leq 7$, while has less effect on other values of $p$.   If one looks at the  analytical expressions for $f_{\rm NL}$ this is clearly explained by the fact the dependence on $Hl_{E}$ is strongly suppressed by the presence of $e^{N-N_*}$ factors outside the range $2<p \leq 7$. Indeed, we study the  role of number of e-folds $N-N_{*}$ in Fig. \ref{fig:3}, where, as expected from the analytical expressions of $f_{NL}$, this factor has less effect in the range $2 \leq p \leq 7 $.}

Since we are dealing with an open quantum system, it would be interesting to constrain the interaction system-environment. To this end, we take advantage of the   observational constrains from {\it Planck} on equilateral bispectrum, from which we have \footnote{Here, for simplicity,  we are smoothing the constraint from $-63<f_{NL}<120$ into $|f_{NL}|<120$.  Also, a completely self-consistent data analysis would entail to compute the non-Gaussian shape of the bispectrum from cosmic-decoherence, and to apply to this shape a proper estimator for primordial non-Gaussianity, which is however beyond the scope of this paper. The constraints we obtain here are  meant as a (conservative) estimate that one could obtain by applying an equilateral primordial non-Gaussianity estimator.

} $|f_{NL}|<120$ at $ 95\% $ CL \cite{akrami2020planck}. We can proceed in two different ways to get such constraints. Either, we constrain the overall factor $(\alpha k_{\gamma}^{2}/k_{*})$ appearing in front of the various expressions of $f_{NL}$, which sets the overall non-Gaussiantiy amplitude, as shown in Fig.~\ref{fig:4}, or we fix $\alpha k_{\gamma}$ and constrain the factor $k_{\gamma}/ k_{*}$. This last choice serves to make a comparison with the previous constraints on $k_{\gamma}/ k_{*}$ from decoherence effects on the curvature power-spectrum obtained in \cite{martin2018observational} possible. Obviously, the downside of second choice is the need to fix $\alpha k_{\gamma}$ with a given value. In Fig.~\ref{fig:5}  we present such constraints on $k_{\gamma}/ k_{*}$ for different values of $\alpha k_{\gamma}$.

Finally, in Fig.~\ref{fig:6}, we shed light on the effect of environment correlation length on the obtained constraints. Here, we fix the value of $\alpha k_{\gamma}$ to $10^{9}$ and present the constraints for two different values of $Hl_{E}$.  Fig.~\ref{fig:5} and Fig.~\ref{fig:6} can be compared directly to, e.g., Fig. 6 of \cite{martin2018observational}. It is interesting to notice that there are regions in the parameter space according to different values of $\alpha k_{\gamma}$ and $Hl_{E}$ for which observational constraints coming from {\it Planck} bispectrum measurements start to be competitive (if not stronger) than those from decoherence effects and constraints on the power spectrum. Of course this possibility is indeed model-dependent. One could still ask whether the values considered for $\alpha k_{\gamma}$ are physically relevant or constrained by other considerations. If one considers the parameter $\alpha$ as an expansion parameter in~(\ref{eq:2}) then typically one has to impose that $|\alpha \delta \varphi| \ll 1$. Therefore in this case one should check whether this might be consistent (and under which conditions) with values, e.g., of $\alpha k_{\gamma} =10^9$. 
One has to account that the constraints from almost-scale invariance of the power spectrum and decoherence implies $k_{\gamma}/k_* < 10^{-1}$~\cite{martin2018observational} (with $k_*=0.05$ Mpc $^{-1}$ $\sim 3 \times   10^{39}$ GeV$^{-1}$) which means imposing a lower bound on $\alpha$. Thus one can check that the perturbativity condition $|\alpha \delta \varphi| \ll 1$ implies an extremely low  values of the slow-roll parameter $\epsilon$ and hence of the standard tensor-to-scalar ratio $r$ (at least if one uses the (leading-order) dimensionless power-spectra such as the curvature power-spectrum $\mathcal{P}_{\zeta}=\mathcal{P}_{\delta \varphi}/(2 \epsilon M_{pl}^2) ~\sim 10^{-10}$).~\footnote{In this case the decoherence induced corrections on the tensor sector discussed in the section~\ref{sec:Decoherence-of-tensor-1} might become the dominant contribution.} 

On the other hand notice that not necessarily the expression~in~(\ref{eq:2}) should always be considered as an expansion, since the operator $\hat{A}$ enters in the interaction terms, thus defining the source to the power-spectra induced by the decoherence effects, and this source can be provided by fields that are different w.r.t. to the system field itself $\zeta$.

\begin{figure}[H]
\centering
\includegraphics[scale=0.6]{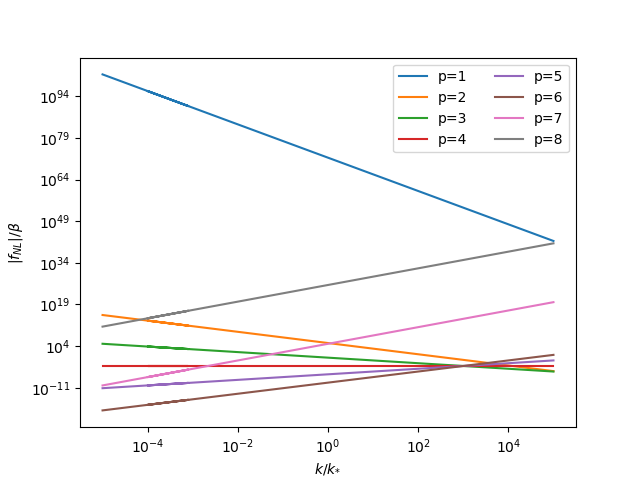}
\caption{Bispectrum parameter $f_{NL}$ absolute value rescaled by $\beta=\frac{\alpha k_{\gamma}^{2}}{k_{*}}$         as function of $\frac{k}{k_{*}}$
          for the various values of parameter $p.$ We choose $Hl_{E}=10^{-3}$, and
          $N-N_{*}=50$.}
\label{fig:1}
\end{figure}
\subsection{Additional corrections to the curvature power spectrum\label{subsec:Additional-correction-to}}
{As we already mentioned in the previous section, ~\cite{martin2018non,martin2018observational} considered
the effects induced by decoherence for two forms of the pointer
observable, linear $\hat{A}\left(\boldsymbol{x}\right)=\hat{v}$ and
pure quadratic $\hat{A}\left(\boldsymbol{x}\right)=\hat{v}^{2}$,
and when it comes to the power spectrum, $P_{vv}$, they obtained
non zero corrections for both cases. However, as we stated before, there are various  motivations to consider a pointer observable which is 
expanded perturbatively in $\hat{v}$ with an expansion coupling $\alpha$. By doing so and
in order to have a consistent and accurate result, we realized that
actually there is an extra correction which must be considered along
the pure quadratic one in the Lindblad term, because they are of the
same order in $\alpha$, namely second order. In what follows, we
show this last point and compute the additional correction that must
be added to the ones already obtained in \cite{martin2018observational}.}{\par}

\begin{figure}[H]
\centering
\includegraphics[scale=0.6]{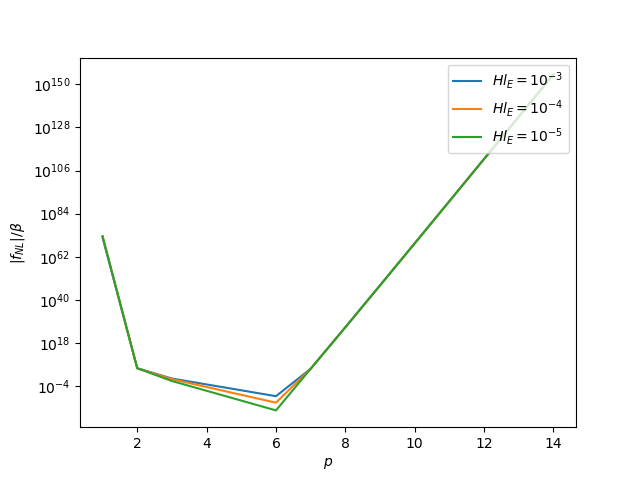}
\caption{Bispectrum parameter $f_{NL}$ absolute value rescaled by $\beta=\frac{\alpha k_{\gamma}^{2}}{k_{*}}$  as function of $p$
       for different values of $Hl_{E}$. We choose $\frac{k}{k_{*}}=1$,  and $N-N_{*}=50$.}
\label{fig:2}
\end{figure}

\begin{figure}[H]
\centering
\includegraphics[scale=0.6]{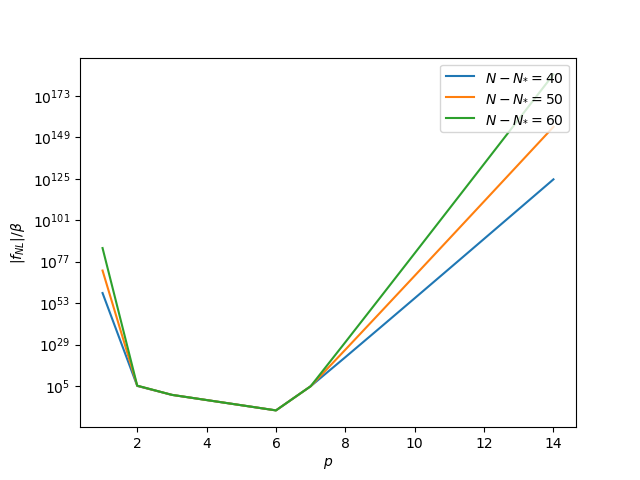}
\caption{Bispectrum parameter $f_{NL}$ absolute value rescaled by $\beta=\frac{\alpha k_{\gamma}^{2}}{k_{*}}$ as     function of $p$
       for different values of $N-N_{*}$. We choose $Hl_{E}=10^{-3}$,   and $\frac{k}{k_{*}}=1$.}
\label{fig:3}
\end{figure}

\begin{figure}[H]
\centering
\includegraphics[scale=0.6]{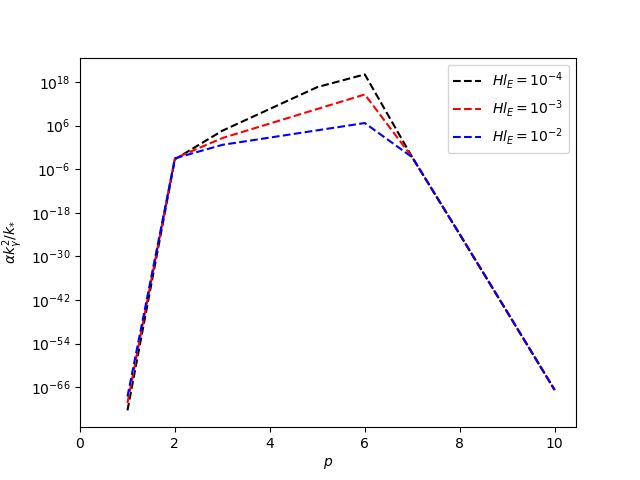}
\caption{Constraints on $\frac{\alpha k_{\gamma}^{2}}{k_{*}}$         corresponding to  $|f_{NL}|<120$ for different values of      $Hl_{E}$. We choose $\frac{k}{k_{*}}=1$, and          $N-N_{*}=50$.}
\label{fig:4}
\end{figure}

\begin{figure}[H]
\centering
\includegraphics[scale=0.6]{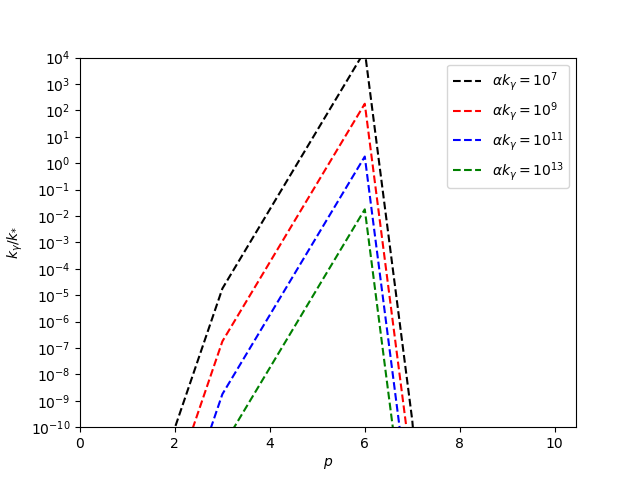}
\caption{Constraints on $\frac{k_{\gamma}}{k_{*}}$         corresponding to  $|f_{NL}|<120$ for different values of      $\alpha k_{\gamma}$. We choose $\frac{k}{k_{*}}=1$, $N-N_{*}=50$, and $Hl_{E}=10^{-3}$.}
\label{fig:5}
\end{figure}

\begin{figure}[H]
\centering
\includegraphics[scale=0.6]{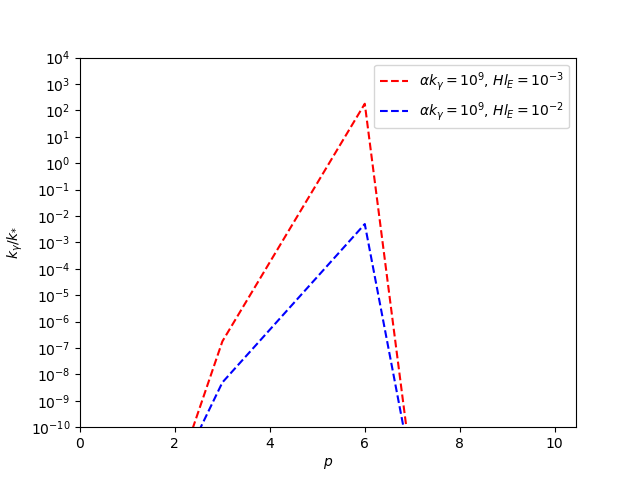}
\caption{Constraints on $\frac{k_{\gamma}}{k_{*}}$           corresponding to  $|f_{NL}|<120$ for different values of  $Hl_{E}$ and fixed value of $\alpha k_{\gamma}=10^{9}$. We choose $\frac{k}{k_{*}}=1$, and $N-N_{*}=50$. }
\label{fig:6}
\end{figure}

{Since we are interested in corrections up to order $\alpha^{2}$,
we adopt the pointer observable }
\begin{equation}
\hat{A}\left(\boldsymbol{x}\right)=\hat{v}+\alpha\hat{v}^{2}+\alpha^{2}\hat{v}^{3},\label{57}
\end{equation}
{and (\ref{eq:1}) becomes up to second order in $\alpha$}
\begin{equation}
\begin{array}{cllcc}
\frac{d\left\langle \hat{O}\right\rangle }{d\eta} & = & \left\langle \frac{\partial\hat{O}}{\partial\eta}\right\rangle -i\left\langle \left[\hat{O},\hat{H_{v}}\right]\right\rangle -\frac{\gamma}{2}\int d^{3}\boldsymbol{x}d^{3}\boldsymbol{y}C_{R}\left(\boldsymbol{x},\boldsymbol{y}\right)\\
\\
 & \times & \left\langle \left[\left[\hat{O},\hat{v}\left(\boldsymbol{x}\right)+\alpha\hat{v}^{2}\left(\boldsymbol{x}\right)+\alpha^{2}\hat{v}^{3}\left(\boldsymbol{x}\right)\right],\hat{v}\left(\boldsymbol{y}\right)+\alpha\hat{v}^{2}\left(\boldsymbol{y}\right)+\alpha^{2}\hat{v}^{3}\left(\boldsymbol{y}\right)\right]\right\rangle \\
\\
 & = & \left\langle \frac{\partial\hat{O}}{\partial\eta}\right\rangle -i\left\langle \left[\hat{O},\hat{H_{v}}\right]\right\rangle -\frac{\gamma}{2}\int d^{3}\boldsymbol{x}d^{3}\boldsymbol{y}C_{R}\left(\boldsymbol{x},\boldsymbol{y}\right)\left\{ \left\langle \left[\left[\hat{O},\hat{v}\left(\boldsymbol{x}\right)\right],\hat{v}\left(\boldsymbol{y}\right)\right]\right\rangle \right.\\
\\
 & + & \alpha\left\langle \left[\left[\hat{O},\hat{v}\left(\boldsymbol{x}\right)\right],\hat{v}^{2}\left(\boldsymbol{y}\right)\right]+\left[\left[\hat{O},\hat{v}^{2}\left(\boldsymbol{x}\right)\right],\hat{v}\left(\boldsymbol{y}\right)\right]\right\rangle \\
\\
 & + & \left.\alpha^{2}\left(\left\langle \left[\left[\hat{O},\hat{v}^{2}\left(\boldsymbol{x}\right)\right],\hat{v}^{2}\left(\boldsymbol{y}\right)\right]\right\rangle +\underbrace{\left\langle \left[\left[\hat{O},\hat{v}\left(\boldsymbol{x}\right)\right],\hat{v}^{3}\left(\boldsymbol{y}\right)\right]+\left[\left[\hat{O},\hat{v}^{3}\left(\boldsymbol{x}\right)\right],\hat{v}\left(\boldsymbol{y}\right)\right]\right\rangle }\right)\right\}\, . \\
\\
\end{array}\label{eq:58}
\end{equation}
{From this last equation we can see in the last line that
the pure quadratic interaction term $\left\langle \left[\left[\hat{O},\hat{v}^{2}\left(\boldsymbol{x}\right)\right],\hat{v}^{2}\left(\boldsymbol{y}\right)\right]\right\rangle $
is accompanied by a cross term coming from a combination of the linear
and cubic interaction $\left\langle \left[\left[\hat{O},\hat{v}\left(\boldsymbol{x}\right)\right],\hat{v}^{3}\left(\boldsymbol{y}\right)\right]+\left[\left[\hat{O},\hat{v}^{3}\left(\boldsymbol{x}\right)\right],\hat{v}\left(\boldsymbol{y}\right)\right]\right\rangle $.
It is this last term that could not be captured in the computations done in
\cite{martin2018observational}. We now compute the corresponding correction to the power spectrum 
$P_{vv}$. Fortunately, all what we will need to
do at the end is to add this correction to the linear and quadratic ones,
because, as we will see in a moment, the solution to the evolution equation of $P_{vv}$ is linear in the source functions which, on their turn, are proportional, apart from a numerical factor, to those  real  terms
in (\ref{eq:58}).}{\par}

{Before doing the computations, we need to Fourier transform
(\ref{eq:58}) using for the cubic terms
}
\begin{equation}
\hat{v}^{3}\left(\boldsymbol{x}\right)=\frac{1}{\left(2\pi\right)^{9/2}}\int d^{3}\boldsymbol{p}d^{3}\boldsymbol{p}_{1}d^{3}\boldsymbol{p}_{2}d^{3}\hat{v}_{\boldsymbol{p}_{1}}\hat{v}_{\boldsymbol{p}_{2}}\hat{v}_{\boldsymbol{p}-\boldsymbol{p}_{1}-\boldsymbol{p}_{2}}e^{i\boldsymbol{p}.\boldsymbol{x}}\, . 
\end{equation}
{Noticing that the term linear in $\alpha$ in (\ref{eq:58}) gives a vanishing contribution
to the correction of $P_{vv}$, and exploiting (\ref{eq:9 real}), the Fourier transform is given
by }

\begin{equation}
\begin{array}{cllcc}
\\
\frac{d\left\langle \hat{O}\right\rangle }{d\eta} & = & \left\langle \frac{\partial\hat{O}}{\partial\eta}\right\rangle -i\left\langle \left[\hat{O},\hat{H_{v}}\right]\right\rangle -\frac{\gamma}{2}\left(2\pi\right)^{3/2}\int d^{3}\boldsymbol{k}\widetilde{C}_{R}\left(\left|\boldsymbol{k}\right|\right)\left\langle \left[\left[\hat{O},\hat{v}_{\boldsymbol{\boldsymbol{k}}}\right],\hat{v}_{\boldsymbol{\boldsymbol{-k}}}\right]\right\rangle \\
\\
 & - & \frac{\gamma\alpha^{2}}{2\left(2\pi\right)^{3/2}}\int d^{3}\boldsymbol{k}d^{3}\boldsymbol{p}_{1}d^{3}\boldsymbol{p}_{2}\widetilde{C}_{R}\left(\left|\boldsymbol{k}\right|\right)\times\left(\left\langle \left[\left[\hat{O},\hat{v}_{\boldsymbol{p}_{1}}\hat{v}_{\boldsymbol{\boldsymbol{k}-p}_{1}}\right],\hat{v}_{\boldsymbol{p}_{2}}\hat{v}_{\boldsymbol{-\boldsymbol{k}-p}_{2}}\right]\right\rangle \right. \\
\\
 & + & \left.2\left\langle \left[\left[\hat{O},\hat{v}_{\boldsymbol{\boldsymbol{k}}}\right],\hat{v}_{\boldsymbol{p}_{1}}\hat{v}_{\boldsymbol{p}_{2}}\hat{v}_{\boldsymbol{-\boldsymbol{k}-\boldsymbol{p}_{1}-p}_{2}}\right]\right\rangle \right).\\
\\
\end{array} \label{eq:60}
\end{equation}
The next step is to derive the equations governing the various two-point
correlators
\begin{equation}
\begin{array}{llc}
\\
\frac{d\left\langle \hat{v}_{\boldsymbol{k}_{1}}\hat{v}_{\boldsymbol{k}_{2}}\right\rangle }{d\eta} & =\left\langle \hat{v}_{\boldsymbol{k}_{1}}\hat{p}_{\boldsymbol{k}_{2}}\right\rangle +\left\langle \hat{p}_{\boldsymbol{k}_{1}}\hat{v}_{\boldsymbol{k}_{2}}\right\rangle \\
\\
\frac{d\left\langle \hat{v}_{\boldsymbol{k}_{1}}\hat{p}_{\boldsymbol{k}_{2}}\right\rangle }{d\eta} & =\left\langle \hat{p}_{\boldsymbol{k}_{1}}\hat{p}_{\boldsymbol{k}_{2}}\right\rangle -\omega^{2}\left(k_{2}\right)\left\langle \hat{v}_{\boldsymbol{k}_{1}}\hat{v}_{\boldsymbol{k}_{2}}\right\rangle \\
\\
\frac{d\left\langle \hat{p}_{\boldsymbol{k}_{1}}\hat{v}_{\boldsymbol{k}_{2}}\right\rangle }{d\eta} & =\left\langle \hat{p}_{\boldsymbol{k}_{1}}\hat{p}_{\boldsymbol{k}_{2}}\right\rangle -\omega^{2}\left(k_{1}\right)\left\langle \hat{v}_{\boldsymbol{k}_{1}}\hat{v}_{\boldsymbol{k}_{2}}\right\rangle \\
\\
\frac{d\left\langle \hat{p}_{\boldsymbol{k}_{1}}\hat{p}_{\boldsymbol{k}_{2}}\right\rangle }{d\eta} & =-\omega^{2}\left(k_{2}\right)\left\langle \hat{p}_{\boldsymbol{k}_{1}}\hat{v}_{\boldsymbol{k}_{2}}\right\rangle -\omega^{2}\left(k_{1}\right)\left\langle \hat{v}_{\boldsymbol{k}_{1}}\hat{p}_{\boldsymbol{k}_{2}}\right\rangle +\gamma\left(2\pi\right)^{3/2}\widetilde{C}_{R}\left(\left|\boldsymbol{k}_{1}\right|\right)\delta\left(\boldsymbol{k}_{1}+\boldsymbol{k}_{2}\right)+\alpha^{2}\frac{\gamma}{\left(2\pi\right)^{3/2}}\\
\\
 & \times\left[4 \int d^{3}\boldsymbol{k}\widetilde{C}_{R}\left(\left|\boldsymbol{k}\right|\right)\left\langle \hat{v}_{\boldsymbol{k}+\boldsymbol{k}_{1}}\hat{v}_{-\boldsymbol{k}+\boldsymbol{k}_{2}}\right\rangle 
 +\underbrace{3\left(\widetilde{C}_{R}\left(\left|\boldsymbol{k}_{1}\right|\right)+\widetilde{C}_{R}\left(\left|\boldsymbol{k}_{2}\right|\right)\right)\int d^{3}\boldsymbol{k}\left\langle \hat{v}_{\boldsymbol{k}}\hat{v}_{\boldsymbol{k}_{1}+\boldsymbol{k}_{2}-\boldsymbol{k}}\right\rangle }
 \right]\, .
 \\
\end{array}\,,\label{eq:61}
\end{equation}
{The term marked with $\underbrace{}$ is the new contribution
to the power spectrum correction. Similarly to what happens for the system of equations governing the various bispectra in Eq.~(\ref{eq:12}), it is worth mentioning that, though decoherence effects appear explicitly only
in the correlator} $\left\langle \hat{p}_{\boldsymbol{k}_{1}}\hat{p}_{\boldsymbol{k}_{2}}\right\rangle $,
they do affect all the correlators since the latter are governed by a coupled
 system of equations, as it is clear from (\ref{eq:61}). Another point
to mention, which we already discussed in the case of the bispectrum,
is the appearance of }$\delta^{(3)}\left(\boldsymbol{k}_{1}+\boldsymbol{k}_{2}\right)$
{in the linear contribution to }$\left\langle \hat{p}_{\boldsymbol{k}_{1}}\hat{p}_{\boldsymbol{k}_{2}}\right\rangle ${, 
which ensures that the interaction with the environment preserves
statistical homogeneity. We do not see the Dirac delta function in the
last two terms $\propto\alpha^{2}$, however, remember that this term
is also $\propto\gamma$ and the system is solved through a perturbative
expansion in $\gamma$. During the first iteration, the last two Lindblad
terms contain the correlators} $\left\langle \hat{v}_{\boldsymbol{k}+\boldsymbol{k}_{1}}\hat{v}_{-\boldsymbol{k}+\boldsymbol{k}_{2}}\right\rangle ${\large{}
and }$\left\langle \hat{v}_{\boldsymbol{k}}\hat{v}_{\boldsymbol{k}_{1}+\boldsymbol{k}_{2}-\boldsymbol{k}}\right\rangle ${
to be computed in the free case, so they turn out to be proportional to }$\delta^{(3)} \left(\boldsymbol{k}_{1}+\boldsymbol{k}_{2}\right)$
{thus guaranteeing indeed that the solution is statistically homogeneous. In addition we adopted
the case }$\widetilde{C}_{R}\left(\boldsymbol{k}\right)\equiv\widetilde{C}_{R}\left(k\right)$,
{i.e an isotropic environmental correlation function. Thus
the system for isotropic solutions reduces to }
\begin{equation}
\begin{array}{rlc}
\\
\frac{dP_{vv}\left(k\right)}{d\eta} & =P_{vp}\left(k\right)+P_{pv}\left(k\right)\\
\\
\frac{dP_{vp}\left(k\right)}{d\eta} & =\frac{dP_{pv}\left(k\right)}{d\eta}=P_{pp}\left(k\right)-\omega^{2}\left(k\right)P_{vv}\left(k\right)\\
\\
\frac{dP_{pp}\left(k\right)}{d\eta} & =-\omega^{2}\left(k\right)\left(P_{vp}\left(k\right)+P_{pv}\left(k\right)\right)+\gamma\left(2\pi\right)^{3/2}\widetilde{C}_{R}\left(k\right)+\alpha^{2}\frac{\gamma}{\left(2\pi\right)^{3/2}}\\
\\
 & \times\left[4\int d^{3}\boldsymbol{k^{\prime}}\widetilde{C}_{R}\left(k^{\prime}\right)P_{vv}\left(\left|\boldsymbol{k^{\prime}}+\boldsymbol{k}\right|\right)+6\widetilde{C}_{R}\left(k\right)\int d^{3}\boldsymbol{k^{\prime}}P_{vv}\left(\left|\boldsymbol{k^{\prime}}\right|\right)\right] \, .\\
\\
\end{array}
\end{equation}

{As it was done for the bispectrum case, one can combine the above
equations in order to get a single differential equation for $P_{vv}$, this time of third order, given by}
\begin{equation}
P_{vv}^{\prime\prime\prime}+4\omega^{2}P_{vv}^{\prime}+4\omega\omega^{\prime}P_{vv}=S_{1}+S_{2}+S_{3}\,,\label{eq:63}
\end{equation}
{where $S_{1}$ is a source function coming from the linear
contribution to the correction}
\begin{equation}
S_{1}=2\left(2\pi\right)^{3/2}\gamma\widetilde{C}_{R}\left(k\right)\,,
\end{equation}
{while $S_{2}$ and $S_{3}$ are source functions arising from 
the pure quadratic interaction correction and the crossed term between
linear and cubic interaction, respectively. They are functions of
time and involve the power spectrum $P_{vv}$ itself evaluated at
all scales, namely}{\par}

\begin{equation}
\begin{array}{ccc}
\\
S_{2}\left(k,\eta\right)= & \alpha^{2}\frac{8\gamma}{\left(2\pi\right)^{3/2}}\int d^{3}\boldsymbol{k^{\prime}}\widetilde{C}_{R}\left(k^{\prime}\right)P_{vv}\left(\left|\boldsymbol{k^{\prime}}+\boldsymbol{k}\right|\right)\, ,\\
\\
S_{3}\left(k,\eta\right)= & \alpha^{2}\frac{12\gamma}{\left(2\pi\right)^{3/2}}\widetilde{C}_{R}\left(k\right)\int d^{3}\boldsymbol{k^{\prime}}P_{vv}\left(\left|\boldsymbol{k^{\prime}}\right|\right) \, .\\
\\
\end{array}
\end{equation}
{\par}

{Guided by the intuition based on previous computations,
both those in \cite{martin2018observational,martin2018non} and that
of the bispectrum, we introduce the following functions which will constitute
a solution to (\ref{eq:63})}
\begin{equation}
\begin{array}{ccc}
\\
P_{vv}^{\left(i\right)}= & -\frac{2}{W^{2}}\intop_{-\infty}^{\eta}d\eta^{\prime}S_{i}\left(k,\eta^{\prime}\right)\mathscr{\mathfrak{Im^{2}\left[\mathrm{v_{\mathbf{k}}\left(\eta^{\prime}\right)v_{\mathbf{k}}^{\ast}\left(\eta\right)}\right]}}\\
\\
\end{array}\,,
\end{equation}
{with $i=1,2,3$. So the general solution to (\ref{eq:63})
is given by}
\begin{equation}
P_{vv}=v_{\boldsymbol{k}}\left(\eta\right)v_{\boldsymbol{k}}^{*}\left(\eta\right)+P_{vv}^{\left(1\right)}+P_{vv}^{\left(2\right)}+P_{vv}^{\left(3\right)}\,,\label{69}
\end{equation}
{where the first term is the standard result, i.e without
decoherence; $P_{vv}^{\left(1\right)}$ and $P_{vv}^{\left(2\right)}$
have already been computed in \cite{martin2018observational}, and
the aim now is to compute explicitly $P_{vv}^{\left(3\right)}$ which represents
the new correction we obtained. However, before starting, it is useful
to make a comparison between $P_{vv}^{\left(2\right)}$ and $P_{vv}^{\left(3\right)}$ expressions,
since they are of the same perturbative order in $\alpha$. It is easy to notice
that $P_{vv}^{\left(2\right)}$ differs from $P_{vv}^{\left(3\right)}$
by the fact that in the former the environment correlation function is inside the integral
while in the latter it is outside of it. Combining this last observation
with the computations carried out for $P_{vv}^{\left(2\right)}$ in
\cite{martin2018observational} in the limit $k\eta\rightarrow0$
and $Hl_{E}\ll1$, we find a relation between the source functions
$S_{2}\left(k,\eta\right)$ and $S_{3}\left(k,\eta\right)$. Indeed, It has been shown in \cite{martin2018observational} that the environment correlation function $ \widetilde{C}_{R}\left(k^{\prime}\right)$ which is inside the integral in $S_{2}\left(k,\eta\right)$ could be replaced by $ \widetilde{C}_{R}\left(k\right)$ } and be factorized outside of the integral, in such case we simply get  
\begin{equation}
\begin{array}{clc}
S_{2}\left(k,\eta\right)= & \alpha^{2}\frac{8\gamma}{\left(2\pi\right)^{3/2}}\int d^{3}\boldsymbol{k^{\prime}}\widetilde{C}_{R}\left(k^{\prime}\right)P_{vv}\left(\left|\boldsymbol{k^{\prime}}+\boldsymbol{k}\right|\right)\\
\\
 & \simeq\alpha^{2}\frac{8\gamma}{\left(2\pi\right)^{3/2}}\widetilde{C}_{R}\left(k\right)\int d^{3}\boldsymbol{k^{\prime}}P_{vv}\left(\left|\boldsymbol{k^{\prime}}+\boldsymbol{k}\right|\right)\\

\\
 & =\frac{2}{3}S_{3}\left(k,\eta\right)\, ,
\end{array} \label{eq:68}
\end{equation}
{where $\eta_{_{IR}}$ is an IR cutoff to make the integral
finite. Thus from (\ref{eq:68}) we deduce that there is no need to
compute $P_{vv}^{\left(3\right)}$ from scratch, since it is related
to $P_{vv}^{\left(2\right)}$, therefore, the overall correction at
order $\alpha^{2}$ is given by $\frac{5}{3}P_{vv}^{\left(2\right)}$
and this factor of $5/3$ represents the correction brought by considering
the pointer observable (\ref{57}) instead of a pure quadratic one.

The second important fact to be deduced from (\ref{eq:68}), is that
$P_{vv}^{\left(3\right)}$ will represent a scale independent correction
to the standard power spectrum for}{
$p=3$ since it was shown in \cite{martin2018observational} that
$P_{vv}^{\left(2\right)}$ is scale independent for this value of
$p$, and this endorses once again the analysis we made below equation (\ref{eq:54}). 
In this case, despite that $P_{vv}^{\left(2\right)}$
and $P_{vv}^{\left(3\right)}$ are originated from different terms
in (\ref{eq:60}), namely }$\alpha^{2}\left\langle \left[\left[\hat{O},\hat{v}_{\boldsymbol{p}_{1}}\hat{v}_{\boldsymbol{\boldsymbol{k}-p}_{1}}\right],\hat{v}_{\boldsymbol{p}_{2}}\hat{v}_{\boldsymbol{-\boldsymbol{k}-p}_{2}}\right]\right\rangle ${
and }$\alpha^{2}\left\langle \left[\left[\hat{O},\hat{v}_{\boldsymbol{\boldsymbol{k}}}\right],\hat{v}_{\boldsymbol{p}_{1}}\hat{v}_{\boldsymbol{p}_{2}}\hat{v}_{\boldsymbol{-\boldsymbol{k}-\boldsymbol{p}_{1}-p}_{2}}\right]\right\rangle $,
respectively, they are both $\propto\alpha^{2}$. Thus, according
to the conjecture we made, they should share the same value of the parameter $p$, see (\ref{eq:73}), 
for which their expressions are scale independent, and this is exactly what happens.{\par}

{Notice that the same line of thoughts should be applied to the computation of the 
trispectrum which receives its first non vanishing correction
at order $\alpha^{2}$. So in addition to the pure quadratic interaction
contribution, computed in \cite{martin2018non}, we must add the new
term marked with $\underbrace{}$ in (\ref{eq:58}). We expect 
that the difference between the pure quadratic pointer observable
and ours will be more relevant in the trispectrum than it is in the case of the power spectrum.}{\par}

\section{Decoherence of tensor perturbations \label{sec:Decoherence-of-tensor-1}}

{The aim of this section is to compute the decoherence induced
corrections to the tensor (i.e. inflationary gravitational waves) power spectrum. 
Since the ultimate goal is to investigate the effect of the interaction with the environment on the scalar
and tensor perturbations we will apply the same Lindblad equation (\ref{eq:1})
to the tensor perturbations that represent, now, our system. In fact the Lindblad
equation (\ref{eq:1}) was derived in full generality regardless of
the nature of the system, provided that the
environment should satisfy some properties and conditions encoded
in the correlation function }
\begin{equation}
C_{R}\left(\boldsymbol{x},\boldsymbol{y}\right)={\rm Tr}_{E}\left(\rho_{E}R\left(\boldsymbol{x}\right)R\left(\boldsymbol{y}\right)\right)\,.
\end{equation}

{We introduce the following Fourier expansion of the tensor perturbations
$h_{ij}$$\left(\eta,\boldsymbol{x}\right)$, which will be useful
in the following computations, }
\begin{equation}
h_{ij}\left(\boldsymbol{x}\right)=\frac{1}{\left(2\pi\right)^{3/2}}\int d^{3}\boldsymbol{k}h_{ij}\left(\boldsymbol{k}\right)e^{i\boldsymbol{k}.\boldsymbol{x}}\,,
\end{equation}
with

\begin{equation}
h_{ij}\left(\boldsymbol{k}\right)=\sum_{\lambda=+,\times}h_{\lambda}\left(\boldsymbol{k}\right)e_{ij}^{\lambda}\,, \,\,\,\,\,\,\,\,\,e_{ij}^{\lambda}e_{ij}^{\lambda^{\prime}}=2\delta^{\lambda\lambda^{\prime}},\label{eq:336}
\end{equation}
{\large{} where we adopted the well-known case with $\boldsymbol{k}$ aligned along z-direction i.e $\boldsymbol{k}\equiv k(0,0,1)$
and the polarization tensors given by}
\begin{equation}
e^{+}=\left(\begin{array}{ccc}
1 & 0 & 0\\
0 & -1 & 0\\
0 & 0 & 0
\end{array}\right)\,\,,\,\,e^{\times}=\left(\begin{array}{ccc}
0 & 1 & 0\\
1 & 0 & 0\\
0 & 0 & 0
\end{array}\right)\,,
\end{equation}
{and for canonical normalization, we introduce}
\begin{equation}
v_{\lambda}=\frac{aM_{pl}}{\sqrt{2}}h_{\lambda}\,.\label{eq:90}
\end{equation}
{Now, we have to choose} $\hat{H}_{int}$ {between
the tensor perturbations and the environment of the following
local form}
\begin{equation}
\hat{H}_{int}=\int d^{3}\boldsymbol{x}A\left(\eta,\boldsymbol{x}\right)\otimes E\left(\eta,\boldsymbol{x}\right)\,.
\end{equation}
{The first obvious choice is to have linear interaction in
tensor perturbations since it is expected to give the dominant contribution
}
\begin{equation}
A\left(\eta,\boldsymbol{x}\right)=h_{ij}\left(\eta,\boldsymbol{x}\right)\,,\label{equt:5.7}
\end{equation}
{\large{}and in order to contract the spatial indices
in $h_{ij}$, the environment interaction operator should be of the type}

\begin{equation}
E\left(\eta,\boldsymbol{x}\right)\equiv E_{ij}\left(\eta,\boldsymbol{x}\right)\,.
\end{equation}
{To make contact with our previous computations of the curvature perturbation case and with those in \cite{martin2018observational,martin2018non}, we take the case where indices refer to spatial derivatives so that one could write}\\
\begin{equation}
E_{ij}\left(\eta,\boldsymbol{x}\right)=\partial{}_{i}R_{1}\left(\eta,\boldsymbol{x}\right)\partial{}_{j}R_{2}\left(\eta,\boldsymbol{x}\right)\, .
\end{equation}
{For example, if we consider a scalar field $\psi$
as environment}\footnote{This is the case we might have mostly in mind but, still, we are not excluding an environment
made of tensor fields themselves, since we can have for example $E_{ij}\left(\eta,\boldsymbol{x}\right)=\partial{}_{i}\partial_{j}\left(\chi_{kl}\chi^{kl}\right)$.  Also note that the scalar 
field environment could be provided by small-scale curvature perturbations $\zeta$ acting on a system of (long-wavelength) tensor perturbation modes.

}{ then among the possible interactions we have}
\begin{equation}
\mathcal{H}_{int}\propto 
h_{ij}\psi\partial{}_{i}\partial_{j}\psi,\,\,h_{ij}\partial{}_{i}\psi\partial_{j}\psi,.....
\end{equation}
{We adopt this convention remembering that the environment
states with respect to which we are taking the trace are usually considered
to be the Bunch Davies vacuum~\cite{boyanovsky2015effective,gong2019quantum}
which is homogeneous and isotropic. Using the linearity of the trace,
then we can exchange between trace and derivatives in}{\par}

\begin{equation}
C_{E}\left(\boldsymbol{x},\boldsymbol{y}\right)={\rm Tr}_{E}\left(\rho_{E}E\left(\boldsymbol{x}\right)E\left(\boldsymbol{y}\right)\right)\,,\label{eq:77}
\end{equation}
{so that}\footnote{{By $\partial_{i}^{x}$we mean $\frac{\partial}{\partial x_{i}},$ and
same for others.}}
\begin{equation}
C_{Eijmn}\left(\boldsymbol{x},\boldsymbol{y}\right)=\partial^{x}{}_{i}\partial_{j}^{x}\partial{}_{m}^{y}\partial_{n}^{y}{\rm Tr}_{E}\left(\rho_{E}R\left(\boldsymbol{x}\right)R\left(\boldsymbol{y}\right)\right)=\partial_{i}^{x}\partial_{j}^{x}\partial{}_{m}^{y}\partial_{n}^{y}C_{R}(\boldsymbol{x},\boldsymbol{y})\,,\label{79}
\end{equation}
{where we will preserve the convention adopted in \cite{martin2018observational}
regarding the environmental correlation function $C_{R}(\boldsymbol{x},\boldsymbol{y})$,
that is given by (\ref{eq:190}). }{\par}

{We will show in a moment that at linear order in $h_{ij}$
the tensor power spectrum does not receive a correction due to fact
that $h_{ij}$ is transverse, i.e, $\partial^{i}h_{ij}=0$.}{\par}

{Since the linear order gives vanishing correction to the tensor power
spectrum, we consider a second possible choice of $A\left(\eta,\boldsymbol{x}\right)$
consisting in a quadratic interaction.  As the simplest example we consider~\footnote{ Such an example, which here serves as a toy-model, can be extended to the case where (spatial) derivatives act on the tensor fields appearing in Eq.~(\ref{eq:80}). Indeed such an interaction term~(\ref{eq:80}) can arise, e.g., from cubic interactions between curvature perturbations $\zeta$ and tensor modes $h_{ij}$ after integration by parts and considering small scale curvature perturbations as the environment.}
\begin{equation}
A\left(\eta,\boldsymbol{x}\right)=h_{ij}\left(\eta,\boldsymbol{x}\right)h^{ij}\left(\eta,\boldsymbol{x}\right)\,.\label{eq:80}
\end{equation}
{Therefore in this case by relabeling (\ref{eq:77}) we have}
\begin{equation}
C_{E}\left(\boldsymbol{x},\boldsymbol{y}\right)\equiv C_{R}\left(\boldsymbol{x},\boldsymbol{y}\right)\,,\label{eq:81}
\end{equation}
{and we will see that the results will differ slightly
from the curvature power spectrum case. However, it is still important
to emphasize the fact that the curvature power spectrum receives corrections
already for an $A\left(\eta,\boldsymbol{x}\right)$ linear in the Mukhanov-Sasaki variable, while the tensor power spectrum starts receiving decoherence induced corrections only at quadratic order. This last remark implies that we expect the tensor corrections
to be suppressed with respect to those of the curvature perturbation case.}{\par}

\subsection{Linear interaction\label{subsec:Linear-interaction}}

{Considering a pure linear pointer observable,} $A\left(\eta,\boldsymbol{x}\right)=h_{ij}\left(\eta,\boldsymbol{x}\right)$,{\large{}
and using our convention in (\ref{79}), then (\ref{eq:1}) is given
by }\footnote{{The indices of $h_{ij}$ are raised and lowered by $\delta_{ij}$
so we do not sharply distinguish the upper and lower indices of $h_{ij}$.}}
\begin{equation}
\frac{d\left\langle \hat{O}\right\rangle }{d\eta}=\left\langle \frac{\partial\hat{O}}{\partial\eta}\right\rangle -i\left\langle \left[\hat{O},\hat{H_{h}}\right]\right\rangle -\frac{\gamma}{2}\int d^{3}\boldsymbol{x}d^{3}\boldsymbol{y}\partial_{i}^{x}\partial_{j}^{x}\partial{}_{k}^{y}\partial_{l}^{y}C_{R}\left(\boldsymbol{x},\boldsymbol{y}\right)\left\langle \left[\left[\hat{O},h_{ij}\left(\boldsymbol{x}\right)\right],h_{kl}\left(\boldsymbol{y}\right)\right]\right\rangle \,,
\end{equation}
{where for a tensor perturbations system $\hat{H}_{h}$ is
given by }
\begin{equation}
\hat{H}_{h}=\frac{M_{pl}^{2}a^{2}}{8}\int d^{3}\boldsymbol{x}\left[h_{ij}^{\prime2}+\left(\nabla h_{ij}\right)^{2}\right]\,. 
\end{equation}
{As in the scalar case, we prefer to work in Fourier space so
transforming this last equation using (\ref{eq:336}) leads to 
\begin{equation}
\frac{d\left\langle \hat{O}\right\rangle }{d\eta}=\left\langle \frac{\partial\hat{O}}{\partial\eta}\right\rangle -i\left\langle \left[\hat{O},\hat{H_{h}}\right]\right\rangle -\frac{\gamma}{2}\int d^{3}\boldsymbol{k}k_{i}k_{j}k_{m}k_{n}\widetilde{C}_{R}\left(\left|\boldsymbol{k}\right|\right)\left\langle \left[\left[\hat{O},h_{ij}\left(\boldsymbol{k}\right)\right],h_{mn}\left(\boldsymbol{-k}\right)\right]\right\rangle \,.
\end{equation}
{However, since }
$\partial^{i}h_{ij}=0$ {in Fourier space gives} $k^{i}h_{ij}=0$,
{we see that the real, non unitary, part of the Lindblad
equation induced by interactions with the environment vanishes, so that the tensor
power spectrum remains unchanged.}\par}

\subsection{Quadratic interaction}
{The result obtained in the previous section forces us to consider the 
next order in $h_{ij}\left(\eta,\boldsymbol{x}\right)$, so having
(\ref{eq:80}) and (\ref{eq:81}), the Lindblad equation now reads}
\begin{equation}
\frac{d\left\langle \hat{O}\right\rangle }{d\eta}=\left\langle \frac{\partial\hat{O}}{\partial\eta}\right\rangle -i\left\langle \left[\hat{O},\hat{H_{h}}\right]\right\rangle -\frac{\gamma\xi^{2}}{2}\int d^{3}\boldsymbol{x}d^{3}\boldsymbol{y}C_{R}(\boldsymbol{x},\boldsymbol{y})\left\langle \left[\left[\hat{O},h_{ij}h^{ij}\left(\boldsymbol{x}\right)\right],h_{mn}h^{mn}\left(\boldsymbol{y}\right)\right]\right\rangle \,,\label{86}
\end{equation}
{where $\xi^{2}$ is an expansion constant that will serve later
to set up the right dimensions. Fourier transforming equation (\ref{86})
using, again, (\ref{eq:336}) and (\ref{eq:90}), yields}
\begin{equation}
\frac{d\left\langle \hat{O}\right\rangle }{d\eta}=\left\langle \frac{\partial\hat{O}}{\partial\eta}\right\rangle -i\left\langle \left[\hat{O},\hat{H_{v}}\right]\right\rangle -\frac{\beta^{2}\gamma}{2\left(2\pi\right)^{3/2}}\sum_{\lambda\lambda^{\prime}}\int d^{3}\boldsymbol{k}d^{3}\boldsymbol{p}_{1}d^{3}\boldsymbol{p}_{2}\widetilde{C}_{R}\left(\left|\boldsymbol{k}\right|\right)\left\langle \left[\left[\hat{O},\hat{v}_{\boldsymbol{p}_{1}}^{\lambda}\hat{v}_{\boldsymbol{\boldsymbol{k}-p}_{1}}^{\lambda}\right],\hat{v}_{\boldsymbol{p}_{2}}^{\lambda^{\prime}}\hat{v}_{\boldsymbol{-\boldsymbol{k}-p}_{2}}^{\lambda^{\prime}}\right]\right\rangle \,,\label{91}
\end{equation}
where now
\begin{equation}
\hat{H}_{v}=\frac{1}{2}\sum_{\lambda}\int d^{3}\boldsymbol{k}\left[\hat{p}_{\boldsymbol{k}}^{\lambda}\hat{p}_{-\boldsymbol{k}}^{\lambda}+\omega^{2}\hat{v}_{\boldsymbol{k}}^{\lambda}\hat{v}_{\boldsymbol{-k}}^{\lambda}\right]\,\,\,\,{\rm with}\,\,\,\,\omega^{2}=k^{2}-\frac{a^{\prime\prime}}{a}\, .
\end{equation}
{Here we have also defined}
\begin{equation}
\beta=\frac{2\xi}{M_{pl}^{2}}\,,
\end{equation}
being a dimensionless coupling constant
so that we can preserve the definition made in (\ref{eq:11-1}) for, the dimensionless coupling, 
$\sigma_{\gamma}$. It may seem that there
is a missing factor of $a^{-4}$ in the real part of (\ref{91}) but
actually it has been absorbed in the definition of $\gamma$ given in (\ref{eq:coupling}).  This last step was made because, on the one hand, we want to compute the decoherence corrected tensor-to-scalar perturbation ratio, $r$, and for this purpose,  as the simplest possibility, we adopt the same form of the scalar case for the various parametrizations, and on the other hand for the scalar case we considered an interaction system operator $A\left(\eta,\boldsymbol{x}\right)$ directly proportional to the Mukhanov-Sasaki variable rather than the inflaton or the metric scalar fluctuation so the $a$  relating the two was already absorbed in the definition (\ref{eq:24-1}), see appendix
B in \cite{martin2018observational} for more details.

{Having derived the Lindblad equation governing the expectation
values of our system observables, all what remains to do is to derive explicitly the equations obeyed by the different correlators, as was done in
(\ref{eq:61}). However, this time we should add the polarization
indices to (\ref{eq:12-1}). To this end, we use the relation }$\left[\hat{v}_{\boldsymbol{p}}^{s},\hat{p}_{\boldsymbol{k}}^{\lambda}\right]=i\delta^{s\lambda}\delta^{\left(3\right)}\left(\boldsymbol{p}+\boldsymbol{k}\right)$. {Thus following the same steps as before we obtain the following system
of equations}
\begin{equation}
\begin{array}{llc}
\\
\frac{d\left\langle \hat{v}_{\boldsymbol{k}_{1}}^{s}\hat{v}_{\boldsymbol{k}_{2}}^{s}\right\rangle }{d\eta} & =\left\langle \hat{v}_{\boldsymbol{k}_{1}}^{s}\hat{p}_{\boldsymbol{k}_{2}}^{s}\right\rangle +\left\langle \hat{p}_{\boldsymbol{k}_{1}}^{s}\hat{v}_{\boldsymbol{k}_{2}}^{s}\right\rangle \\
\\
\frac{d\left\langle \hat{v}_{\boldsymbol{k}_{1}}^{s}\hat{p}_{\boldsymbol{k}_{2}}^{s}\right\rangle }{d\eta} & =\left\langle \hat{p}_{\boldsymbol{k}_{1}}^{s}\hat{p}_{\boldsymbol{k}_{2}}^{s}\right\rangle -\omega^{2}\left(k_{2}\right)\left\langle \hat{v}_{\boldsymbol{k}_{1}}^{s}\hat{v}_{\boldsymbol{k}_{2}}^{s}\right\rangle \\
\\
\frac{d\left\langle \hat{p}_{\boldsymbol{k}_{1}}^{s}\hat{v}_{\boldsymbol{k}_{2}}^{s}\right\rangle }{d\eta} & =\left\langle \hat{p}_{\boldsymbol{k}_{1}}^{s}\hat{p}_{\boldsymbol{k}_{2}}^{s}\right\rangle -\omega^{2}\left(k_{1}\right)\left\langle \hat{v}_{\boldsymbol{k}_{1}}^{s}\hat{v}_{\boldsymbol{k}_{2}}^{s}\right\rangle \\
\\
\frac{d\left\langle \hat{p}_{\boldsymbol{k}_{1}}^{s}\hat{p}_{\boldsymbol{k}_{2}}^{s}\right\rangle }{d\eta} & =-\omega^{2}\left(k_{2}\right)\left\langle \hat{p}_{\boldsymbol{k}_{1}}^{s}\hat{v}_{\boldsymbol{k}_{2}}^{s}\right\rangle -\omega^{2}\left(k_{1}\right)\left\langle \hat{v}_{\boldsymbol{k}_{1}}^{s}\hat{p}_{\boldsymbol{k}_{2}}^{s}\right\rangle \\
\\
 & +\beta^{2}\frac{4\gamma}{\left(2\pi\right)^{3/2}}\int d^{3}\boldsymbol{k}\widetilde{C}_{R}\left(\left|\boldsymbol{k}\right|\right)\left\langle \hat{v}_{\boldsymbol{k}+\boldsymbol{k}_{1}}^{s}\hat{v}_{-\boldsymbol{k}+\boldsymbol{k}_{2}}^{s}\right\rangle \, .
\\
\end{array}
\end{equation}
{Apart from the polarization indices and the expansion constant
($\beta$ vs $\alpha$ compared to the scalar case), the previous system of equations
does not differ from the scalar case. Therefore, all what we need
to do is to exchange $\alpha$ by $\beta$ in the solutions of the scalar
case and multiply the result by 2 to account for the two possible
polarization modes $\text{\ensuremath{\left(+,\times\right)}}$. The
tensor power spectrum }$P_{vv}=\sum_{s}\left\langle \hat{v}_{\boldsymbol{k}_{1}}^{s}\hat{v}_{\boldsymbol{k}_{2}}^{s}\right\rangle ${
is governed by }
\begin{equation}
P_{vv}^{\prime\prime\prime}+4\omega^{2}P_{vv}^{\prime}+4\omega\omega^{\prime}P_{vv}=S\left(\boldsymbol{k},\eta\right)\,,
\end{equation}
{with }
\begin{equation}
\label{seq}
S\left(\boldsymbol{k},\eta\right)=\beta^{2}\frac{16\gamma}{\left(2\pi\right)^{3/2}}\int d^{3}\boldsymbol{k^{\prime}}\widetilde{C}_{R}\left(k^{\prime}\right)P_{vv}\left(\left|\boldsymbol{k^{\prime}}+\boldsymbol{k}\right|\right)\,,
\end{equation}
{which, as we know, admits as solution
\begin{equation}
P_{vv}=\sum_{s}v_{\mathbf{k}}^{s}\left(\eta\right)v_{\mathbf{k}}^{s\ast}\left(\eta\right)+2\sum_{s}\intop_{-\infty}^{\eta}d\eta^{\prime}S\left(k,\eta^{\prime}\right)\mathscr{\mathfrak{Im^{2}\left[\mathrm{v_{\mathbf{k}}^{s}\left(\eta^{\prime}\right)v_{\mathbf{k}}^{s\ast}\left(\eta\right)}\right]}}\,.\label{97}
\end{equation}
{The first term is  the standard result while the
second term is the correction induced by the interaction with the environment.
Since we are interested in the power spectrum of tensor perturbations
$h$ and not the variable $v$, and in order to facilitate the task
of computing the corrected tensor-to-scalar ratio $r$, then, as was
done for the scalar peturbations in \cite{martin2018observational}, we adopt
the following notation for the total dimensionless tensor power spectrum
$\mathcal{P}_{T}$},\footnote{By total we mean the standard result plus the correction induced by
decoherence.} 
\begin{equation}
\mathcal{P}_{T}=\frac{k^{3}}{2\pi^{2}}\frac{2P_{vv}}{M_{pl}^{2}a^{2}}=\mathcal{P}_{T}\left|_{standard}\right.\left[1+\Delta\mathcal{P}_{T}\right]\,,\label{eq:488}
\end{equation}
{\large{}with }
\begin{equation}
\mathcal{P}_{T}\left|_{standard}\right.=\frac{8}{M_{pl}^{2}}\left(\frac{H}{2\pi}\right)^{2}\left(\frac{k}{aH}\right)^{-2\epsilon}\simeq\frac{8}{M_{pl}^{2}}\left(\frac{H}{2\pi}\right)^{2}\,,
\end{equation}
{and} 
\begin{equation}
\Delta\mathcal{P}_{T}=\frac{2\sum_{s}\intop_{-\infty}^{\eta}d\eta^{\prime}S\left(k,\eta^{\prime}\right)\mathscr{\mathfrak{Im^{2}\left[\mathrm{v_{\mathbf{k}}^{s}\left(\eta^{\prime}\right)v_{\mathbf{k}}^{s\ast}\left(\eta\right)}\right]}}}{\sum_{s}v_{\mathbf{k}}^{s}\left(\eta\right)v_{\mathbf{k}}^{s\ast}\left(\eta\right)}\, .
\end{equation}
{The computation of the integral in (\ref{97}) was already
done in \cite{martin2018observational} for curvature perturbations,
so we need just to make the modifications already mentioned above
to get the results corresponding to the tensor case.} {We will report the expressions for $\Delta\mathcal{P}_{T}$
as a function of the parameter $p$ defined in (\ref{eq:coupling}). We have
three regimes in addition to two singular cases,}{\par}
\begin{itemize}
\item {for $p>6$}

\begin{equation}
\Delta\mathcal{P}_{T}\simeq\frac{8}{27\pi}\beta^{2}\sigma_{\gamma}\left(\frac{k}{k_{*}}\right)^{3}\left(\frac{\eta}{\eta_{*}}\right)^{6-p}\left[\frac{1}{p^{2}}-\frac{2}{\left(p-3\right)^{2}}+\frac{1}{\left(p-6\right)^{2}}+\frac{18}{p\left(p-3\right)\left(p-6\right)}\ln\left(\frac{\eta_{IR}}{\eta}\right)\right]\, ,
\label{t1}
\end{equation}
{where the expression of $\sigma_{\gamma}$ is given by (\ref{eq:11-1}), ($-1/\eta_{IR}$) is an IR cutoff in the integral (\ref{seq})
and $k_{*}$ refers to a pivot scale; $\ln\left(\eta_{IR}/\eta\right)=N-N_{IR}$
gives the number of e-folds elapsed since the beginning of inflation. We notice that in this regime the power spectrum correction scales
as $k^{3}$, in addition to be not frozen on large scales and continues
to increase, leading to a very large enhancement of the correction
compared to the standard power spectrum at late time. This behavior of the tensor power spectrum correction corresponds to the analogous case of the curvature case in \cite{martin2018observational}. }{\par}

\item {For $2<p<6$}
\begin{equation}
\begin{array}{clc}
\Delta\mathcal{P}_{T} & \simeq\frac{2^{p-1}\left(p-4\right)}{3p\varGamma\left(p-1\right)\sin\left(\pi p/2\right)}\beta^{2}\sigma_{\gamma}\left(\frac{k}{k_{*}}\right)^{p-3}\left[\ln\left(\frac{\eta_{IR}}{\eta_{*}}\right)+\frac{1}{p-4}-\frac{2\left(p-1\right)}{p\left(p-2\right)}\right.\\
\\
 & \left.-\frac{\pi}{2}\cot\left(\frac{\pi p}{2}\right)+\ln\left(2\right)-\psi\left(p-2\right)+\ln\left(\frac{k}{k_{*}}\right)\right]\, ,\\
\\
\end{array}\label{eq:450}
\end{equation}
{where here $\psi$ is the digamma function. In this case we obtain a
scale invariant correction for $p=3$ which was also the case for
the correction to the curvature power spectrum coming from the quadratic
interaction,  i.e $P_{vv}^{\left(2\right)}+P_{vv}^{\left(3\right)}$
in (\ref{69}). However let us remember that, in contrast to the 
tensor case, the dominant correction to the curvature power spectrum comes
from the linear interaction, providing a scale invariant result for $p=5$. Therefore at leading
order, the scale invariant corrections to the scalar and tensor power
spectra correspond to different values of $p$.  In particular notice that, in the cases where the parameter $p$ would be the same for the scalar and tensor cases, then by imposing a scale invariant scalar power spectrum, implies a strong blue-tilted tensor power spectrum.}
\item {For $p<2$}
\begin{equation}
\Delta\mathcal{P}_{T}\simeq\frac{4\left(H_{*}l_{E}\right)^{p-2}}{3\pi\left(2-p\right)}\beta^{2}\sigma_{\gamma}\left(\frac{k}{k_{*}}\right)^{p-3}\left[\frac{1}{2-p}+N_{*}-N_{IR}+\ln\left(H_{*}l_{E}\right)+\ln\left(\frac{k}{k_{*}}\right)\right]\, .
\label{t3}
\end{equation}
{Notice that in this case the decoherence induced corrections to the power spectrum become negligible on very small scales.
}{\par}
\item {For $p=2$ and $p=6$ which are singular we find} 
\begin{equation}
\begin{array}{clc}
\\
\Delta\mathcal{P}_{T}\left|_{p=2}\right. & \simeq\frac{1}{18\pi}\beta^{2}\sigma_{\gamma}\left(\frac{k}{k_{*}}\right)^{-1}\left[12-\pi^{2}+12C\left(2+C\right)-12\ln^{2}\left(H_{*}l_{E}\right)+24\left[C+1-\ln\left(H_{*}l_{E}\right)\right]\right.\\
\\
 & \left.\times\left[\left(N_{*}-N_{IR}\right)+\ln\left(\frac{k}{k_{*}}\right)\right]\right]\\
\\
\Delta\mathcal{P}_{T}\left|_{p=6}\right. & \simeq\frac{1}{162\pi}\beta^{2}\sigma_{\gamma}\left(\frac{k}{k_{*}}\right)^{3}\left[2\pi^{2}-21-12C\left(1+2C\right)-12\left(3+4C\right)\left(N-N_{IR}\right)\right.\\
\\
 & +12\left(1+4C\right)\left(N-N_{*}\right)+24\left(N-N_{*}\right)\left[2\left(N-N_{IR}\right)-\left(N-N_{*}\right)\right]\\
\\
 & \left.-12\ln\left(\frac{k}{k_{*}}\right)\left[1+4\left(C+N_{*}-N_{IR}\right)2\ln\left(\frac{k}{k_{*}}\right)\right]\right]\\
\label{t4}
\end{array}
\end{equation}
where $C$ is a constant.{\par}

\end{itemize}

\subsection{Decoherence induced Corrections to the tensor-to-scalar perturbation ratio $r$ }

{In Section \ref{sec:J.Martin-et-al}, we presented  the computations made in \cite{martin2018observational}  to get the decoherence induced corrections to} \textbf{$n_{s}$ }{and $r$. However, in their computations of $r$
they assumed the tensor power spectrum $\mathcal{P}_{T}$ to remain
unaffected by the environment.  However, as we showed with the computations we just carried out in the previous section,  the tensor sector can be affected by decoherence induced contributions, which therefore should be accounted for.} Therefore, our aim now is
to give the decoherence corrected expression of $r$ taking into account
our results regarding the effect of environment on tensor modes. 

{Once we come to compute the
corrected $r$, the scale independence of the leading
decoherence induced corrections to the curvature and tensor power
spectra correspond to different values of $p$, namely $p=5$ and
$p=3$, respectively. However, it is important to remember that curvature
corrections are dominant with respect to tensor ones, since the latter receive their first non vanishing correction only at quadratic
level. Therefore, if we restrict ourselves to linear interactions,
then, the decoherence corrected $r$ is still given by the expression
computed by J.Martin et al in \cite{martin2018observational}}
\begin{equation}
r=\frac{r\left|_{standard}\right.}{1+\frac{\pi}{6}\frac{k_{\gamma}^{2}}{k_{*}^{2}}}\,. \label{eq:5.34}
\end{equation}
{However if we consider the leading corrections of both power
spectra, curvature and tensor, regardless of their order, then, the
previous equation will be modified. Since curvature power spectrum
has been well confirmed to be quasi scale independent, at least up
to certain sensitivity, then we choose the value for which its correction
is also scale independent, namely $p=5$ (see in any case the comment in footnote~\ref{f26}}). 
This will lead to a tensor-to-scalar perturbation ratio $r$ which in full generality will be given by 
\begin{equation}
r=\frac{r\left|_{standard}\right.}{1+\frac{\pi}{6}\frac{k_{\gamma}^{2}}{k_{*}^{2}}} \left[1+\Delta\mathcal{P}_{T}\right]\Bigg|_{k=k_*}\,,
\label{generalr}
\end{equation}
where one should consider the various cases for $\Delta\mathcal{P}_{T}$ obtained above for the tensor perturbations in Eqs.(\ref{t1}), (\ref{eq:450}), (\ref{t3}) and (\ref{t4}). For example, at least in the cases where the parameter $p$ for the tensor sector is taken to be the same as for the scalar power-spectra, i.e. $p=5$, then the tensor power spectrum correction is blue titled with a spectral
index $n_{T}\approx2$, as seen from (\ref{eq:450}), and is given
up to leading order by\footnote{By leading order we mean to neglect the constant numbers that are much smaller than $\ln\left(\frac{\eta_{IR}}{\eta_{*}}\right)$. } 
\begin{equation}
\begin{array}{clc}
\Delta\mathcal{P}_{T} & \approx\frac{2}{45}\beta^{2}\sigma_{\gamma}\left(\frac{k}{k_{*}}\right)^{2}\left[\ln\left(\frac{\eta_{IR}}{\eta_{*}}\right)+\ln\left(\frac{k}{k_{*}}\right)\right]\\
\\
\end{array}\,.\label{eq:455}
\end{equation}
Substituting (\ref{eq:455}) in (\ref{eq:488}) yields }
\begin{equation}
\mathcal{P}_{T}\approx\mathcal{P}_{T}\left|_{standard}\right.\left[1+\frac{2}{45}\beta^{2}\sigma_{\gamma}\left(\frac{k}{k_{*}}\right)^{2}\left[\ln\left(\frac{\eta_{IR}}{\eta_{*}}\right)+\ln\left(\frac{k}{k_{*}}\right)\right]\right]\,.\label{eq:457}
\end{equation}

{So combining (\ref{generalr})
with (\ref{eq:457}) in this specific case leads to a decoherence corrected $r$, evaluated at the pivot scale i.e $k=k_{*}$, given by}
\begin{equation}
r\simeq r\left|_{standard}\right.\frac{1+\frac{2}{45}\beta^{2}\sigma_{\gamma}(N_{*}-N_{IR})}{1+\frac{\pi}{6}\frac{k_{\gamma}^{2}}{k_{*}^{2}}}\,.\label{eq:459}
\end{equation}
{ Notice that in principle $(N_{*}-N_{IR})=\ln(a_*/a_i)$ (where $a_*$ is the scale-factor when the reference mode $k_*=0.05$ Mpc$^{-1}$ crosses the horizon during inflation and $a_i$ is the scale factor at the beginning of inflation)  can be of the order of $10^4$, even if this is model dependent. Therefore,  even very low values of the coupling-parameter $\beta^2 \sigma_\gamma$ between the tensor fluctuation system and the environment, can indeed lead to a non-negligible correction to the tensor-to-scalar perturbation ratio. For example, given that the interaction between tensor and environment considered here resembles the quadratic interaction case of~\cite{martin2018observational}, and indeed we get similar results, as a first estimate one could use their  constraints. So, choosing $p=5$ to be consistent with (almost) scale-invariance of the scalar perturbations, such value of $p$ then leads to a un upper bound $\beta^2 \sigma_\gamma < 10^{-5}$ from their results in the quadratic interactions (see, e.g. their Fig. 8) which still, as discussed above, leads in any case to observable consequences in the tensor-to-scalar perturbation ratio. }

{A high precise measurement of the tensor
power spectrum $\mathcal{P}_{T}$ and the tensor-to-scalar perturbation ratio $r$
will provide tight constraints on the interaction strength, encoded in
$\beta^{2}\sigma_{\gamma}$, between the primordial tensor fluctuations and their environment.
Those constraints can be compared or used in combination with those obtained
from the scalar correlation functions, namely, power spectrum, bispectrum and trispectrum.}{\par}

\section{Conclusions}
\label{Conclusions}

In this paper we addressed the question of the quantum to
classical transition in the early universe. In particular, we aimed
at the identification of observational connections between the quantum
initial state and the classical universe we observe today through
the completion of the inflation formalism with a model that accounts for
the quantum to classical transition of the primordial quantum fluctuations.
A well motivated model for such a {\bf }goal} is quantum decoherence, where
many attempts have been made to apply it on large cosmological scales yielding
the so called cosmic decoherence. Several works in this direction
were carried out ~\cite{polarski1996semiclassicality, kiefer2007pointer, martineau2007decoherence, Burgess:2006jn, kiefer2009cosmological, nelson2016quantum, martin2018observational, martin2018non},
which inspired the work done in this paper. In particular, our work  builds up, extends, and to some extent completes,  works done in \cite{boyanovsky2015effective, martin2018observational, martin2018non}.

{We started by presenting briefly the derivation of the Lindblad
equation which is the corner stone in studying the quantum to classical
transition of an open system. Then, for the sake of smoothing the
flow of ideas, we summarized in Section \ref{sec:J.Martin-et-al}
the main results of J.Martin et al. in ~\cite{martin2018observational,martin2018non}, based on which we built our model.
In particular we discussed the approach adopted by them in studying
the corrections induced by cosmic decoherence to the various curvature
correlation functions. Confronting the corrections obtained with cosmological
data, they were able to constrain the set of the possible environments
which could decohere the primordial fluctuations. Since cosmological
data has well confirmed the quasi scale independence of curvature
power spectrum, a particular and quite remarkable environment was
a massive scalar field which induces scale independent corrections
to the curvature power spectrum and trispectrum. Another remarkable
result obtained by J.Martin et al. was computing the correction induced
by decoherence into the scalar spectral index $n_{s}$ and the tensor-to-scalar perturbation ratio $r$, assuming the tensor perturbations to remain
unaffected. They showed how the corrected $\left(n_{s},r\right)$
affect the status of various inflation models when compared to cosmological
data. An intriguing result was also obtained by J.Martin
et al., where they showed that decoherence induces vanishing bispectrum
along with non vanishing trispectrum.Therefore,  decoherence was stated 
as one of the rare examples where the bispectrum is perturbatively
suppressed compared to the trispectrum, which therefore would contain the
relevant non-Gaussian signal. }{\par}
{This last result was one of the reasons that sparked the work done in this paper, where
we realized that its validity is not as general as it may seems. Indeed,
adopting a different approach than that of ~\cite{martin2018observational,martin2018non} we were
able to obtain interesting results which we summarize as follow:}{\par}
\begin{enumerate}
\item {First of all, through (\ref{eq:352}) we were able not only
to reproduce all the results obtained in~\cite{martin2018observational,martin2018non}, as can be seen using (\ref{356}), but we also showed in \ref{subsec:Additional-correction-to}
that there were missing terms in the power spectrum
and trispectrum due to the restricted choice of the form of the
pointer observable. Therefore, through the pointer observable (\ref{eq:352}) we have
chosen, we generalized the model initiated by J.Martin et al in \cite{martin2018observational,martin2018non}.
In addition, we motivated the form of such a pointer observable by a physical
process that could produce it. It turned out to be, again, a generalization
of the physical process considered by J.Martin et al with many new
implications. }{\par}
\item {We showed in \ref{subsec:Computation-of-bispectrum} that
decoherence does induce a non vanishing bispectrum. Moreover, it is
dominant with respect to the trispectrum. 
Indeed the highest
the order of correlation function is, then  the highest is the
order of $\alpha$ at which it receives its first non vanishing contributions.
In particular, we saw in the
first part of our work that the first non vanishing correction
to power spectrum, bispectrum, and trispectrum was found at the order
$\alpha^{0}$, $\alpha^{1}$, and $\alpha^{2}$, respectively. Another interesting remark consists
on the possibility of the bispectrum induced by decoherence 
to be dominant with respect to the one arising in standard single field
inflation.}{\par}
\item  Generally the decoherence induced bispectrum can have a specific scale dependence. On the other hand, similarly to the power spectrum and trispectrum, we showed that the bispectrum could be, at leading order, scale independent
in the case of a massive scalar field as environment. In this respect we conjectured
that the scale independence of the corrections induced by cosmic decoherence
is governed by a criterion which surpasses the type of environment. As
an attempt to find such criterion, we conjectured that obtaining a
scale independent corrections of a given correlation function is related
to the order in the expansion parameter $\alpha$ at which the correction arises, independently
from any choice of the environment including the massive scalar field.

\item {Another quite interesting result, that we derived in this
paper, consisted in computing cosmic decoherence effects on primordial
tensor perturbations. We have shown in \ref{subsec:Linear-interaction}
that by considering a linear interaction, decoherence does not affect
tensor fluctuations due to the transversality property
of tensor modes. Therefore,
we considered the next order and studied a quadratic interaction
which indeed modifies the tensor power spectrum. In its turn this 
implies that the tensor-to-scalar perturbation ratio $r$ computed in \cite{martin2018observational} is modified, see (\ref{eq:5.34}) and, e.g., (\ref{eq:459}). By doing so, we saw that scalar
and tensor power spectra corrections are scale independent for different
values of the free parameter(s) $p$.  So, e.g., in the case where the parameter  $p$ is the same for scalar and tensor perturbations, by adopting the value $p=5$ that
makes (the dominant) scalar correction scale independent, leads to a strongly blue tilted
tensor correction. At the same time, interestingly enough, for
the same value of $p=5$, the curvature bispectrum turns out to be also linearly
scale dependent.}{\par}
\end{enumerate}
{The work done so far about cosmic decoherence shows that
the question of quantum to classical transition in the early universe
is not a mere foundational question, but its answer could contain
important observational signatures which could, in their turn, bring new
insights into our current standard predictions. }{\par}

\begin{acknowledgments}
We would like to thank Sabino Matarrese for various stimulating comments and discussions. We thank Vincent Vennin and J\'er\^{o}me Martin for useful discussions and clarifications. N. B. would like to thank Francesco Paolo Lopez for useful discussions. N. Bartolo acknowledges support from the COSMOS network (www.cosmosnet.it)  
through the ASI (Italian Space Agency) Grants 2016-24-H.0, 2016-24-H.1-2018 and 2020-9-HH.0. This work is supported in part by the MUR Departments of Excellence grant "Quantum Frontiers".
\end{acknowledgments}

\appendix
\section{Computation of the non linearity parameter $f_{NL}$  \label{sec:Appendix}}

\selectlanguage{English}%

{In this appendix we give more details on the computation
of the non linear term $f_{NL}$.} {Before doing so, let
us remind some definitions which will be needed to evaluate (\ref{eq:17}).
First, we will put ourselves in the de-Sitter limit, thus, the scale
factor is given by $a=$-1/$\left(H\eta\right)$ and the MS variable
, $v_{\mathbf{k}}\left(\eta\right)$, by}
\begin{equation}
v_{\mathbf{k}}\left(\eta\right)=\frac{e^{-ik\eta}}{\sqrt{2k}}\left(1-\frac{i}{k\eta}\right)\,,
\end{equation}
{which by turn yields }
\begin{equation}
\begin{array}{llc}
\\
\mathfrak{Im^{3}\left[\mathrm{v_{\mathbf{k}}\left(\eta^{\prime}\right)v_{\mathbf{k}}^{\ast}\left(\eta\right)}\right]}=\frac{\cos\left(k\left(\eta-\eta^{\prime}\right)\right)}{2k}\left(\frac{1}{k\eta}-\frac{1}{k\eta^{\prime}}\right)+\frac{\sin\left(k\left(\eta-\eta^{\prime}\right)\right)}{2k}\left(1+\frac{1}{k^{2}\eta\eta^{\prime}}\right)\\
\\
\left\langle \hat{v}_{\boldsymbol{\boldsymbol{k}}}\hat{v}_{-\boldsymbol{\boldsymbol{k}}}\right\rangle =\frac{1}{2k}\left[1+\frac{1}{\left(-k\eta\right)^{2}}\right]\\
\\
\mathfrak{Re}\left\langle \hat{p}_{\boldsymbol{\boldsymbol{k}}}\hat{v}_{-\boldsymbol{\boldsymbol{k}}}\right\rangle =\frac{1}{2\left(-k\eta\right)^{3}}\\
\\
\end{array}\,.
\end{equation}

{For convenience, we remind also }
\begin{equation}
\tilde{C}_{R}\left(k\right)=\sqrt{\frac{2}{\pi}}\bar{C}_{R}\frac{l_{E}}{3a^{3}}\varTheta\left(\frac{kl_{E}}{a}\right)\,,
\end{equation}
{so using $k_{*}=a_{*}H_{*}$ in addition to the fact that
$H\simeq H_{*}\simeq constant$, then, (\ref{eq:17}) could be written
as}
\begin{equation}
B_{vvv}=B_{vvv}^{\left(1\right)}+B_{vvv}^{\left(2\right)}+B_{vvv}^{\left(3\right)}\,,
\end{equation}
{which }\footnote{Notice that the in $B_{vvv}^{\left(2\right)}$ we have $\delta\left(1+zHl_{E}\right)$
instead of $\delta\left(zHl_{E}\right)$ and this is due to our definition
of top hat function as being non zero if $-zHl_{E}<1$, therefore
we shifted the argument and used the definition of the derivative
of Heaviside function.}{ by adopting the new variable $z$ defined as $z=k\eta^{\prime}$
leads to }
\begin{equation}
\begin{array}{clc}
B_{vvv}^{\left(1\right)}= & 7\alpha\frac{k_{\gamma}^{2}}{k_{*}^{4}}\left(\frac{k}{k_{*}}\right)^{p-7}\intop_{-\infty}^{k\eta}dz\left\{ \varTheta\left(-zHl_{E}\right)\left(-z\right)^{-p}\right.\\
\\
 & \times\left.\left[\cos\left(k\eta-z\right)\left(\frac{1}{k\eta}-\frac{1}{z}\right)+\sin\left(k\eta-z\right)\left(1+\frac{1}{k\eta z}\right)\right]^{3}\right\} \\
\\
B_{vvv}^{\left(2\right)}= & \alpha\frac{k_{\gamma}^{2}}{k_{*}^{4}}\left(\frac{k}{k_{*}}\right)^{p-7}Hl_{E}\intop_{-\infty}^{k\eta}dz\left\{ \delta\left(1+zHl_{E}\right)\left(-z\right)^{3-p}\left(1+\frac{1}{(-z)^{2}}\right)\right.\\
\\
 & \times\left.\left[\cos\left(k\eta-z\right)\left(\frac{1}{k\eta}-\frac{1}{z}\right)+\sin\left(k\eta-z\right)\left(1+\frac{1}{k\eta z}\right)\right]^{3}\right\} \\
\\
B_{vvv}^{\left(3\right)}= & \alpha\frac{k_{\gamma}^{2}}{k_{*}^{4}}\left(\frac{k}{k_{*}}\right)^{p-7}\left(p-3\right)\intop_{-\infty}^{k\eta}dz\left\{ \varTheta\left(-zHl_{E}\right)\left(-z\right)^{2-p}\left(1+\frac{1}{(-z)^{2}}\right)\right.\\
\\
 & \times\left.\left[\cos\left(k\eta-z\right)\left(\frac{1}{k\eta}-\frac{1}{z}\right)+\sin\left(k\eta-z\right)\left(1+\frac{1}{k\eta z}\right)\right]^{3}\right\} \\
\\
\end{array}\,,
\end{equation}
{Notice that the presence of $\varTheta\left(-zHl_{E}\right)$
in the first and third integral sets a lower bound on our variable
$z$, where $z_{min}=$$-\left(Hl_{E}\right)^{-1}$, and this ensures
that those integrals are finite, it is also obvious that the presences
of delta function in second integral is due to the derivative of Heaviside
function. The above integrals are not exactly analytically computable,
but the fact that we are interested on super Horizon modes $-k\eta\rightarrow0$
will help to simplify them, the behavior of the correlators can be
obtained by identifying the region in the integration domain from
where the integral receives its main contribution. }{\large\par}

{1- For $B_{vvv}^{\left(1\right)}$ we see that for} \footnote{Supposing p positive is due the fact that classicalization of our
perturbations becomes more efficient as our modes exit horizon, which
is equivalent to saying that decoherence becomes more efficient as the coupling system-environment increases.}{ $p>0$ the main contribution is always coming from the upper
bound where $-k\eta\ll1$ and $-z\ll1$ (with $Hl_{E}\ll1)$ thus
we can expand the integrand in this limit and $B_{vvv}^{\left(1\right)}$
becomes }
\begin{equation}
\begin{array}{ccc}
\\
B_{vvv}^{\left(1\right)}= & -7\alpha\frac{k_{\gamma}^{2}}{k_{*}^{4}}\left(\frac{k}{k_{*}}\right)^{p-7}\intop_{-\left(Hl_{E}\right)^{-1}}^{k\eta}dz\dfrac{\left(z-k\eta\right)^{3}}{\left(-z\right)^{p}}\\
\\
\end{array}\,,\label{eq:27 goves-1}
\end{equation}

{\large{}For $p\neq1,\,2,\,3,\,4$, which are singular cases to be
computed apart, the above integral gives for $4>p$}
\begin{equation}
\begin{array}{llc}
\\
B_{vvv}^{\left(1\right)} & =-7\alpha\frac{k_{\gamma}^{2}}{k_{*}^{4}}\left(\frac{k}{k_{*}}\right)^{p-7}\left(-k\eta\right)^{4-p}\text{\ensuremath{\left[\frac{1}{p-4}-\frac{3}{p-3}+\frac{3}{p-2}-\frac{1}{p-1}\right]}}\\
\\
 & =-7\alpha\frac{k_{\gamma}^{2}}{k_{*}^{4}}\left(\frac{k}{k_{*}}\right)^{p-7}\left(-k\eta\right)^{4-p}\frac{3p-4}{\left(p-4\right)\left(p-3\right)\left(p-2\right)\left(p-1\right)}\\
\\
\end{array}\,,
\end{equation}

\begin{itemize}
\item For $p=1$, (\ref{eq:27 goves-1}) gives 
\begin{equation}
B_{vvv}^{\left(1\right)}=7\alpha\frac{k_{\gamma}^{2}}{k_{*}^{4}}\left(\frac{k}{k_{*}}\right)^{-6}\left(k\eta\right)^{3}\left(\ln\left(\left|k\eta\right|\right)-\dfrac{11}{6}\right)\,,
\end{equation}
\item For $p=2$, (\ref{eq:27 goves-1}) gives 
\begin{equation}
B_{vvv}^{\left(1\right)}=21\alpha\frac{k_{\gamma}^{2}}{k_{*}^{4}}\left(\frac{k}{k_{*}}\right)^{-5}\left(k\eta\right)^{2}\left(\ln\left(\left|k\eta\right|\right)-\dfrac{1}{2}\right)\,,
\end{equation}
\item For $p=3$, (\ref{eq:27 goves-1}) gives 
\begin{equation}
B_{vvv}^{\left(1\right)}=21\alpha\frac{k_{\gamma}^{2}}{k_{*}^{4}}\left(\frac{k}{k_{*}}\right)^{-4}\left(k\eta\right)\left(\ln\left(\left|k\eta\right|\right)+\dfrac{1}{2}\right)\,,
\end{equation}
\item For $p=4$, (\ref{eq:27 goves-1}) gives
\begin{equation}
B_{vvv}^{\left(1\right)}=7\alpha\frac{k_{\gamma}^{2}}{k_{*}^{4}}\left(\frac{k}{k_{*}}\right)^{-3}\left(\ln\left(\left|k\eta\right|\right)+\dfrac{11}{6}\right)\,,
\end{equation}
{ We see that in the singular cases and in super horizon limit
the first term $\propto$$\ln\left(\left|k\eta\right|\right)$ gives
the largest contribution , we notice also that in all cases, singular
and non singular, $B_{vvv}^{\left(1\right)}\propto k^{-3}$. }{\large\par}
\end{itemize}
{\large{}}
{2- For $B_{vvv}^{\left(2\right)}$ the delta function rends
the integration task simple where considering only the dominant terms
for in the limit $-k\eta\ll1$ and $\left(Hl_{E}\right)^{-1}\gg1$
we get }
\begin{equation}
\begin{array}{ccc}
\\
B_{vvv}^{\left(2\right)}= & \alpha\frac{k_{\gamma}^{2}}{k_{*}^{4}}\left(\frac{k}{k_{*}}\right)^{p-7}\frac{1}{\left(k\eta\right)^{3}}\left(Hl_{E}\right)^{p-2}\cos^{3}\left(Hl_{E}\right)^{-1}\\
\\
\end{array}\,,
\end{equation}

{3- For $B_{vvv}^{\left(3\right)}$ in the case $p>2$ and
we find ourselves again in the limit where the main contribution comes
from the upper bound thus for $-k\eta\ll1$, $-z\ll1$ $B_{vvv}^{\left(3\right)}$
could be written as }
\begin{equation}
\begin{array}{clc}
\\
B_{vvv}^{\left(3\right)}= & -\alpha\frac{k_{\gamma}^{2}}{k_{*}^{4}}\left(\frac{k}{k_{*}}\right)^{p-7}\left(p-3\right)\intop_{-\left(Hl_{E}\right)^{-1}}^{k\eta}dz\dfrac{\left(z-k\eta\right)^{3}\left(z^{2}+1\right)}{\left(-z\right)^{p}}\\
\\
\end{array}\,,\label{eq:34-1}
\end{equation}
{\large{}which gives }
\begin{equation}
\begin{array}{clc}
B_{vvv}^{\left(3\right)}= & -\alpha\frac{k_{\gamma}^{2}}{k_{*}^{4}}\left(\frac{k}{k_{*}}\right)^{p-7}\left(p-3\right)\left\{ \left(k\eta\right)^{6-p}\left[\dfrac{1}{p-6}-\dfrac{3}{p-5}+\dfrac{3}{p-4}-\dfrac{1}{p-3}\right]\right.\\
 & +\left.\left(k\eta\right)^{4-p}\left[\frac{1}{p-4}-\dfrac{3}{p-3}+\dfrac{3}{p-2}-\dfrac{1}{p-1}\right]\right\} \\
\\
\end{array}\,,
\end{equation}
{simplifying this last we get}
\begin{equation}
B_{vvv}^{\left(3\right)}=-6\alpha\frac{k_{\gamma}^{2}}{k_{*}^{4}}\left(\frac{k}{k_{*}}\right)^{p-7}\left[\left(-k\eta\right)^{6-p}\frac{1}{\left(p-6\right)\left(p-5\right)\left(p-4\right)}+\left(-k\eta\right)^{4-p}\frac{1}{\left(p-4\right)\left(p-2\right)\left(p-1\right)}\right]\,,
\end{equation}
{as in the previous case we got some singular cases which
need to be computed apart. }{\large\par}
\begin{itemize}
\item {\large{}For $p=4$ (\ref{eq:34-1}) gives}
\begin{equation}
\begin{array}{llc}
B_{vvv}^{\left(3\right)} & =-\alpha\frac{k_{\gamma}^{2}}{k_{*}^{4}}\left(\frac{k}{k_{*}}\right)^{-3}\intop_{-\left(Hl_{E}\right)^{-1}}^{k\eta}dz\dfrac{\left(z-k\eta\right)^{3}\left(z^{2}+1\right)}{z^{4}}\\
\\
 & =-\alpha\frac{k_{\gamma}^{2}}{k_{*}^{4}}\left(\frac{k}{k_{*}}\right)^{-3}\left[\left(k\eta\right)^{2}\left(3\ln\left(\left|k\eta\right|\right)-\frac{5}{2}\right)+\ln\left(\left|k\eta\right|\right)-\frac{17}{6}\right]\\
\\
\end{array}\,,
\end{equation}
\item {For $p=5$ (\ref{eq:34-1}) gives}
\begin{equation}
\begin{array}{clc}
B_{vvv}^{\left(3\right)} & =2\alpha\frac{k_{\gamma}^{2}}{k_{*}^{4}}\left(\frac{k}{k_{*}}\right)^{-3}\intop_{-\left(Hl_{E}\right)^{-1}}^{k\eta}dz\dfrac{\left(z-k\eta\right)^{3}\left(z^{2}+1\right)}{z^{5}}\\
\\
 & =-2\alpha\frac{k_{\gamma}^{2}}{k_{*}^{4}}\left(\frac{k}{k_{*}}\right)^{-3}\left[3\left(k\eta\right)\left(\frac{7}{2}+\ln\left(\left|k\eta\right|\right)\right)+\frac{1}{4k\eta}\right]
\end{array}\,,
\end{equation}
\item {\large{}For $p=6$ (\ref{eq:34-1}) gives}
\begin{equation}
\begin{array}{clc}
B_{vvv}^{\left(3\right)} & =-3\alpha\frac{k_{\gamma}^{2}}{k_{*}^{4}}\left(\frac{k}{k_{*}}\right)^{-3}\intop_{-\left(Hl_{E}\right)^{-1}}^{k\eta}dz\dfrac{\left(z-k\eta\right)^{3}\left(z^{2}+1\right)}{z^{6}}\\
\\
 & =3\alpha\frac{k_{\gamma}^{2}}{k_{*}^{4}}\left(\frac{k}{k_{*}}\right)^{-3}\left[\ln\left(\left|k\eta\right|\right)+\frac{3}{2}+\frac{17}{60}\frac{1}{\left(k\eta\right)^{2}}\right]
\end{array}\,,
\end{equation}
\end{itemize}
{Now we turn to the cases $p=2\,,1$}{\large\par}

{\large{}$p=2$ given by }
\begin{equation}
\begin{array}{clc}
B_{vvv}^{\left(3\right)} & =-\alpha\frac{k_{\gamma}^{2}}{k_{*}^{4}}\left(\frac{k}{k_{*}}\right)^{-5}\intop_{-\left(Hl_{E}\right)^{-1}}^{k\eta}dz\left\{ \left(1+\frac{1}{(-z)^{2}}\right)\right.\\
\\
 & \times\left.\left[\cos\left(k\eta-z\right)\left(\frac{1}{k\eta}-\frac{1}{z}\right)+\sin\left(k\eta-z\right)\left(1+\frac{1}{k\eta z}\right)\right]^{3}\right\} 
\end{array}\,,
\end{equation}
{we see two contributions to the above integral }$B_{vvv}^{\left(3\right)}=B_{vvv}^{\left(3\right)}\left|_{1}\right.+B_{vvv}^{\left(3\right)}\left|_{2}\right.$
{\large{}, the first is dominated by the upper bound therefore expanding
the integrand again around $-k\eta\ll1$, $-z\ll1$ leads to}
\begin{equation}
\begin{array}{llc}
B_{vvv}^{\left(3\right)}\left|_{1}\right.=\alpha\frac{k_{\gamma}^{2}}{k_{*}^{4}}\left(\frac{k}{k_{*}}\right)^{-5}\intop_{-\left(Hl_{E}\right)^{-1}}^{k\eta}dz\dfrac{\left(z-k\eta\right)^{3}}{z^{2}}\\
\\
\,\,\,\,\,\,\,\,\,\,\,\,=-3\alpha\frac{k_{\gamma}^{2}}{k_{*}^{4}}\left(\frac{k}{k_{*}}\right)^{-5}\left(k\eta\right)^{2}\left[\ln\left(\left|k\eta\right|\right)-\dfrac{1}{2}\right]\\
\\
\end{array}\,,
\end{equation}
{second contibution is given by}
\begin{equation}
\begin{array}{clc}
B_{vvv}^{\left(3\right)}\left|_{2}\right.= & \alpha\frac{k_{\gamma}^{2}}{k_{*}^{4}}\left(\frac{k}{k_{*}}\right)^{-5}\intop_{-\left(Hl_{E}\right)^{-1}}^{k\eta}dz\left(\cos\left(z-k\eta\right)\left(\dfrac{1}{k\eta}-\dfrac{1}{z}\right)-\sin\left(z-k\eta\right)\left(1+\dfrac{1}{zk\eta}\right)\right)^{3}\\
\\
\end{array}\,,
\end{equation}
{this last could be computed exactly, and since the antiderivative
is too involved we will pick only the leading terms }
\begin{equation}
\begin{array}{llc}
\\
B_{vvv}^{\left(3\right)}\left|_{2}\right.= & \alpha\frac{k_{\gamma}^{2}}{k_{*}^{4}}\left(\frac{k}{k_{*}}\right)^{-5}\left[\frac{3}{4}\frac{Si\left(-\left(Hl_{E}\right)^{-1}\right)}{\left(k\eta\right)^{3}}-\frac{11}{16}\frac{Ci\left(-k\eta\right)}{\left(k\eta\right)^{3}}\right]\\
\\
\end{array}\,,
\end{equation}
{where $Ci\left(x\right)$ and $Si\left(x\right)$ are the
CosIntegral and SinIntegral, respectively, Thus we see that in the
limit $k\eta\rightarrow0$ the $B_{vvv}^{\left(3\right)}\left|_{1}\right.$is
subdominant with respect to $B_{vvv}^{\left(3\right)}\left|_{2}\right.$,
therefore we may safely conclude that }
\begin{equation}
\begin{array}{ccc}
\\
B_{vvv}^{\left(3\right)}= & \alpha\frac{k_{\gamma}^{2}}{k_{*}^{4}}\left(\frac{k}{k_{*}}\right)^{-5}\left[\frac{3}{4}\frac{Si\left(-\left(Hl_{E}\right)^{-1}\right)}{\left(k\eta\right)^{3}}-\frac{11}{16}\frac{Ci\left(-k\eta\right)}{\left(k\eta\right)^{3}}\right]\\
\\
\end{array}\,.
\end{equation}

{Finally we compute $B_{vvv}^{\left(3\right)}$ for $p=1$
which is given by }
\begin{equation}
\begin{array}{clc}
B_{vvv}^{\left(3\right)} & =2\alpha\frac{k_{\gamma}^{2}}{k_{*}^{4}}\left(\frac{k}{k_{*}}\right)^{-6}\intop_{-\left(Hl_{E}\right)^{-1}}^{k\eta}dz\left\{ \left(z-\frac{1}{z}\right)\right.\\
\\
 & \times\left.\left[\cos\left(k\eta-z\right)\left(\frac{1}{k\eta}-\frac{1}{z}\right)+\sin\left(k\eta-z\right)\left(1+\frac{1}{k\eta z}\right)\right]^{3}\right\} 
\end{array}\,,
\end{equation}
{the first integral could be computed exactly, but again for
simplicity we write down the leading terms }
\begin{equation}
\begin{array}{clc}
 & \intop_{-\left(Hl_{E}\right)^{-1}}^{k\eta}dzz\left[\cos\left(k\eta-z\right)\left(\frac{1}{k\eta}-\frac{1}{z}\right)+\sin\left(k\eta-z\right)\left(1+\frac{1}{k\eta z}\right)\right]^{3}\\
\simeq & \frac{16}{9}\frac{\cos\left(Hl_{E}\right)^{-1}}{\left(k\eta\right)^{3}}-\frac{5}{6}\frac{\left(Hl_{E}\right)^{-1}\sin\left(Hl_{E}\right)^{-1}}{\left(k\eta\right)^{3}}\\
\\
\end{array}\,,
\end{equation}
{the second integral receives its main contribution from the
upper bound thus as usual expanding the integrand in the limit $-k\eta\ll1,-z\ll1$
we end up with }
\begin{equation}
\begin{array}{ccc}
 & \intop_{-\left(Hl_{E}\right)^{-1}}^{k\eta}dz\dfrac{\left(z-a\right)^{3}}{z}\\
\simeq & \left(k\eta\right)^{3}\left[\frac{5}{3}-\ln\left(\left|k\eta\right|\right)\right]\\
\\
\end{array}\,,\label{47-1}
\end{equation}
{so in total, and considering only the leading term in (\ref{47-1}),
namely the one $\propto\ln\left(\left|k\eta\right|\right)$, which
becomes now subdominant it the total $B_{vvv}^{\left(3\right)}$}
\begin{equation}
\begin{array}{ccc}
B_{vvv}^{\left(3\right)}= & 2\alpha\frac{k_{\gamma}^{2}}{k_{*}^{4}}\left(\frac{k}{k_{*}}\right)^{-6}\left[\frac{16}{9}\frac{\cos\left(Hl_{E}\right)^{-1}}{\left(k\eta\right)^{3}}-\frac{5}{6}\frac{\left(Hl_{E}\right)^{-1}\sin\left(Hl_{E}\right)^{-1}}{\left(k\eta\right)^{3}}-\left(k\eta\right)^{3}\ln\left(\left|k\eta\right|\right)\right]\\
\\
\\
\end{array}\,,
\end{equation}
{Having finished the computations of of the different $B_{vvv}^{\left(i\right)}$,
we substitute the results found for the different values of $p$ in
(\ref{eq:Fnl}) in order to obtain the decoherence induced\footnote{We are using in the following expressions of $f_{NL}$ the dimensionless curvature power
spectrum $\mathcal{P_{\zeta}}=H^{2}/\left(8\pi^{2}M_{\text{pl}}^{2}\epsilon\right)\simeq2.2\times10^{-9}$} $f_{NL}$. }{\par}
\begin{itemize}
\item $p=1$, in this case the contribution coming from $B_{vvv}^{\left(1\right)}$
is negligible compared to the other two \footnote{We are using in the following expressions $k\eta=\frac{k}{k_{*}}k_{*}\eta=-\frac{k}{k_{*}}e^{-\left(N-N_{*}\right)}$}
\begin{equation}
\begin{alignedat}{1}f_{NL}= & \frac{5}{9\pi\sqrt{\mathcal{P}_{\zeta}}}\frac{\alpha k_{\gamma}^{2}}{k_{*}}\left(\frac{k}{k_{*}}\right)^{-3}\left[\frac{5}{3}\left(Hl_{E}\right)^{-1}\sin\left(Hl_{E}\right)^{-1}\left(\frac{k}{k_{*}}\right)^{-3}e^{3\left(N-N_{*}\right)}\right.\\
\\
 & \left.-\frac{32}{9}\cos\left(Hl_{E}\right)^{-1}\left(\frac{k}{k_{*}}\right)^{-3}e^{\left(N-N_{*}\right)}+\left(Hl_{E}\right)^{-1}\cos^{3}\left(Hl_{E}\right)^{-1}\right]\\
\\
\end{alignedat}
\end{equation}
\item $p=2$, in this case the contribution coming from $B_{vvv}^{\left(1\right)}$
is negligible compared to the other two
\begin{equation}
f_{NL}=\frac{5}{9\pi\sqrt{\mathcal{P}_{\zeta}}}\frac{\alpha k_{\gamma}^{2}}{k_{*}}\left(\frac{k}{k_{*}}\right)^{-2}\left[\cos^{3}\left(Hl_{E}\right)^{-1}+\frac{3}{4}Si\left(-\left(Hl_{E}\right)^{-1}\right)-\frac{11}{16}Ci\left(\frac{k}{k_{*}}e^{-\left(N-N_{*}\right)}\right)\right]
\end{equation}
\item $3\leq p\leq5$, in this case the contribution coming from both $B_{vvv}^{\left(1\right)}$
and $B_{vvv}^{\left(3\right)}$ is negligible compared to the $B_{vvv}^{\left(2\right)}$,
thus we get
\begin{equation}
\label{4.30new}
f_{NL}=\frac{5}{9\pi\sqrt{\mathcal{P}_{\zeta}}}\frac{\alpha k_{\gamma}^{2}}{k_{*}}\left(\frac{k}{k_{*}}\right)^{p-4}\left(Hl_{E}\right)^{p-2}\cos^{3}\left(Hl_{E}\right)^{-1}
\end{equation}
\item $p=6$

\begin{equation}
\begin{alignedat}{1}f_{NL}= & \frac{5}{9\pi\sqrt{\mathcal{P}_{\zeta}}}\frac{\alpha k_{\gamma}^{2}}{k_{*}}\left[\left(\frac{k}{k_{*}}\right)^{2}\left(Hl_{E}\right)^{4}\cos^{3}\left(Hl_{E}\right)^{-1}+\frac{49}{30}\left(\frac{k}{k_{*}}\right)^{3}e^{-\left(N-N_{*}\right)}\right.\\
\\
 & \left.-\frac{17}{60}\frac{k}{k_{*}}e^{-\left(N-N_{*}\right)}-3\left(\frac{k}{k_{*}}\right)^{3}e^{-3\left(N-N_{*}\right)}\left(\text{ln}\left(\frac{k}{k_{*}}\right)+N_{*}-N\right)-\frac{3}{2}\left(\frac{k}{k_{*}}\right)^{3}e^{-3\left(N-N_{*}\right)}\right]\\
\\
\\
\end{alignedat}
\end{equation}

\item for $p>6$, taking into account all the terms, including the subleading,
we get 

\begin{equation}
\begin{alignedat}{1}f_{NL}= & \frac{5}{9\pi\sqrt{\mathcal{P}_{\zeta}}}\frac{\alpha k_{\gamma}^{2}}{k_{*}}\left(\frac{k}{k_{*}}\right)^{3}\left[\left(Hl_{E}\right)^{p-2}\left(\frac{k}{k_{*}}\right)^{p-7}\cos^{3}\left(Hl_{E}\right)^{-1}-7\frac{\left(3p-4\right)e^{-\left(7-p\right)\left(N-N_{*}\right)}}{\left(p-4\right)\left(p-3\right)\left(p-2\right)\left(p-1\right)}\right.\\
\\
 & \left.-6\left(\frac{k}{k_{*}}\right)^{2}\frac{e^{-\left(9-p\right)\left(N-N_{*}\right)}}{\left(p-6\right)\left(p-5\right)\left(p-4\right)}-6\frac{e^{-\left(7-p\right)\left(N-N_{*}\right)}}{\left(p-6\right)\left(p-5\right)\left(p-4\right)}\right]\\
\\
\\
\end{alignedat}
\end{equation}
\end{itemize}

{\bibliographystyle{unsrt}
\bibliography{Paper_draft}
}{\par}
\end{document}